\documentclass[a4paper,11pt]{article}
\pdfoutput=1 
\usepackage{jheppub} 
\usepackage{bm}
\usepackage[T1]{fontenc} 
\usepackage[dvipsnames]{xcolor}
\usepackage{tikz-cd}
\usepackage{comment}
\usepackage{caption}
\usepackage{subcaption}
\usepackage{mathrsfs,bbm}
\usepackage{slashed}
\usepackage{amssymb}
\usepackage[mathscr]{euscript}
\usepackage{accents}
\usepackage[makeroom]{cancel}
\usepackage{empheq}

\usepackage[hyphens]{url} 
\definecolor{linkc}{rgb}{0,0,1}
\usepackage[pagebackref=true,colorlinks=true,linkcolor=linkc,citecolor=linkc,urlcolor=linkc,linktoc=all,pdfusetitle=true]{hyperref}
\renewcommand*{\backref}[1]{}
\renewcommand*{\backrefalt}[4]{{%
		\ifcase #1 
		\or [Cited: pg.~#2.]%
		\else [Cited: pgs. #2.]%
		\fi%
	}}

\usepackage{lineno,enumerate} 
\usepackage[all,2cell]{xy}

\newcommand{\Z}{\mathbb{Z}}

\newcommand{\F}{\mathbb{F}}
\newcommand{\im}{\mathsf{i}} 
\newcommand{\tsf}[1]{\textsf{#1}}
\newcommand{\ov}[1]{\overline{#1}}
\newcommand{\wt}[1]{\widetilde{#1}}
\newcommand{\wh}[1]{\widehat{#1}}
\usepackage{enumerate}
\usepackage{float}
\usepackage{dsfont}
\usepackage{soul}
\usepackage[most]{tcolorbox}
\usepackage{longtable}
\usepackage[bottom,hang,flushmargin]{footmisc}
\usepackage{mdframed}

\newcommand*{\dt}[1]{%
  \accentset{\mbox{\large\bfseries .}}{#1}}

\def\n{\bm{\nabla}}
\def\be{ \begin{equation} }
\def\ee{ \end{equation}}
\usepackage[all,2cell]{xy} 

\newcommand\Tstrut{\rule{0pt}{2.6ex}} 

\def\bmod{\mathsf{\,mod\,}} 

\def\nn{\nonumber\\}

\makeatletter
\newbox\LT@firstfoot
\def\endfirstfoot{\LT@end@hd@ft\LT@firstfoot}
\newdimen\LT@footdiff
\def\LT@start{%
  \let\LT@start\endgraf
  \endgraf\penalty\z@
  \vskip\LTpre\endgraf
  \LT@footdiff-\ht\LT@foot
  \advance\LT@footdiff\ht\LT@firstfoot
  \dimen@\pagetotal
  \advance\dimen@ \ht\ifvoid\LT@firsthead\LT@head\else\LT@firsthead\fi
  \advance\dimen@ \dp\ifvoid\LT@firsthead\LT@head\else\LT@firsthead\fi
  \advance\dimen@ \ht\ifvoid\LT@firstfoot\LT@foot\else\LT@firstfoot\fi
  \dimen@ii\vfuzz
  \vfuzz\maxdimen
  \setbox\tw@\copy\z@
  \setbox\tw@\vsplit\tw@ to \ht\@arstrutbox
  \setbox\tw@\vbox{\unvbox\tw@}%
  \vfuzz\dimen@ii
  \advance\dimen@ \ht
      \ifdim\ht\@arstrutbox>\ht\tw@\@arstrutbox\else\tw@\fi
  \advance\dimen@\dp
      \ifdim\dp\@arstrutbox>\dp\tw@\@arstrutbox\else\tw@\fi
  \advance\dimen@ -\pagegoal
  \ifdim \dimen@>\z@\vfil\break\fi
  \global\@colroom\@colht
  \ifvoid\LT@firstfoot
    \ifvoid\LT@foot
    \else
      \advance\vsize-\ht\LT@foot
      \global\advance\@colroom-\ht\LT@foot
      \dimen@\pagegoal\advance\dimen@-\ht\LT@foot\pagegoal\dimen@
      \maxdepth\z@
    \fi
  \else
    \advance\vsize-\ht\LT@firstfoot
    \global\advance\@colroom-\ht\LT@firstfoot
    \dimen@\pagegoal\advance\dimen@-\ht\LT@firstfoot\pagegoal\dimen@
    \maxdepth\z@
  \fi
  \ifvoid\LT@firsthead\copy\LT@head\else\box\LT@firsthead\fi\nobreak
  \output{\LT@output}%
}
\def\LT@output{%
  \ifnum\outputpenalty <-\@Mi
    \ifnum\outputpenalty > -\LT@end@pen
      \LT@err{floats and marginpars not allowed in a longtable}\@ehc
    \else
      \setbox\z@\vbox{\unvbox\@cclv}%
      \ifdim \ht\LT@lastfoot>\ht\LT@foot
        \dimen@\pagegoal
        \advance\dimen@-\ht\LT@lastfoot
        \ifdim\dimen@<\ht\z@
          \setbox\@cclv\vbox{\unvbox\z@\copy\LT@foot\vss}%
          \@makecol
          \@outputpage
          \setbox\z@\vbox{\box\LT@head}%
        \fi
      \fi  
      \global\@colroom\@colht
      \global\vsize\@colht   
      \vbox
        {\unvbox\z@\box\ifvoid\LT@lastfoot\LT@foot\else\LT@lastfoot\fi}%
    \fi
  \else
    \ifvoid\LT@firstfoot
      \setbox\@cclv\vbox{\unvbox\@cclv\copy\LT@foot\vss}%
      \@makecol
      \@outputpage
      \global\vsize\@colroom
    \else
      \setbox\@cclv\vbox{\unvbox\@cclv\box\LT@firstfoot\vss}%
      \@makecol
      \@outputpage
      \global\advance\@colroom\LT@footdiff
      \global\vsize\@colroom
    \fi
    \copy\LT@head\nobreak
  \fi
}

\newenvironment{claim}{  \begin{mdframed}[linecolor=black!0,backgroundcolor=black!10]\noindent\itshape\ignorespaces}{\end{mdframed}}

\def\CA{{\cal A}}
\def\CB{{\cal B}}
\def\CC {{\cal C}}
\def\CD {{\cal D}}
\def\CE {{\cal E}}
\def\CF {{\cal F}}

\def\CH {{\cal H}}
\def\CI {{\cal I}}

\def\CL {{\cal L}}

\def\CN {{\cal N}}
\def\CO {{\cal O}}
\def\CP {{\cal P}}
\def\CR {{\cal R}}
\def\CV {{\cal V}}
\def\CW {{\cal W}}

\def\CO {{\cal O}}
\def\CZ {{\cal Z}}
\def\CE {{\cal E}}

\def\CH {{\cal H}}
\def\CI {{{\cal I}}}
\def\CB {{\cal B}}
\def\CQ {{\cal Q}}
\def\CS {{\cal S}}
\def\CT {{\cal T}}

\def\CZ{{\cal Z}}

\def\IC{\mathbb{C}}

\def\IP{\mathbb{P}}

\def\IR{{\mathbb{R}}}
\def\IS{{\mathbb{S}}}
\def\IT{{\mathbb{T}}}

\def\IX{{\mathbb{X}}}
\def\IZ{{\mathbb{Z}}}

\def\fg{\mathfrak{g}}

\def\fs{\mathfrak{s}}
\def\ft{\mathfrak{t}}

\def\fs{\mathfrak{s}}
\def\ft{\mathfrak{t}}
\def\fu{\mathfrak{u}}

\def\F\IX{\mathfrak{x}}

\def\F\IX{\mathfrak{X}}

\def\ua{\underline{a}}
\def\ub{\underline{b}}

\def\Ker{{\mr{Ker}}}
\def\Cok{{\mr{Cok}}}

\title{\textsc{Topological Twisting of 4d $\mathcal{N}=2$ Supersymmetric Field Theories}}

\author[a]{Gregory W. Moore,}
\author[a,b]{Vivek Saxena,}
\author[a]{Ranveer Kumar Singh}

\affiliation[a]{New High Energy Theory Center and Department of Physics and Astronomy, Rutgers University, 126 Frelinghuysen Rd., Piscataway, NJ 08855, USA}
\affiliation[b]{C.N. Yang Institute for Theoretical Physics, Stony Brook University, Stony Brook, NY 11794, USA}
\emailAdd{gwmoore@physics.rutgers.edu}
\emailAdd{vivek.hepth@gmail.com}
\emailAdd{ranveer.singh@rutgers.edu}

\abstract{We discuss what topological data must be provided to define topologically twisted partition functions of four-dimensional $\mathcal{N}=2$ supersymmetric field theories. The original example of Donaldson-Witten theory depends only on the diffeomorphism type of the spacetime and 't Hooft fluxes (characteristic classes of background gerbe connections, a.k.a. ``one-form symmetry connections.'') The example of $\mathcal{N}=2^*$ theories shows that, in general, the twisted partition functions depend on further topological data. We describe topological twisting for general four-dimensional $\mathcal{N}=2$ theories and argue that the topological partition functions depend on (a): the diffeomorphism type of the spacetime, (b): the characteristic classes of background gerbe connections and (c): a ``generalized spin-c structure,'' a concept we introduce and define. The main ideas are illustrated with both Lagrangian theories and class $\mathcal{S}$ theories. In the case of class $\mathcal{S}$ theories of $A_1$ type, we note that the different $S$-duality orbits of a theory associated with a fixed UV curve $C_{g,n}$ can have different topological data. 
\\\\
November 21, 2024
}

\begin{document}
\maketitle
\flushbottom


\section{Introduction}\label{sec:Introduction}

This paper discusses topological twisting of $d=4$ $\CN=2$ field theories. 
The first example of such a theory was discovered by Witten \cite{Witten:1988ze}, an example which had momentous implications. The study of 
topologically twisted $d=4$ $\CN=2$ field theories is interesting in physics because it provides examples of computable partition functions and correlation functions within nontrivial $d=4$ QFTs. These ``topological correlators'' have been interesting to mathematicians because they provide invariants of smooth four-manifolds.\footnote{In this paper we work with ``standard'' four-manifolds: They are smooth, compact manifolds without boundary. They will be orientable and oriented, although a choice of orientation is an example of a ``field,'' as discussed below. We will denote a typical standard four-manifold by $\IX$.}
More precisely, they provide invariants of smooth four-manifolds equipped 
with extra topological data. One goal of this paper is to clarify exactly what that extra topological data is. 

In Donaldson-Witten theory (based on $\mr{SU(2)}$ $\CN=2$ supersymmetric Yang-Mills theory) the topological correlators only depend on the orientation and the diffeomorphism type of $\IX$ and a choice of 't Hooft flux for the gauge group. However, the example of $\CN=2^*$ theories shows that in general, the correlators depend on further topological data. In this example, topological twisting \underline{requires} the introduction of an ``ultraviolet'' spin$^\mr{c}$ structure \cite{Labastida:1997rg,Manschot:2021qqe}. (This UV spin$^\mr{c}$ structure should not be confused with the IR spin$^\mr{c}$ structures which are used to express the answer derived from an IR viewpoint as in \cite{Moore:1997pc}). This raises the question: 
\begin{claim}
How does one topologically twist an arbitrary $d=4$ $\CN=2$ field theory, and what topological data do the correlators depend on?
\end{claim}
This paper answers that question for all $d=4$ $\CN=2$ theories and illustrates the general construction for renormalizable Lagrangian theories and for some theories of class $\CS$.

There are many viewpoints on what should be meant by the general notion of ``topological twisting.'' 
They all generalize the original example in \cite{Witten:1988ze}. It would be interesting to relate the twistings we describe to the general viewpoint advocated in \cite{Elliott:2018cbx,Elliott:2020ecf,Elliott:2024jvw}. 
 
We will view \emph{topological twisting} as a specification of background fields such that a supersymmetry,  denoted $\CQ$,    remains unbroken.
Since we do not work in a Hamiltonian framework, we view the supersymmetry as an odd vectorfield on the supermanifold of field configurations. 
We assume $\CQ$ is a scalar since general manifolds $\IX$ do not possess covariantly constant vectors, spinors, etc. $\CQ$   must square to zero on physical operators (and, in the Hamiltonian formulation, on states). For example, in a gauge theory, ``physical operators'' are the gauge invariant operators. In general, there can be different ``inequivalent''  topological twistings of a given theory.  
For example, there are, famously, several inequivalent twistings of $\CN=4$ super Yang-Mills (SYM) \cite{Vafa:1994tf} (which is a special case of an $\CN=2$ SYM), as we recall in section \ref{sec:3twist_n=4}. In our view, these are 3 inequivalent choices of background fields effecting a topological twisting. 

The specification of background fields effecting a topological twisting will be given using the mathematical concept of  \emph{transfer of structure group}.\footnote{In an unpublished note,
\cite{Freed:TopTwistUnpublished}, D. Freed interprets Witten's topological twisting in the framework of the ideas of Cartan and Klein, a viewpoint close in spirit to that taken in this paper.} 
Briefly, given a continuous homomorphism $\varphi: G_1 \to G_2$ between topological groups one can define a functor $\varphi_*$   between categories (in fact, groupoids) of principal bundles with connection over $\IX$ for the domain and codomain groups $G_1$ and $G_2$, respectively. Thus, if $(P_1,\n_1)$ is a principal $G_1$-bundle over $\IX$ with connection $\n_1$ then $\varphi_*(P_1,\n_1)$ will be a principal $G_2$-bundle over $\IX$ with connection $\n_2 = \varphi_*\n_1$. We explain this concept with various examples in Appendix \ref{app:transredstructuregroup}.

In section \ref{sec:GeneralSetup} we describe our general approach to topological twisting of a $d=4$ $\CN=2$ field theory. We explain the implementation for any renormalizable Lagrangian theory in section \ref{sec:gen4dn=2} and for some class $\CS$ theories in sections  
\ref{sec:GenClassS-Twisting} and \ref{sec:a1classS}. We describe a twisting valid for 
generic masses (the background scalar vevs of flavor symmetry vectormultiplets). 
For special values of the background VM's other topological twistings will exist. 

In order to sharpen the main question above, we organize the correlators into a ``function'' (more precisely formal series) defined on the $\CQ$-cohomology of the space of operators. This function will depend on continuous parameters and what we call the 
\emph{topological data}.  The topological data only depends on the (smooth) topology of $\IX$.  

In more technical terms,    the $\CQ$-cohomology of the space of (extended) observables will be denoted $\CO$.  Choosing a basis $O_i$ for the 
vector space $\CO$ with dual basis $x^i$ for $\CO^\vee$, we 
denote the generating function of the correlators as:\footnote{The proper formulation of the generating series involves nontrivial issues related to contact terms \cite{Losev:1997tp,Moore:1997pc}. But these will not affect the present considerations. }
\be 
\mr{Z}^{\mr{tw}}(x) = \big\langle  e^{ \sum_i x^i O_i} \big\rangle  ~.
\ee
We view this as a ``function'' (actually, formal power series) of $x\in \CO^\vee$. 
As we just mentioned, $\mr{Z}^{\mr{tw}}(x)$ will also depend on continuous parameters. 
Examples of such parameters include the superconformal couplings $\tau_{\mr{uv}} \in \CC$, where $\CC$ is the conformal manifold, and the mass parameters 
$m \in (\mathfrak{t}^{\mr{f}})^\vee$, where $\mathfrak{t}^{\mr{f}}$
is the Lie algebra of the group of global ``flavor'' symmetries.\footnote{In Lagrangian theories these can be viewed as equivariant parameters in the equivariant cohomology of moduli spaces of non-Abelian monopole equations.}
Such parameters should be viewed as background fields in the theory, and we will be interested  
in the $\CQ$-invariant background parameters $\CF^{\CQ}$.  
The net result is that the generator of topological correlation functions is a function\footnote{The domain could be a fibration of $\CO^{\vee}$ over $\CF^{\CQ}$.}
\be 
\mr{Z}^{\mr{tw}}: \CO^\vee \times \CF^{\CQ}  \to \mathbb{C} ~. 
\ee
We will show that the function $\mr{Z}^{\mr{tw}}$ depends on the topological data:\footnote{We are assuming there is no BRST anomaly. (See the discussion near \eqref{eq:Q-exact_stress-tensor}.)  For manifolds with $b_2^+=1$, there is such an anomaly. The partition function does depend on the metric but only through the period point, i.e., the self-dual two-form $J$ such that  $J=\star J$ and  $J^2=1$ (the sign of $J$ is resolved by a choice of orientation).  It is piecewise constant as a function of $J$, as has been analyzed in 
detail in special cases  \cite{Moore:1997pc,Manschot:2021qqe}. In the case $b_2^+=0$, one expects continuous metric dependence, but that dependence should be computable. Unfortunately, this has not been investigated in the literature. For $b_2^+>1$, these anomalies do not appear.   }

\begin{enumerate}\itemsep -2pt

\item\label{it:top_data1}  The orientation and diffeomorphism type of $\IX$. 

\item\label{it:top_data2} 't Hooft fluxes (i.e., background connections for ``discrete 1-form symmetries'') 

\item\label{it:top_data3} Generalized spin$^\mr{c}$ structures. 

\end{enumerate}
and \underline{only} depends on these data. No other choices of topological data are needed. 

In Appendix \ref{app:generalized-spin-c}, we define the notion of a generalized spin$^\mr{c}$ structure. 
(We focus on four dimensions. The generalization to other dimensions is obvious.) It is a (mild) generalization of the notion of a spin$^\mr{c}$ structure. One  
replaces the $\mr{U(1)}$ ``factor'' in the definition of the spin$^\mr{c}$ group by a general torus $\IT$, and quotients by a finite Abelian subgroup of $\mr{Spin(4)} \times \IT$ that projects to the $\IZ_2$ subgroup of $\mr{Spin(4)}$ acting trivially on vectorial representations. 

The general arguments of section \ref{sec:GeneralSetup}   apply to all $d=4$ $\CN=2$ field theories, and thus support our main statement. Sections \ref{sec:NonSpin}, \ref{sec:TopData},   \ref{sec:gen4dn=2}, \ref{sec:backind}, and \ref{sec:GenClassS-Twisting} can be viewed as a check on the general reasoning in some important special cases. 
In section \ref{sec:backind} (which is largely a review of known results), we will show that the topological data for twisted Lagrangian theories indeed only consists of the above list of three. 
In sections \ref{sec:GenClassS-Twisting} and \ref{sec:a1classS}, we argue that the same is true of the class $\CS$ theories we consider.

It is interesting to compare these considerations to those of the IR effective theory. Just the way global symmetries and their 't Hooft anomalies are important RG invariants, we would propose that the topological data of twisted partition functions are likewise RG invariants and must be the same in the IR and the UV. After all, from the functorial viewpoint, the IR theory is simply the functor of the UV theory with the metric scaled to infinity. This invariance is tacitly assumed in derivations of topologically twisted partition functions from the IR viewpoint, but it seems worth stating explicitly. It is an interesting question whether it will prove a useful observation in other contexts. 

The above paragraph raises an interesting question: Given the IR effective theory, how could one deduce the topological data needed to define the UV topological theory? Traditionally, the IR effective theory is presented in terms of a prepotential, or, more globally, in terms of special K\"ahler manifold. (See \cite{Freed:1997dp}  for a careful mathematical definition of special K\"ahler geometry.) We therefore ask: How could one deduce the topological data from this geometry? We offer the thought that one place to look is in the measure of the Coulomb branch integral (a.k.a. the ``$u$-plane integral''). One must use several facts about the topological data to check the single-valuedness of the Coulomb branch measure in the examples that have been studied in the literature 
\cite{Moore:1997pc,Marino:1998bm,Manschot:2019pog,Manschot:2021qqe,Aspman:2022sfj}. One may ask if, conversely, the single-valuedness of the Coulomb branch measure is sufficiently strong to determine the topological data. Posing this question in a precise way, to say nothing of answering it, is a subtle and interesting question, but one beyond the scope of the present work. 

Another interesting direction for future research is whether one can discuss topological twisting of 
class $\CS$ theories more thoroughly than we do here using a geometric approach based on branes in 
string/M-theory and supergravity.\footnote{We thank E. Witten for this suggestion.}
This quickly leads one into the theory of $G_2$-manifolds and raises many interesting questions. 

Here is a brief summary of the remainder of the paper:  In section \ref{sec:GeneralSetup}, we describe our approach to topological twisting in very general terms. In section \ref{sec:NonSpin}, we begin an explication of this approach for Lagrangian field theories, first defining with some (but not complete) care the space of fields of the theories. The field content is usually described in terms of the metric, the vector- and hyper- multiplets, the background gauge fields for flavor symmetries, and the background gauge fields for $\mr{R}$-symmetries. But these are all related to each other by a choice of total structure group, which depends on a 
\underline{choice} of Abelian subgroup of the center of the covering group. 
We stress the interconnection in section \ref{sec:cohcond}, where we give cohomological conditions relating cohomology classes associated with the principal bundles associated with the above components. Section \ref{sec:TopData} contains a brief discussion of the distinction between dynamical and background fields in the context of Lagrangian theories. Section \ref{sec:gen4dn=2} gives a prescription for twisting arbitrary Lagrangian theories, and in section \ref{sec:backind}, we review the standard arguments that the partition function is independent of the connections on the bundles associated with background fields (in particular, the metric). In section \ref{sec:GenClassS-Twisting}, we turn to class $\CS$ theories and how these can be topologically twisted. To keep the discussion under control we only consider theories of $A$-type with simple and full punctures. Along the way, we noticed some interesting questions about Gaiotto gluing which do not seem to have been addressed in the literature. In section \ref{sec:a1classS}, we focus on class $\CS$ theories of type $A_1$, which are also Lagrangian theories. We consider how the cohomological conditions of section \ref{sec:cohcond} behave under $S$-duality and show the expected result that they are the same for all theories in a fixed $S$-duality orbit. Curiously, a given theory based on an ultraviolet curve $C_{g,n}$ has several distinct $S$-duality orbits and hence has several different choices of topological twisting data since in general, the data for the different $S$-duality orbits will be different. 

In Appendix \ref{app:Symbols}, we provide a list of symbols used in this paper, together with a brief description and where they first appear. We hope this will make the paper more readable. In Appendix  \ref{app:transredstructuregroup}, we review the notions of transfer and reduction of structure groups of bundles with connection. This simple mathematical concept is of crucial importance here.  Then Appendix \ref{app:generalized-spin-c} gives a careful definition of what we mean by a ``generalized spin$^\mr{c}$ structure'' together with some examples. On several occasions, we need to refer to superconformal multiplets and the superconformal index, so these well-known topics are briefly reviewed in Appendices \ref{app:SCIMultiplets} and \ref{app:SupSymIndex}. In section \ref{sec:GenClassS-Twisting}, we make heavy use of the (Schur limit of the) superconformal index for trinion theories. However, one important class of trinion theories is free Lagrangian theories. Therefore in Appendix 
\ref{app:Lag-ffs-Trinion-Analysis}, we check our statements using standard Lagrangian field theory. 
Finally, in section \ref{sec:GenClassS-Twisting}, we mention an analog, for class $\CS$, of the cohomological conditions of section \ref{sec:cohcond} for Lagrangian theories. Some examples of these are illustrated in Appendix \ref{app:Class-S-Coho-Examples}.


\acknowledgments
We thank M. Albanese, A. Banerjee, C. Closset, A. Debray, D. Freed, I. Garc{\'i}a Etxebarria, J. Manschot, M. Mari{\~n}o, A. Neitzke, S. Razamat, M. Ro\v{c}ek, N. Seiberg, S. Stubbs, Y. Tachikawa, R. Tao, A. Tomasiello, and E. Witten for valuable comments, discussions and comments on the draft. 
The work of  G.M., V.S., and R.K.S. was supported by the US Department of Energy under grant DE-SC0010008.


\section{General Setup For Topological Twisting}\label{sec:GeneralSetup}

Since we wish our discussion to apply to non-Lagrangian theories, it is very useful to adopt the ``functorial'' formulation of QFT going back to the work of Segal in 2d conformal field theory \cite{Segal:1988zk,Segal1988,Segal:2002ei} and Atiyah in topological field theory \cite{Atiyah:1989vu}.\footnote{Atiyah and Segal quote Feynman path integrals as their main motivation. However, as pointed out to us by Y. Tachikawa, the original papers of Schwinger \cite{Schwinger1948} and Tomonaga \cite{Tomonaga:1946zz} formulated relativistic field theory in terms of families of unitary operators $U(\sigma, \sigma')$ associated to spatial slices $\sigma,\sigma'$ and satisfying the basic axioms of the functorial formulation. One important difference is that in the older works, the theories are formulated in Minkowski spacetime, hence on invertible bordisms with a globally defined time direction, and without Wick rotation.} 
The reason is that these formulations make use of the \underline{answer} to a path integral, without presupposing some representation in terms of dynamical fields governed by an action principle. 
For recent reviews, see \cite{FreedCobordism,FreedCBMS}. 
In this approach, one defines a set of ``fields'' $\mathfrak{F}$ - to be thought of as background fields.\footnote{Technically, we should consider ``fields'' to be a sheaf on $\mr{Man}_4^{\mr{op}}$, as explained in \cite{Freed:2012bs,Freed:2013gc}.  }
One considers a bordism category $\mr{Bord}^{\mathfrak{F}}_{\leq 4}$ of manifolds of dimension $\leq 4$ equipped with such fields. Then a ``theory'' is a \underline{choice} of $\mathfrak{F}$ and a monoidal functor 
\be \label{eq:monoidalfunc}
\mr{Z}: \mr{Bord}^{\mathfrak{F} }_{\leq 4} \to  \mr{VECT} ~,
\ee
where $\mr{VECT}$ is a category of vector spaces over $\IC$.\footnote{In modern parlance we are considering a two-step theory. Eventually one wishes to have a fully extended, or fully local theory, but such considerations are far beyond the scope of this paper. Moreover, a fully rigorous formulation for the kinds of interacting field theories we will consider has not been formulated. Rather this is a topic of current research. See \cite{Kontsevich:2021dmb} for a significant contribution to this current research. We are being deliberately vague about the precise nature of the vector spaces. They will need to be infinite-dimensional.}
We stress that we regard the choice of fields, and hence the domain bordism category for $\mr{Z}$ as a \underline{choice} that one makes in defining the theory.

Specializing now to the class of $d=4$ $\CN=2$ field theories the most obvious choice for the fields, which we denote as $\wt{\CF}$,
evaluated on a fixed manifold $\IX$ includes the supermanifold of field configurations of $d=4$ $\CN =2$ conformal supergravity coupled to $\CN=2$ vectormultiplets for the ``flavor'' symmetries.\footnote{For general backgrounds involving $\CN=2$ conformal supergravity, it is natural to also consider couplings to background $\CN=2$ hypermultiplets, but we will not do so.} 
This viewpoint (which was important in \cite{Cushing:2023rha}) is too general for our present purposes, so we will restrict attention to a submanifold where many of the fields in the conformal supergravity multiplet have been set to zero.  For our purposes, the fields will include the orientation of $\IX$ together with the groupoid of principal $\wt{G}$-bundles with connection for the Lie group 
\be\label{eq:3factorGroup}
\wt{G} = \mr{Spin(4)} \times \mr{SU(2)_R} \times G^\mr{f} ~,
\ee
where the group $G^\mr{f}$ is often called the  ``flavor'' group. It is the global internal symmetry group that commutes with supersymmetry.

Let $p_i$, $i=1,2,3$ denote the projection to the three factors 
of the group \eqref{eq:3factorGroup}: 
\begin{equation}\label{eq:defproj}
    \begin{tikzcd}
	        &  \wt{G} \ar[ld, "p_1", swap] \ar[d, "p_2"] \ar[rd, "p_3"] & \\
	         \mathsf{Spin(4)} & \mathsf{SU(2)_R} & G^{\mathsf{f}}
\end{tikzcd}
\end{equation}

Denote the principal $\wt{G}$-bundle with connection over $\IX$ by $(\wt{P},\wt{\n})$. The $G^\mr{f}$ component of the connection, that is,\footnote{We remind the reader that the transfer of structure group and connection is described in Appendix \ref{app:transredstructuregroup}.}
the connection $p_{3,*}(\wt{P},\wt{\n})$, should be enhanced to a full $d=4$ $\CN=2$ vectormultiplet.  The factor $\mr{Spin(4)}$ is to be regarded as the spin covering of the $\mr{SO(4)}$ structure group of the bundle of oriented frames of $T\IX$, 
so $\pi_{\IX,*}(\wt{P}) = \mr{OFr}(T\IX)$ where $\pi_{\IX}$ is a projection on the first factor followed by the standard quotient homomorphism $\mr{Spin(4)} \to \mr{SO(4)}$. Thus we will begin 
with $d=4$ $\CN=2$ theories viewed as a functor  
\be\label{eq:TildeZee}
\wt{\mr{Z}}: \mr{Bord}^{\wt{\CF}}_{\leq 4}  \to \mr{VECT} ~.
\ee
Since this is a functor, $\wt{\mr{Z}}$ will define state spaces. On $\IR^3$ with Euclidean metric and zero background fields for  $\mr{SU(2)_R} \times G^\mr{f}$, the state space will be a unitary representation of the $d=4$ $\CN=2$ Poincar\'e group. Related to this, the local operators of the theory should be in $d = 4$ $\CN=2$ Poincar\'e multiplets.

Note that the domain category of \eqref{eq:TildeZee} only admits 4-manifolds $\IX$ with a spin structure.
The domain category of the theory can be enlarged. Put differently, the theory can be extended to include non-spin manifolds as follows. Consider a subgroup of the center of $\wt{G}$, written  
$C \subset Z(\wt{G})$. Define $\CF$ to be the groupoid of principal $G:= \wt{G}/C$ bundles with connection. We say the theory ``extends'' to a theory  
\be 
\mr{Z}: \mathsf{Bord}_{\leq 4}^{\CF} \to \mathsf{VECT} ~,
\ee
if the diagram
\be\label{eq:FactorFunctor}
    \begin{tikzcd}
        \mathsf{Bord}_{\leq 4}^{\wt{\CF}} \ar[rr, "\wt{\mr{Z}}"]  \ar[rd, "\pi_*", swap] & &
        \mathsf{VECT} \\
        & \mathsf{Bord}_{\leq 4}^{\CF} \ar[ru, "\mr{Z}", swap, blue, dashed] & 
    \end{tikzcd}
\ee
commutes (up to isomorphism of functors). Here $\pi_*: \mathsf{Bord}_{\leq 4}^{\wt{\CF}}\to \mathsf{Bord}_{\leq 4}^{\CF}$  is transfer of structure group using the natural quotient homomorphism 
$$
\pi:\wt{G} \to G ~ . 
$$

A necessary and sufficient condition for this factorization is that the full functor $\wt{\mr{Z}}$ be $C$-invariant. If we just examine the partition function then we note that $C$ acts as a group of automorphisms on the principal $\wt{G}$-bundles with connection over $\IX$. The action on the connection is via a center symmetry: Holonomies are multiplied by central elements. Therefore it acts on the partition function $\wt{\mr{Z}}$ considered as a function of the bundle with connection. The condition is 
\be 
c\cdot \wt{\mr{Z}} = \wt{\mr{Z}} \qquad  \forall ~ c \in C ~ . 
\ee
There can be many choices of $C$. Each choice of $C$ defines a different ``theory.'' A maximal choice is one such that all theories extend to that choice. If $G^\mr{f}$ is semisimple, the center is a finite Abelian group, so there exists a maximal subgroup. 

\begin{remark}
An analogy might clarify our reasoning here. Suppose we are given the partition function of a theory of chiral fermions on $2n$-dimensional Riemannian spin manifolds where $n$ is an integer. But we are given this mathematical structure without being told what fields were used to compute that partition function. How could we deduce that the theory does not extend to non-spin manifolds? Thus, we are given,  for every $2n$-dimensional spin manifold a Hermitian line bundle over the space of Riemannian metrics on that manifold,  and a section of that line bundle. (In fact, the line bundle is the determinant line bundle $\mr{DET}(\slashed{D}^+)$ of the chiral Dirac operator with its Quillen metric, but we are not given that information because we do not know the fields.) We are given the information that the line bundle depends on a spin structure and we are given the action of the center symmetry.  The center of the spin group acts as a group of automorphisms on the spin bundles, and this induces an action with character $\chi$ acting on sections of the line bundle. If the action is nontrivial then we cannot extend to nonspin manifolds.  Indeed, for  $\mr{DET}(\slashed{D}^+)$ the action of $-1$ is given by $\chi(-1)= (-1)^{\mathsf{Ind}(\slashed{D}^+)}$ and is nontrivial. In this way, we see that for a chiral fermion, the functor does not factorize through non-spin manifolds, even though we have not been informed of the nature of the fields. (A more elaborate argument applies to Dirac fermions.) The point of these remarks is that the same reasoning applies to theories where we do not know the fundamental fields, such as the non-Abelian 6d (2,0) theory or generic 4d $\CN=2$ theories of class $\CS$. 
%
%
\end{remark}
\begin{remark}\label{rmk:invonpfvsfunctor}
It is important to remark that the invariance condition on the partition function  
is in principle less stringent than the invariance condition on the full functor. In this paper, we are primarily concerned with the partition function on compact manifolds. 
\end{remark}

In the application to four-manifold invariants, we definitely wish to include manifolds $\IX$ which are not spinnable. Therefore we wish to have principal $G$-bundles over an $\IX$ which is nonspinnable. This implies the necessary condition:  
\be\label{eq:p1-C-condition}
p_1(C) = \langle (-1,-1) \rangle \subset Z(\mr{Spin(4)}) ~,
\ee
where we have used $\mr{Spin(4)} \cong \mr{SU(2)} \times \mr{SU(2)}$. 

Different theories will have different local operator content, but one supermultiplet that is present in all $d=4$ $\CN=2$ local QFTs is the energy-momentum supermultiplet. See \cite{Cordova:2016xhm,Cordova:2016emh,Eberhardt:2020cxo} for details
and Appendix \ref{app:SCIMultiplets}  for a quick summary. 
In order for this multiplet to be invariant under $C$ we must also have 
\be\label{eq:p1p2-C-condition}
(p_1\times p_2)(C) = \langle (-1,-1,-1) \rangle \subset Z(\mr{Spin(4)} \times \mr{SU(2)_R}) ~.
\ee

The original example \cite{Witten:1988ze} had $G^\mr{f}=1$ and was based on a choice of background metric and \tsf{R}-symmetry gauge field obtained by transfer of structure group using a crucial homomorphism 
\begin{align}\label{eq:WittenHomomorphism}
  \varphi^\mr{W}: \mr{SO(4)} &\longrightarrow (\mr{Spin(4)} \times \mr{SU(2)_R})/\langle (-1,-1,-1) \rangle ~ . 
\end{align}
To write it, consider $\mr{SO(4)}$ as the quotient $\mr{Spin(4)}/\IZ_2$ where 
$\mr{Spin(4)} \cong \mr{SU(2)} \times \mr{SU(2)}$ and 
\begin{equation}
 \IZ_2\cong\langle(-1,-1)\rangle\cong\mr{diag}(\IZ_2\times\IZ_2)\subset \IZ_2\times\IZ_2\cong Z(\mr{Spin(4)})~.   
\end{equation}
Thus, an element of $\mr{SO(4)}$ can be denoted as $[(u_1, u_2)]$ where 
square brackets indicate an equivalence class and $u_1, u_2\in \mr{SU(2)}$.
Then we can write the homomorphism \eqref{eq:WittenHomomorphism} as: 
\be \label{eq:varphiW}
 \varphi^\mr{W}\big([ (u_1,u_2)] \big) := [ (u_1,u_2),  u_2] ~,
\ee
where the superscript $\mr{W}$ stands for Witten and will be called the \emph{Witten homomorphism} because it  was essential in the original construction of 
\cite{Witten:1988ze}.

The specification of background fields will be done via 
transfer of structure group along a homomorphism 
\be \label{eq:varphi_tw_bck}
\varphi^{\mr{tw,bck}}:  \underbrace{(\mr{Spin(4)} \times \IT^{\mr{f}})/C^{\mr{tw,bck}}}_{G^{\mr{tw,bck}}} \rightarrow 
(\mr{Spin(4)} \times \mr{SU(2)_R} \times G^\mr{f} )/C  ~,
\ee
where $C^{\mr{tw,back}}$ is a subgroup of the center of $\mr{Spin(4)} \times \IT^{\mr{f}}$.  
If we take   $\IT^\mr{f}$ to be a maximal torus of $G^\mr{f}$ then we can consider a homomorphism 
\be \label{eq:tildevarphi_tw_bck}
\wt \varphi^{\mr{tw,bck}}:  \mr{Spin(4)} \times \IT^{\mr{f}} \rightarrow 
(\mr{Spin(4)} \times \mr{SU(2)_R} \times G^\mr{f} )/C  ~,
\ee
defined by 
\be
\wt \varphi^{\mr{tw,bck}}( (u_1, u_2), \mu) = [ (u_1 , u_2), u_2, \mu] ~.
\ee
and then simply take $C^{\mr{tw,bck}} $ to be the kernel of $\wt \varphi^{\mr{tw,bck}}$.
This is sufficient to be able to define a twisting for all theories.
(For special values of the masses, there will be special twistings.)
Examples such as the case of $\CN=2^*$ theories teach us that 
the existence of a homomorphism   to a quotient group such that 
$C\cdot \wt{\mr{Z}} = \wt{\mr{Z}}$ requires that, in general, $C$  must intersect $G^\mr{f}$ nontrivially.

Other twisting homomorphisms are possible but they are constrained by the requirement that the diagram 
\be\label{eq:WittenHomConstraint}
    \begin{tikzcd}
        G^{\mr{tw,bck}} \ar[r, "\varphi^{\mr{tw,bck}}"] \ar[d, "p_1",swap]   &   \wt{G}/C \ar[d,"p_1\times p_2"] \\
       \mr{Spin(4)}/\langle (-1,-1)\rangle    \ar[r, "\varphi^{\mr{W}}"] &  (\mr{Spin(4)} \times \mr{SU(2)_R})/\langle (-1,-1,-1)\rangle 
    \end{tikzcd} 
\ee
commutes. This is indeed  satisfied by the construction above.

Note that if $p_1$ is the projection on the first factor of $G^{\mr{tw},\mr{bck}}$ then we should impose 
\be \label{eq:gen_spinc_structure}
(p_1)_* (P^{\mr{tw},\mr{bck}}, \n^{\mr{tw},\mr{bck}})\cong  (\mr{OFr}(T\IX), \n^{\mr{LC}} ) ~. 
\ee
When $G^{\mr{tw},\mr{bck}}$ is of the form given in \eqref{eq:varphi_tw_bck}
with $C^{\mr{tw},\mr{bck}}$ a finite abelian subgroup of the center 
we say that such a principal bundle $P^{\mr{tw},\mr{bck}}$ with structure group $G^{\mr{tw},\mr{bck}}$ defines a \emph{generalized spin$^\mr{c}$ structure } on $\IX$. Morever, a connection $\n^{\mr{tw},\mr{bck}}$ satisfying \eqref{eq:gen_spinc_structure}
 is a \emph{generalized spin$^\mr{c}$ connection}. We elaborate on this definition and give some examples in Appendix \ref{app:generalized-spin-c}. We will meet this condition again, in the context of Lagrangian theories in equation \eqref{eq:metric-restriction}. 

Returning to the $d=4$ $\CN=2$ energy-momentum multiplet, the relation (see \eqref{eq:Q-tranformation_stress-tensor-multiplet}) 
\be \label{eq:Q-action_supercurrent}
\{ \CQ_{\alpha A} , G^{B}_{\mu\dt \beta } \} = \sigma^{\nu}_{\alpha\dt\beta} \delta_A{}^{B} T_{\mu\nu} ~,
\ee
implies that, in the presence of background fields obtained by transfer of structure group $\varphi^{\mr{tw,bck}}$ satisfying 
\eqref{eq:WittenHomConstraint}, there will be a vector-field $\CQ$ on the super-manifold of field configurations such that\footnote{Note that \eqref{eq:Q-exact_stress-tensor} is obtained from the second equation of \eqref{eq:Q-tranformation_stress-tensor-multiplet} because in our conventions, the scalar supercharge $\CQ$ is obtained from the \textit{right-handed} spinor supercharge of the $d=4$ $\CN=2$ supersymmetry algebra. We are being slightly schematic here to avoid distracting notation.} 
\be \label{eq:Q-exact_stress-tensor}
T_{\mu\nu} = \{ \CQ, \Lambda_{\mu\nu} \} ~.
\ee
A standard argument, going back to \cite{Witten:1988ze}, shows 
that \eqref{eq:Q-exact_stress-tensor}  leads to the ``topological invariance'' of $\mr{Z}$. ``Topological invariance'' means independence of the choice of Riemannian metric.     The reason is that the derivative of $\mr{Z}$ with respect to the metric $g_{\mu\nu}$ is given by $\mr{Z}$ with an insertion of $T_{\mu\nu}$. If there is no BRST anomaly then insertions of $\CQ$-exact operators must vanish. (As noted above, there \underline{can} be BRST anomalies, leading to metric dependence, when $b_2^+(\IX) \leq 1$.) 

We will refer to theories with domain $\mathsf{Bord}_{\leq 4}^{\CF}$ such that, for all manifolds and background fields in the domain there is a $\CQ$-symmetry, as ``twisted theories.'' 

Similar reasoning shows that a twisted theory has a partition function that is invariant under continuous deformation of the other components of the $G^{\mr{tw,bck}}$ connection. Using the general structure of the conserved current multiplets (which couple to the flavor symmetry connections) the general equation (see \eqref{eq:Q-action_flavor1} and \eqref{eq:Q-action_flavor2})
\be 
\delta_{A}^{~B} \sigma^\mu_{\alpha\dt\beta} J^a_\mu = 
\big\{ \CQ_{A\alpha } , \wt{\psi}^{Ba}_{\dt \beta} \big\} ~,
\ee
becomes, after topological twisting 
\be 
J_\mu^a = \{ \CQ, \Lambda^a_\mu \} ~.
\ee
Here the subscript $a$ is an adjoint index in the Lie algebra of the symmetry. 
Therefore, in the absence of a BRST anomaly, the derivative of the partition function with respect to the connection vanishes. Therefore, the twisted partition function only depends on the three pieces of topological data mentioned in the introduction. 

\begin{remark}\emph{On the terminology ``physical theory'' vs. ``twisted theory.''}
We will use the (common) terminology of ``twisted theory'' and ``physical theory'' in our discussion below. Since we are regarding a ``theory'' as a functor $\mr{Z}$, part of the very definition of a ``theory'' involves a specification of the domain and codomain categories. As we have just mentioned, a ``twisted theory'' has a $\CQ$-symmetry for all objects in the domain category. By a ``physical theory'' we mean a theory where this is not true for all objects in the domain category, although whether these should be considered more ``physical'' is a debatable point, best left to philosophers.

\end{remark} 

\begin{remark}
In the interest of simplicity, our general discussion here has been limited to ``zero-form symmetries.'' It should be extended to include other generalized symmetries. We will indeed do this in the special case of Lagrangian theories in 
sections \ref{sec:NonSpin} -- \ref{sec:backind}.

\end{remark}

\section{Fields For Lagrangian Theories}\label{sec:NonSpin}

In this section, we describe the fields that are needed to define the exponentiated action $e^{-S}$. 
We regard the exponentiated action on $\IX$ as the  partition function on $\IX$
of an invertible field theory \cite{Freed:2004yc}.
We restrict attention to general renormalizable   $d = 4$ $\CN=2$ Lagrangian gauge theories. To define these one makes the following choice of basic data: 
\begin{enumerate}
\item $\wt{G}^{\mr{gauge}}$ : A semisimple simply connected compact Lie group. For simplicity, we will assume it is connected. The gauge group $G^{\mr{gauge}}$ will be a finite quotient of $\wt{G}^{\mr{gauge}}$, as explained below. 

\item $\rho: \wt{G}^{\mr{gauge}} \to \mr{End}_{\IR}(\CR)$ : a quaternionic representation on a real 
vector space $\CR$  equipped with a $G^{\mr{gauge}}$-invariant positive inner product.
The group of orthogonal hyperk\"ahler isometries is denoted $\mr{O}(\rho)$. This data defines a compact Lie group $G^{\mr{f}}$ -- the \emph{flavor group} -- which is defined to be the commutant of $\rho(\wt{G}^{\mr{gauge}})$ in $\mr{O}(\rho)$. 

\item $C^{\mr{phys}}$ : an ``admissible'' subgroup 
 of the center of
\be 
\wt{G}^{\mr{phys}}:= \mr{Spin(4)} \times \mr{SU(2)}_{\mr{R}} \times G^{\mr{f}}\times \wt{G}^{\mr{gauge}}  ~.
\ee
The term ``admissible'' is defined by conditions \eqref{eq:G-constraint2c} and \eqref{eq:InclusionGroup} below.
There are different choices of admissible groups $C^{\mr{phys}}$, and they correspond to different theories.

\end{enumerate}

The set of fields for the theory will include a  principal bundle with connection over $\IX$  for the structure group\footnote{Our discussion at this point has some tangential overlap with recent discussions \cite{Heckman:2022suy,Brennan:2023vsa}.} 
\be 
\label{eq:Gphys}
G^{\mr{phys}} := \big(\mr{Spin(4)} \times \mr{SU(2)}_{\mr{R}} \times G^{\mr{f}} \times \widetilde G^{\mr{gauge}}\big)/C^{\mr{phys}} = \wt{G}^{\mr{phys}}/C^{\mr{phys}} ~.
\ee
We denote the principal bundle with connection by 
\be\label{eq:phys-bundle}
(P^{\mr{phys}} \to \IX, \n^{\mr{phys}}) ~.
\ee
If we think of the invertible field theory (a.k.a. the exponentiated action $e^{-S}$) as a functor 
\be 
\mr{Z}: \mr{Bord}_{\leq 4}^{\CF} \to \mr{VECT} ~,
\ee
then  $\CF(\IX)$ includes the groupoid of principal   $G^{\mr{phys}}$-bundles with connection.

The bundle with connection  $(P^{\mr{phys}} \to \IX, \n^{\mr{phys}})$ contains several pieces of physical data which are usually thought of as independent, but they are not completely independent because of the quotient by the finite group $C^{\mr{phys}}$. 

%

Let $p_i$,  $i=1,2,3,4$ denote the projection to each of the factors of $\wt{G}^{\mr{phys}}$. We have a transfer of structure group for bundles for each of the image groups $p_i(G^{\mr{phys}})$. We denote these by: 
\begin{align}
\label{eq:p-star-bundles}
\begin{split}  
\mr{Fr}(\IX) & := (p_1)_*(P^{\mr{phys}})  ~,\\ 
P^{\mr{R}} & := (p_2)_*(P^{\mr{phys}}) ~,\\ 
P^{\mr{f}} & := (p_3)_*(P^{\mr{phys}}) ~,\\
P^{\mr{gauge}} & := (p_4)_*(P^{\mr{phys}}) ~,
\end{split}
\end{align}
and refer to them respectively as the frame bundle, \tsf{R}-symmetry bundle, flavor bundle, and gauge bundle.

We are now ready to define the notion that  $C^{\mr{phys}}$ is \emph{admissible} as follows:

\begin{enumerate}

\item\label{it:admissible-1}
We can rederive the condition \eqref{eq:p1p2-C-condition} as follows. First of all, we have: 
\be\label{eq:G-constraint2c} 
(p_1 \times p_2)(C^{\mr{phys}}) \subset \langle (-1,-1,-1) \rangle 
\subset Z(\mr{Spin(4)} \times \mr{SU(2)_R} )  \cong \IZ_2^3 ~.
\ee
The reason is that vectormultiplet fields must be singlets under 
the action of $C^{\mr{phys}}$. They are in the adjoint representation of the gauge group and hence are singlets under the center of that group. Moreover, they are singlets under the action of the flavor group. 
Finally, we will want to identify $\mr{Spin(4)}/p_1(C^{\mr{phys}}) \cong \mr{SO(4)}$ with the structure group of $\mr{OFr}(\IX)$, the bundle of oriented frames on $T\IX$. If we wish the domain bordism category to include non-spin manifolds then we must have that 
\be 
(p_1 \times p_2)({C^{\mr{phys}}}) = \langle (-1,-1,-1) \rangle  ~ . 
\ee

\item \label{it:G-const4} If there are hypermultiplets in the theory then 
$C^{\mr{phys}}$ must also be a subgroup of the group: 
\begin{align}\label{eq:InclusionGroup}\begin{split} 
C^{\mr{phys}}_{\mr{max}} &:= 
\{ (\zeta,\zeta,\zeta,\mu,g)~ \vert ~ \zeta\mu\rho(g)= \mathbbm{1}_{{\rm End}_{\IR}(\CR)} \} \\
& \qquad \subset Z(\mr{Spin(4)} ) \times Z(\mr{SU(2)}_{\mr{R}} ) \times Z(G^{\mr{f}})\times Z(\widetilde G^{\mr{gauge}})  ~,
\end{split}
\end{align}
where $\zeta \in \{ \pm 1 \} \cong \IZ_2$ acts on $\CR$ by scalar multiplication. 
The reason for this condition is that fields in the hypermultiplet representation $\rho$ must be singlets under 
$C^{\mr{phys}}$ (and we have used   condition \eqref{eq:G-constraint2c}).  Note that this forces   $C^{\mr{phys}}$ to be a finite Abelian group. We will call $C^{\mr{phys}}_{\mr{max}}$ the \textit{maximal admissible subgroup}. 
\end{enumerate} 

We make a few remarks on the bundles appearing in \eqref{eq:p-star-bundles}: 

\begin{itemize}

\item The gauge bundle $P^{\mr{gauge}}$ is a principal $G^{\mr{gauge}} \cong \wt{G}^{\mr{gauge}}/C^{\mr{grb}}$ bundle, where we define
\be \label{eq:Agrb-def}
C^{\mr{grb}} := p_4(C^{\mr{phys}}) ~.
\ee
The reason for this notation is explained in item \ref{item:gerbecond} below. 

\item The   bundle $\mr{Fr}(\IX)$ is a principal $\mr{SO(4)}$ bundle because
\be \label{eq:spin/p2=so4}
\mr{Spin(4)}/p_1(C^{\mr{phys}}) \cong \mr{SO(4)} ~. 
\ee
Moreover, if we use the transfer of structure group associated to the canonical homomorphism\footnote{\label{foot:pi-def} We define $p_i(G^{\mr{phys}}):= p_i(\widetilde G^{\mr{phys}})/p_i(C^{\mr{phys}})$.}
$p_1: G^{\mr{phys}} \to \mr{SO(4)} \cong \mr{Spin(4)}/\langle (-1,-1)\rangle $ then we require: 
\be\label{eq:metric-restriction}
(p_1)_* (P^{\mr{phys}} \to \IX, \n^{\mr{phys}})  \cong  (\mr{OFr}(T\IX), \n^{\mr{LC}} ) ~,
\ee
where $\n^{\mr{LC}}$ is the torsion-free (Levi-Civita) connection associated to a Riemannian metric on $\IX$.

\item The \tsf{R}-symmetry bundle $P^{\mr{R}}$ is a principal $\mr{SO(3)}_{\mr{R}}$ bundle. 

\item The flavor bundle $P^{\mr{f}}$ is a principal $G^{\mr{f}}/p_{3}(C^{\mr{phys}})$ bundle. 

\end{itemize}

\begin{remark}
\text{}    

\begin{enumerate}

\item For generic hypermultiplet masses it is natural to replace $G^{\mr{f}}$ 
with a maximal torus, so that that $P^{\mr{f}}$ is in fact a $\IT^{\mr{f}}/p_3(C^{\mr{phys}}) $-bundle where $\IT^{\mr{f}}\subset G^{\mr{f}}$ is a maximal torus. In this case, we have restricted the background fields to reducible $G^{\mr{f}}$-connections and assumed a reduction of the structure group to 
its maximal torus.  

\item The existence of $P^{\mr{phys}}$ implies a restriction on characteristic classes of the ``frame,'' ``\tsf{R}-symmetry,'' ``flavor,'' and ``gauge''  bundles derived from $P^{\mr{phys}}$. See section \ref{sec:cohcond} below. 

\end{enumerate}
\end{remark}
 
The exponentiated action of the theory $e^{-S}$,  regarded as an invertible field theory, 
can be defined on a bordism category equipped with a set of fields $\CF^{\mr{phys}}$ that we informally write as:  
\be\label{eq:InformalFieldSpace}
\CF^{\mr{phys}} := \{ \tau, m, (P^{\mr{phys}} \to \IX, \n^{\mr{phys}}), \Phi^{\mr{phys}}, b , \Phi^{\mr{grav}}  \} ~,
\ee 
where 

\begin{enumerate} 

\item $\tau$ is a $\IC$-valued $\mathsf{Ad}(G^{\mr{gauge}} \times G^{\mr{f}})$-invariant bilinear form on $\mr{Lie}(G^{\mr{gauge}} \times G^{\mr{f}})$ subject to the condition that the imaginary part is positive definite.\footnote{This follows from unitarity of the theory evaluated on Minkowski space.} 
$\tau$ enters the classical Lagrangian by defining the Yang-Mills couplings of the gauge theory. ($\tau$ will be replaced by $\tau_{\mr{uv}}$ or $\Lambda$ below in the quantum theory.) 

\item $m \in \mr{Lie}(G^{\mr{f}}) \otimes \IC$, subject to the condition that $[m, m^\dagger]=0$. 
The field $m$ enters the Lagrangian as $\langle m, \mu \rangle$ where $\mu$ is the hyperk\"ahler 
moment map for $G^{\mr{f}}$. 
%
%
One should view $m$ as the background vectormultiplet scalars 
for $G^{\mr{f}}$

\item\label{it:matter-fieldsrep} $\Phi^{\mr{phys}}   \in \Gamma(E_\rho)$ where $E_\rho$ is an associated bundle to $P^{\mr{phys}}$ constructed using $\rho$ and the rules of supersymmetry for constructing vectormultiplets and hypermultiplets. Its sections include fermions, and hence, we should view the space of fields on $\IX$ as an infinite-dimensional supermanifold.\footnote{We will see later that, for suitable background fields, there is a fermionic vectorfield on this supermanifold derived from a scalar supersymmetry with $\CQ^2 = 0$. The space of fields that define such an odd vectorfield is called the nilpotence variety. We will describe a subvariety of the nilpotence variety.}

\item\label{item:gerbecond} The $b$-field is a flat $C^{\mr{grb}}$-gerbe connection\footnote{A gerbe with connection is (up to isomorphism) an element of a suitable differential cohomology group of degree 3.   For a recent pedagogical introduction, see \cite{Hitchin:1999fh,TASIGMVS:2024}.}
in a 5d topological field theory. 
In order to handle properly the conditions on how one sums over 't Hooft fluxes it is important to invoke the quiche picture of finite 
``categorical symmetry'' in field theory. See \cite{Freed:2022qnc} and references therein.
Example 3.15 of \cite{Freed:2022qnc} is particularly relevant, and an elaboration of the relevant issues can be found in   \cite{Freed:2022iao}. One must consider the 4d theory we discuss in this paper as a right boundary theory (on $\{ 1\} \times \IX$)  for a 5d $\pi$-finite topological field theory with target $B^2 C^{\mr{grb}}$ on $[0,1]\times \IX$. There is a field in the right boundary theory which is an isomorphism of the 
restriction of $b$ with the gerbe describing the obstruction to lifting $(p_4)_*(P^{\mr{phys}})$ to a principal $\wt{G}^{\mr{gauge}}$-bundle over $\IX$. This is the origin of the notation $C^{\mr{grb}}$ introduced in equation \eqref{eq:Agrb-def} above. We denote the characteristic class by   $\mu(b) \in H^2(\IX, C^{\mr{grb}})$. 
There is a topological boundary condition $\rho$ (possibly with the insertion of topological defects) 
at the boundary $\{0\} \times \IX$. Depending on the choice of $\rho$ we might or might 
not sum over 't Hooft fluxes in computing physical quantities.\footnote{The pullback of  $b$  to $\IX$ itself can have automorphism so there can be several inequivalent isomorphisms to the gauge theory gerbe on the right (physical) boundary. Therefore we should also sum over $H^1(\IX,C^{\mr{grb}})$. }
We note that we see no obvious reason why one should not take $b$ to be a gerbe for the full group $C^{\mr{phys}}$
and indeed this generalization will be needed in Appendix \ref{app:Class-S-Coho-Examples}. We have not investigated this generalization systematically and doing so might be an interesting line of future enquiry. It might well streamline the presentation in this paper. 

\item\label{rmk:1p2}   $\Phi^{\mr{grav}}$ is short hand for various ``gravitational'' fields. The Riemannian metric is already encoded in the 
connection $\n^{\mr{phys}}$, but we must also choose an orientation of $\IX$, and a truly complete story would include all the fields of background supergravity, including the odd partners of the metric. We will not go into details here, although some relevant constructions can be found in \cite{Cushing:2023rha}. 

\item Explicit formulae for $e^{-S}$ as a function of these fields can be extracted from the standard references:  \cite{Gates:1983nr,Deligne:1999ur,deWit:1984rvr,Freedman:2012zz}.
\end{enumerate}
%
%
\begin{remark}\label{rmk:1p1}
It is sometimes said that, since the supersymmetric gauge theory contains fermions, it must be defined on a spin manifold. Then, after twisting, the twisted theory can also be defined on non-spin manifolds. From the viewpoint of the present paper that reasoning is incorrect. 
In defining a theory one must specify the domain bordism category. Different choices of group $C^{\mr{phys}}$ lead to different choices of bordism category. For example, if we choose $C^{\mr{phys}}$ to be trivial then the domain bordism category must contain only spin manifolds. However, if we choose $C^{\mr{phys}}$ so that $p_1\times p_2$ in equation 
\eqref{eq:G-constraint2c} is surjective then the domain category might include all standard four-manifolds. 
This is just as true for the ``physical theory'' with general background fields as it is for the twisted theory defined below.  (Related remarks can be found in 
section 2.4.2 of \cite{Cordova:2018acb}, and in \cite{Brennan:2023vsa}.)
\end{remark}
\begin{remark}
We have not been careful with the global structure of the space of fields in 
\eqref{eq:InformalFieldSpace}. In fact, the criterion \eqref{eq:BackgroundGerbeCondition} which 
connects the characteristic class of the (4d) boundary value of the (5d) gerbe $b$ with a characteristic class of the flavor gauge fields suggests that there is a nontrivial fibration, and perhaps even a useful 2-group structure \cite{Cordova:2018cvg}. We leave this for future investigation. 
\end{remark}

\section{Cohomological Conditions For The Existence Of $P^{\mr{phys}}$}\label{sec:cohcond}

In this section, we show that the existence of the   bundle $P^{\mr{phys}}$ in equation \eqref{eq:phys-bundle} above implies cohomological conditions linking the background gravity, flavor, and gerbe data. 

Let us assume that the $G^{\mr{phys}}$-bundle $P^{\mr{phys}}\to \IX$ exists. We choose a good cover  of $\IX$, denoted by $\left\{U_p\right\}_{p \in \CP}$, where $\CP$ is some index set. 
Therefore, on double overlaps $U_{pq} := U_{p}\cap U_{q}$, we have transition functions,
%
%
\begin{equation}
t_{p q}^{\mr {phys }}: U_{p q} \longrightarrow G^{\mr {phys }}~,
\end{equation}
that satisfy the cocycle condition on triple overlaps $U_{p q r}:=U_p \cap U_q \cap U_r$,
\begin{equation}\label{eq:cocytpqphys}
t_{p q}^{\mr {phys }}(x) t_{q r}^{\mr {phys }}(x) t_{r p}^{\mr {phys }}(x)=\mathds{1}_{G^{\mr {phys }}}~ ,\quad \forall\, x\in U_{pqr}~.
\end{equation}

We may choose the following transition functions for the bundles
$(p_i)_*(P^{\mr{phys}})$ of equation \eqref{eq:p-star-bundles} on overlaps: 
\begin{equation}\label{eq:choice-t-tilde}
\begin{split}
 \left(\wt{t}_{p q}^{\,+}, \wt{t}_{p q}^{\,-}\right): U_{p q} 
 & \longrightarrow \mr{Spin(4)} \cong \mr{SU(2)} \times \mr{SU(2)} ~, \\
\wt{t}_{p q}^{\,\mr{R}}: U_{p q} &\longrightarrow \mr{SU(2)_{R}}~, \\
 \wt{t}_{p q}^{\,\mr{f}}: U_{p q} & \longrightarrow \mathbb{T}^{\mr{f}} ~,\\
 \wt{t}_{p q}^{\,g}: U_{p q}  & \longrightarrow \wt{G}^{\mr{gauge}} ~, 
\end{split}
\end{equation}
such that\footnote{The equivalence class $[\cdot]$ in \eqref{eq:tphys} arises from the quotient by $C^{\mr{phys}}$, see \eqref{eq:Gphys}.}
\begin{equation}\label{eq:tphys}
t_{p q}^{\mr{phys}}=\left[\left(\wt{t}_{p q}^{\,+}, \wt{t}_{p q}^{\,-}, \wt{t}_{p q}^{\,\mr{R}}, \wt{t}_{p q}^{\,\mr{f}},\wt{t}_{p q}^{\,g}\right)\right]~.
\end{equation}
Note that our choice is ambiguous by 
\be 
\left(\wt{t}_{p q}^{\,+}, \wt{t}_{p q}^{\,-}, \wt{t}_{p q}^{\,\mr{R}}, \wt{t}_{p q}^{\,\mr{f}},\wt{t}_{p q}^{\,g}\right) \sim  
\left(\wt{t}_{p q}^{\,+}, \wt{t}_{p q}^{\,-}, \wt{t}_{p q}^{\,\mr{R}}, \wt{t}_{p q}^{\,\mr{f}},\wt{t}_{p q}^{\,g}\right) \cdot \alpha ~,
\ee
where $\alpha \in C^{\mr{phys}}$. Therefore, the functions \eqref{eq:choice-t-tilde}
need not individually satisfy the cocycle conditions. We can therefore define \v{C}ech-cocycles
\begin{equation}\label{eq:coh-cech-cocycles}
\begin{aligned}
& \zeta_{p q r}^{ \pm}:=\wt{t}_{p q}^{\,\pm} \wt{t}_{q r}^{\, \pm} \wt{t}_{r p}^{\,\pm}~, \\
& \zeta_{p q r}^{\,\mr{R}}:=\wt{t}_{p q}^{\,\mr{R}} \wt{t}_{q r}^{\,\mr{R}} \wt{t}_{r p}^{\,\mr{R}}~, \\
& \zeta_{p q r}^{\,\mr{f}}:=\wt{t}_{p q}^{\,\mr{f}} \wt{t}_{q r}^{\,\mr{f}} \wt{t}_{r p}^{\,\mr{f}}~,\\
& \zeta_{p q r}^g:=\wt{t}_{p q}^{\,g} \wt{t}_{q r}^{\,g} \wt{t}_{r p}^{\,g}~.
\end{aligned}
\end{equation}
Then the cocycle condition \eqref{eq:cocytpqphys} is equivalent to
\begin{equation}\label{eq:cohcondAphys}\boxed{
\big(\zeta_{p q r}^+, \zeta_{p q r}^{-}, \zeta_{p q r}^{\mr{R}}, \zeta_{p q r}^{\mr{f}},\zeta_{p q r}^g\big) \in C^{\mr{phys}}~ .
}
\end{equation}
From \eqref{eq:G-constraint2c},
\begin{equation}\label{eq:R=spincocy}
\zeta_{p q r}^{-}=\zeta_{p q r}^{+} =\zeta_{p q r}^{\mr{R}} ~.
\end{equation}
We will also define $\zeta_{p q r}^s :=\zeta_{p q r}^{-}=\zeta_{p q r}^{+}$ for the cocycle corresponding to the $\mr{Spin(4)}$ factor of $\wt{G}^{\mr{phys}}$.
To derive cohomological conditions from \eqref{eq:cohcondAphys}, we need to identify what each cocycle represents. By definition,
\begin{equation} \label{eq:transfunc-gauge}
t_{p q}^g:=\left[\,\wt{t}_{p q}^{\,g}\right]: U_{p q} \longrightarrow \wt{G}^{\mr{gauge}} / C^{\mr{grb}} \cong G^{\mr{gauge}} ~,
\end{equation}
represents a choice of transition functions on the gauge bundle $P^{\mr{gauge}} \to \IX$. The cocycle $\zeta_{p qr}^g$ represents the 't Hooft flux of the gauge bundle, i.e., the obstruction to lifting the structure group from $ G^{\mr{gauge}} \cong \wt{G}^{\mr{gauge}} /C^{\mr{grb}}$ to $\wt{G}^{\mr{gauge}}$:
\begin{equation}\label{eq:b-char-class}
\mu(b) := \left[\zeta_{pqr}^g\right] \in H^2\big(\IX, C^{\mr{grb}}\big) ~.
\end{equation}
The requirement just below \eqref{eq:G-constraint2c} implies that $\zeta_{pqr}^s$ represents the second Stiefel-Whitney class of $T\IX$:
\begin{equation}\label{eq:sw-class-TX}
w_2(\IX):= w_2(T\IX) = \left[\zeta_{pqr}^s\right] \in H^2\left(\IX, \mathbb{Z}_2\right) ~,
\end{equation}
which is the obstruction to $\IX$ being spin. Similarly, $\zeta_{p q r}^{\mr{R}}$ represents the obstruction to lifting the structure group of the $\mr{R}$-symmetry bundle $P^{\mr{R}}\to \IX$ from $\mr{SO(3)_R}$ to $\mr{S U(2)_R}$: 
\begin{equation}\label{eq:sw-class-PR}
w_2\big(P^{\mr{R}}\big):=\big[\zeta_{pqr}^{\mr{R}}\big] \in H^2\left(\IX, \mathbb{Z}_2\right) ~.
\end{equation}

From \eqref{eq:R=spincocy}, we immediately get the first cohomological condition, namely
\begin{equation}\label{eq:cohcond-w2pr-w2x} \boxed{
w_2(P^{\mr{R}})=w_2(\IX) ~.}
\end{equation}

Whenever we have hypermultiplets in the theory, the condition \eqref{eq:cohcondAphys} 
can be written using item \ref{it:G-const4} above as
\begin{equation}\label{eq:cohcondrho}
\begin{aligned}
& \rho\left(\zeta_{p q r}^g\right) \zeta_{p q r}^s \zeta_{p q r}^{\mr{f}}=1 
\iff  \zeta_{p q r}^{\mr{f}}=\rho\left(\zeta_{p q r}^g\right)^{-1} \zeta_{p q r}^s \in \IT^{\mr{f}} ~ .
\end{aligned}
\end{equation}
Note that \eqref{eq:cohcondrho} is equivalent to \eqref{eq:cohcondAphys} when $C^{\mr{phys}}$ is the maximal admissible subgroup.\footnote{We emphasize that the condition \eqref{eq:cohcondrho} is \underline{not} equivalent to \eqref{eq:cohcondAphys} when $C^{\mr{phys}}$ is \underline{not} the maximal admissible subgroup. 
The simplest counterexample is when $Z(\wt{G}^{\mr{gauge}})$ is a cyclic group with generator $g$. Consider 
\begin{equation*}
C^{\mr{phys}}=\langle (-1,-1,-1,-\rho(g)^{-1},g)\rangle~.    
\end{equation*}
In this case, $C^{\mr{phys}}$ is not the maximal admissible subgroup, and clearly,
\begin{equation*}
    p_4(C^{\mr{phys}}) \cong Z(\wt{G}^{\mr{gauge}}) ~,\quad (p_1\times p_2)(C^{\mr{phys}})\cong \mr{diag}(\Z_2^3)~.
\end{equation*}
Now, if $g$ is not a square in $Z(\wt{G}^{\mr{gauge}})$,  the element $(1,1,1,\rho(g)^{-1},g)\in \Z_2^3\times Z(G^{\mr{f}})\times Z(\wt{G}^{\mr{gauge}})$ satisfies \eqref{eq:cohcondrho} but is not an element of $C^{\mr{phys}}$. More generally, whenever $C^{\mr{phys}}$ is a proper subgroup of the maximal admissible subgroup, \eqref{eq:cohcondrho} is not equivalent to \eqref{eq:cohcondAphys}.}

We can turn the condition \eqref{eq:cohcondrho}
into a cohomological restriction on the background fields as follows. 
Recall that we are choosing generic masses so we are reducing the flavor symmetry 
to its maximal torus. 
Note that as a representation of $\wt{G}^{\mr{gauge}}$ over $\IC$, there is an isotypical decomposition 
(here we \underline{choose} a complex structure on $\CR$): 
\be \label{eq:rho-isotyp}
\rho \cong  \bigoplus_{u} \rho_u ~,
\ee
where each summand $\rho_u$ is an irreducible representation.  
 Also, note that the irreducible representations for $u\not= u'$ might be isomorphic. 
Choose an isomorphism: 
\be \label{eq:ITf-u}
\IT^{\mr{f}} \cong \prod_u \mr{U(1)}_{u} ~.
\ee
We let $p_{4,u}$ denote the projection of $\wt{G}^{\mr{phys}}$ to the $u^{\mathrm{th}}$ component of \eqref{eq:ITf-u}.

Consider any irreducible subrepresentation $\rho_u$ of $\rho$. 
Since $C^{\mr{grb}}$ is finite Abelian, Schur's Lemma\footnote{We are working over $\IC$.} 
implies that the image 
$\rho_u(C^{\mr{grb}} )$ in $\mr{Aut}(\rho_u)$ is 
always a cyclic group acting by scalars. 
This image, together with $-1$,  generates a cyclic subgroup  $\CZ_u \subset \mr{U(1)}$. Note there is a canonical injective homomorphism $\rho_s: \IZ/2\IZ \to \CZ_u$.   Moreover, the condition  \eqref{eq:InclusionGroup} implies that there is an inclusion
$\iota: p_{3,u}( C^{\mr{phys}}) \hookrightarrow    \CZ_u$ and hence $p_{3,u}( C^{\mr{phys}})$ is isomorphic to $\IZ/n_u\IZ$, thus defining a set of integers $\{n_u\}$. 
Next, $(p_{3,u})_*(P^{\mr{phys}},\n^{\mr{phys}})=(P^{\mr{f},u},\n^{\mr{f},u})$ is a well-defined principal $\mr{U(1)}/p_{3,u}(C^{\mr{phys}})$-bundle over $\IX$ with connection. We should think of $\n^{\mr{f},u}$ as the ``background gauge fields for the 
flavor symmetry.''  Let $c_1(u)\in H^2(\IX,\IZ)$ be the first Chern class of the $\mr{U}(1)$-bundle obtained by transfer of structure group of $P^{\mr{f},u}$ along the isomorphism\footnote{\label{foot:power_n_map}Note that this isomorphism is \underline{not} the $n_u^{th}$-power map. More precisely, it is the usual isomorphism obtained from the short exact sequence 
\begin{equation*}
    1\longrightarrow p_{3,u}(C^{\mr{phys}})\stackrel{\iota}{\longrightarrow}\mr{U(1)}\stackrel{P_{n_u}}{\longrightarrow}\mr{U(1)}\longrightarrow 1 ~,
\end{equation*}
where $\iota$ is the inclusion map and $P_{n_u}$ is the $n_u^{th}$-power map.} 
\be 
\mr{U(1)}/p_{3,u}(C^{\mr{phys}})\stackrel{\cong}{\longrightarrow} \mr{U}(1) ~ . 
\ee
Let $r_n: \IZ\to \IZ/n\IZ$ denote reduction modulo $n$. The cocycle $[\zeta^{\mr{f}}_{pqr}] $ is then a representative of $r_{n_u}(c_1(u))$ and is the obstruction to lifting $P^{\mr{f},u}$ to
a $\mr{U(1)}$-bundle along the quotient map $\mr{U(1)} \to \mr{U(1)}/p_{3,u}(C^{\mr{phys}})$.  The existence of $P^{\mr{phys}}$ implies the following cohomological restriction on background fields
(``background 1-form symmetry connections''):\footnote{The second term in \eqref{eq:BackgroundGerbeCondition} is the image of $w_{2}(\IX)$ under the map $\rho_{s}: H^{2}(\IX, \IZ_2) \to H^{2}(\IX, \CZ_u)$ induced by the canonical homomorphism $\rho_{s}: \IZ/2\IZ \to \CZ_u$.}
\be\label{eq:BackgroundGerbeCondition}
\boxed{\rho_{u}(\mu(b) ) + \rho_{s}(w_2(\IX)) + \iota \circ r_{n_u}(c_1(u)) = 0 \in H^2(\IX, \CZ_u)} ~,
\ee
which must hold for all the irreducible subrepresentations of $\rho$.

We now show that \eqref{eq:BackgroundGerbeCondition} is equivalent to the existence of $P^{\mr{phys}}$ when 
$C^{\mr{phys}}$ is the maximal admissible subgroup and the flavor symmetry is $G^{\mr{f}} = \mr{U(1)}$. 
In the case that $G^{\mr{f}} = \mr{U(1)}^d$, we simply take a fiber product. 

Consider now a principal $(\wt{G} / C^{\mr{grb}})$-bundle $P^{\mr{gauge}}$ with 't Hooft flux $\mu(b) \in H^2(\IX, C^{\mr{grb}})$, a $\mr{R}$-symmetry bundle $P^{\mr{R}}$ which is an $\mr{S O(3)_R}$-bundle with second Stiefel-Whitney class $w_2(P^{\mr{R}}) \in H^2\left(\IX, \mathbb{Z}_2\right)$ and a flavor symmetry bundle $P^{\mr{f}}$ which is a $\mr{U(1)} / p_3(C^{\mr{phys}})$-bundle. The condition  \eqref{eq:InclusionGroup} implies that $p_3(C^{\mr{phys}})\hookrightarrow \CZ \subset\mr{U(1)}$ where $\CZ$ is a subgroup generated by $\rho(C^{\mr{grb}})$ and $-1$. There is an isomorphism,
\begin{equation} \label{eq:ndef}
    p_3(C^{\mr{phys}})\cong \IZ/n\IZ~,
\end{equation}
for some $n \in \IZ_{>0}$. We also have the frame bundle $\mr{Fr}(\IX)$ which is an $\mr{S O(4)}$-bundle with second Stiefel-Whitney class $w_2(\IX) \in H^2\left(\IX, \mathbb{Z}_2\right)$. Moreover, assume that the cohomological conditions \eqref{eq:cohcond-w2pr-w2x} and \eqref{eq:BackgroundGerbeCondition} are satisfied. We now show that these bundles can be assembled into a $G^{\mr{phys}}$-bundle $P^{\mr{phys}} \rightarrow \IX$. Indeed, take transition functions
\begin{equation}\label{eq:choice-t}
\begin{aligned}
& t_{p q}^s: U_{p q} \longrightarrow \mr{SO(4)}~, \\
& t_{p q}^{\mr{R}}: U_{p q} \longrightarrow \mr{SO(3)_R}~, \\
& t_{p q}^{\mr{f}}: U_{p q} \longrightarrow \mr{U(1)} / p_3(C^{\mr{phys}})~,\\
& t_{p q}^g: U_{p q} \longrightarrow \wt{G}^{\mr{gauge}} / C^{\mr{grb}}~,
\end{aligned}
\end{equation}
on $P^{\mr{g a u g e}}, \mr{Fr}(\IX), P^{\mr{R}}$ and $P^{\mr{f}}$ respectively, and consider local lifts -- as defined in \eqref{eq:choice-t-tilde} -- to
\begin{equation}\label{eq:local-lifts}
\begin{split}
\left(\wt{t}_{p q}^{\,+}, \wt{t}_{p q}^{\,-}\right): U_{p q} &\longrightarrow \mr{Spin(4)} \cong \mr{SU(2)} \times \mr{SU(2)}~, \\
\wt{t}_{p q}^{\,\mr{R}}: U_{p q} &\longrightarrow \mr{SU(2)_R}~, \\
\wt{t}_{p q}^{\,\mr{f}}: U_{p q} &\longrightarrow \IR~ ,\\
\wt{t}_{p q}^{\,g}: U_{p q} &\longrightarrow \wt{G}^{\mr{gauge}} ~, 
\end{split}
\end{equation}
where the lift $\wt{t}_{p q}^{\,\mr{f}}$ is along the homomorphism
\begin{equation}
\begin{aligned}
\pi: \IR & \longrightarrow \mr{U(1)} / p_3(C^{\mr{phys}}) \\
x & \longmapsto\left[\exp\left(\frac{2 \pi \im x}{ n}\right)\right]~ ,
\end{aligned}
\end{equation}
where $n$ is given by \eqref{eq:ndef}. 
Thus
\begin{equation}
\left[\exp\left(\frac{2 \pi \im \wt{t}_{p q}^{\,\mr{f}}}{n}\right)\right]=t_{p q}^{\mr{f}}~.
\end{equation}
The first Chern class $c_1(\mr{f})\in H^2(\IX,\IZ)$ of the $\mr{U}(1)$-bundle obtained by transfer of structure group of $P^{\mr{f}}$ along the isomorphism $\mr{U(1)} / p_3(C^{\mr{phys}})\cong \mr{U}(1)$ (see footnote \ref{foot:power_n_map}) is represented by the cocycle
\begin{equation}\label{eq:cocycle-c1f}
\eta_{p q r}^{\mr{f}}:=\wt{t}_{p q}^{\,\mr{f}}+\wt{t}_{q r}^{\,\mr{f}}+\wt{t}_{r p}^{\,\mr{f}}: U_{p q r} \longrightarrow \mathbb{Z} ~.
\end{equation}
Also, define
\begin{equation}\label{eq:cocycle-mod-n-c1f}
\zeta_{p q r}^{\mr{f}}:=\exp \left(\frac{2 \pi \im \eta_{p q r}^{\mr{f}}}{n}\right),
\end{equation}
which is the cocycle representing the $\bmod~ n$ reduction of $c_1(\mr{f})$. The cocycles for $\left(\wt{t}_{p q}^{\,-}, \wt{t}_{p q}^{\,+}\right),\wt{t}_{p q}^{\,\mr{R}}$ and $\wt{t}_{p q}^{\,g}$ are denoted by
\begin{equation}
\begin{aligned}
\left(\zeta_{p q r}^{+}, \zeta_{p q r}^{-}\right): U_{p qr} & \longrightarrow \mr{diag}(\mathbb{Z}_2^2)~, \\
\zeta_{p q r}^{\mr{R}}: U_{p qr} & \longrightarrow \mathbb{Z}_2~,\\
\zeta_{p q r}^g: U_{p q r} & \longrightarrow C^{\mr{grb}}=p_4(C^{\mr{p h y s}})~, 
\end{aligned}
\end{equation}
 respectively. 
The condition \eqref{eq:cohcond-w2pr-w2x} implies that, perhaps after a change of local trivialization, $\zeta_{p q r}^{\mr{R}}=\zeta_{p q r}^{ \pm}=\zeta_{p q r}^s$. 
If we take
\begin{equation}
t_{p q}^{\mr {phys }}=\left[\left(\wt{t}_{p q}^{\,+}, \wt{t}_{p q}^{\,-}, \wt{t}_{p q}^{\mr{R}}, t_{p q}^{\mr{f}},\wt{t}_{p q}^{\,g}\right)\right]~ ,
\end{equation}
it follows that
\begin{equation}
t_{p q}^{\mr {phys }} t_{q r}^{\mr {phys }} t_{r p}^{\mr{phys }}=\left[\left(\zeta_{p q r}^{+}, \zeta_{p q r}^{-}, \zeta_{p q r}^{\mr{R}}, \zeta_{p q r}^{\mr{f}},\zeta_{p q r}^g\right)\right]~.
\end{equation}
Note that $\zeta_{p q r}^g$ is valued in $C^{\mr{grb}}$ and $\left(\zeta_{p q r}^{-}, \zeta_{p q r}^{+}, \zeta_{p q r}^{\mr{R}}\right)$ is valued in $\left(p_1 \times p_2\right)\left(C^{\mr {phys }}\right)$ and \eqref{eq:BackgroundGerbeCondition} implies that
\begin{equation}
\rho\left(\zeta_{pqr }^g\right) \zeta_{p q r}^s \zeta_{pqr}^{\mr{f}}=1~ .
\end{equation}
As noted above, for maximal admissible subgroup $C^{\mr {phys }}$, this is equivalent to the condition
\begin{equation}
\left(\zeta_{p q r}^{+}, \zeta_{p q r}^{-}, \zeta_{p q r}^{\mr{R}}, \zeta_{p q r}^{\mr{f}},\zeta_{p q r}^g\right) \in C^{\mr {phys }}~.
\end{equation}
Thus $t_{p q}^{\mr {phys }}$ satisfies the cocycle condition for $G^{\mr{phys}}$ and hence can be used to construct a principal $G^{\mr{phys}}$-bundle $P^{\mr{phys}}$.

We now show that in specific examples, \eqref{eq:BackgroundGerbeCondition} reduces to familiar conditions known from the literature.

\paragraph{1. Pure $\mathcal{N}=2$ Super Yang-Mills (SYM) Theory with gauge group $G^{\mr{gauge}}$:} In this case, $C^{\mr {phys }}$ is constrained to be of the form
\begin{equation}
C^{\mr {phys }}=  \mr{diag}\left(\mathbb{Z}_2^3\right)\times Z ~,
\end{equation}
where $Z$ is some subgroup of $Z(\wt{G}^{\mr{gauge}})$. The only condition that we get from the existence of $P^{\mr {phys }}$ is that
\begin{equation}
w_2(\IX)=w_2(P^{\mr{R}})~. \label{eq:pureN=2SYM}
\end{equation}

\paragraph{2. $\mr{SU(2)}$ $\CN=2$ SYM with fundamental flavors:} For fundamental hypermultiplets, 
\begin{equation}
\rho_u(\mu(b))=: w_2(P^{\mr{gauge}})\in H^2(\IX,\Z_2)~ .
\end{equation}
 If we choose not to ``turn on the flavor bundle'' so that $c_1\left(u\right)=0$, \eqref{eq:BackgroundGerbeCondition} yields\footnote{\label{foot:w2P=w2X}Note that we get the same condition if the physical theory corresponds to the choice 
 \begin{equation*}
C^{\mr{phys}}\cong\langle(-1,-1,-1,1,-1)\rangle~.     
 \end{equation*}
Indeed, in this case \eqref{eq:cohcondrho} implies that $\rho\left(\zeta_{p q r}^g\right)= \zeta_{p q r}^s$ which implies $w_2(P^{\mr{gauge}})=w_2(\IX)$.}
\begin{equation}
w_2(P^{\mr{gauge}})=w_2(\IX)=w_2(P^{\mr{R}}) ~, 
\end{equation}
which was already noted in \cite{Moore:1997pc} for the twisted theory. If background $\mr{U(1)}$ flavor gauge bundles are turned on, \eqref{eq:BackgroundGerbeCondition} reduces to
\begin{equation}
c_1(u) \equiv w_2(\IX) + w_2(P^{\mr{gauge}}) \,\,\,\mr{mod\,} 2 ~,
\end{equation}
which is precisely the condition proposed in \cite{Aspman:2022sfj}. (The condition \eqref{eq:pureN=2SYM} still applies.)
We stress again that -- contrary to the viewpoint found in 
\cite{Aspman:2022sfj,Moore:1997pc} -- these are conditions for the physical theory, 
not just specific to topologically twisted theories.

\paragraph{3. $\mathcal{N}=2^*$ Theory:} Matter in the $\mathcal{N}=2^*$ theory consists of a massive adjoint hypermultiplet. Since the center of the gauge group acts trivially on the adjoint representation, it follows that
\begin{equation}
\rho_u(\mu(b))=0~.
\end{equation}
Thus, \eqref{eq:BackgroundGerbeCondition} yields
\begin{equation}\label{eq:N2*physcohcond}
c_1(u) \equiv w_2(\IX) \,\,\, \mr{mod\,} 2 ~.
\end{equation}
Therefore, $c_1(u)$ can be identified with the characteristic class of a spin$^{\mr{c}}$-structure\footnote{\label{foot:spinc4}Recall that a spin$^{\mr{c}}$ structure on $\mr{Fr}(\IX)$ is a reduction of structure group of $\mr{Fr}(\IX)$ from $\mr{SO(4)}$ to $\mr{Spin^c(4)}=(\mr{Spin(4)}\times\mr{U(1)})/\Z_2$ along the homomorphism $\mr{Spin^c(4)}\to \mr{Spin(4)}/\Z_2\cong\mr{SO(4)}$.}
on $\IX$. This condition was obtained in \cite{Labastida:1997rg,Manschot:2021qqe}. (The condition \eqref{eq:pureN=2SYM} still applies.)
\paragraph{4. $d=4$, $\mathcal{N}=4$ Super Yang-Mills Theory:} \label{ex:4dN=4} Consider the 4d, $\CN=4$ super Yang-Mills theory. The field content consists of an $\CN=2$ vectormultiplet and a massless adjoint hypermultiplet. The theory has an $\mr{S U(4)_R ~R}$-symmetry. So,
\begin{equation}
\wt{G}^{\mr {phys }}=\mr{Spin(4)} \times \mr{SU(4)_R}\times \wt{G}^{\mr{gauge}} ~,
\end{equation}
and now $C^{\mr {phys }}$ is a subgroup of the center of $\wt{G}^{\mr{phys}}$, which is
\begin{equation}
Z(\wt{G}^{\mr {phys }})= \mathbb{Z}_2 \times \mathbb{Z}_2 \times \mathbb{Z}_4\times Z(\wt{G}^{\mr{gauge}}) ~.
\end{equation}
The structure group of the physical theory is $G^{\mr {phys }}=\wt{G}^{\mr {phys }} / C^{\mr {phys }}$. The gauge, frame, and $\mr{R}$-symmetry bundles can be obtained via pushforward by the pertinent projection maps as in the general discussion. 
We also have the identification \eqref{eq:metric-restriction}.

The condition that the fields of the theory be singlets under $C^{\mr{phys}}$ along with \eqref{eq:spin/p2=so4} implies that $C^{\mr {phys }}$ be a subgroup of the maximal admissible group defined by\footnote{The fields in the $\mathcal{N}=4$ vectormultiplet are: a $\wt{G}$-connection, an $\mathsf{Ad}(\wt{G})$-valued Weyl fermion in the \underline{\textbf{4}} of $\mr{S U(4)_R}$ and an $\mathsf{Ad}(\wt{G})$-valued scalar in the \underline{\textbf{6}} of $\mr{SU(4)_R}$. Thus, the invariance of the Weyl fermion under $C^{\mr {phys }}$ dictates the form of $C_{\mr{max}}^{\mr {phys }}$. Scalars are automatically singlets under $C_{\mr{max}}^{\mr {phys }}$.}
\begin{equation}
C_{\mr {max }}^{\mr {phys }}:=\{(\zeta, \zeta, \zeta, g) \in \mathbb{Z}_2 \times \mathbb{Z}_2 \times \mathbb{Z}_4\times Z(\wt{G}^{\mr{gauge}}) \}~,    
\end{equation}
where in the last component, $\zeta$ is understood to be the image of the inclusion $\mathbb{Z}_2 \hookrightarrow \mathbb{Z}_4$. Let us now derive the cohomological condition from the existence of the $G^{\mr {phys }}$-bundle $P^{\mr {phys }}$. Note that the structure group of the $\mr{R}$-symmetry bundle $P^{\mr{R}}$ is $\mr{SU(4)_R} / p_3(C^{\mr {phys }}) \cong \mr{S U(4)_R} / \mathbb{Z}_2$. Thus there is an associated obstruction $w_2(P^{\mr{R}}) \in H^2(\mathbb{X}, \mathbb{Z}_2)$ to lifting this bundle to an $\mr{S U(4)_R}$-bundle. Following the same procedure as above, we find the cohomological condition
\begin{equation}
w_2(\IX)=w_2(P^{\mr{R}}) ~.
\end{equation}
In Section \ref{sec:3twist_n=4}, we will see that this theory admits three different twistings associated with three pairs $\left(G_i^{\mr{t w}}, \varphi_i^{\mr{tw}}: G_i^{\mr{tw}} \longrightarrow G^{\mr {phys }}\right)$ of twisting homomorphisms.

\section{Dynamical And Background Fields In Lagrangian Theories}\label{sec:TopData}

In order to define a quantum theory we need to choose a set of background fields $\CF^{\mr{bck}}$ and a fibration
\be \label{eq:bck-fibration}
\pi: \CF^{\mr{phys}} \to \CF^{\mr{bck}} ~.
\ee
 The map $\pi$ is not necessarily surjective. In principle, some background fields cannot be consistently coupled to the quantum theory. However, for simplicity, we will write subsequent formulae as if  $\pi$ were surjective. 
The \emph{dynamical fields} are the fields in the fiber of $\pi$. Using the exponentiated action $e^{-S}$ to define a (formal) measure 
we (formally)  integrate over the dynamical fields to produce functions of the 
background fields.  Thus the   process of path integration defines a pushforward so that the partition function 
$\mr{Z} := (\pi)_*(e^{-S})$ defines a function on the space of background fields:
\be\label{eq:fibration-dep-lambda}
\begin{tikzcd}
    \CF^{\mr{phys}} \ar[d,"\pi_{\Lambda}",swap] \ar[r, "e^{-S}"] & \IC \\
    \CF^{\mr{bck}} \ar[r, "Z"] & \IC ~.
\end{tikzcd}
\ee
Note that the process of path integration usually introduces a scale $\Lambda$. We can regard $\Lambda$ as one of the background fields, and if we wish to emphasize the scale dependence we write $(\pi_{\Lambda})_*$ for integration above background fields with fixed $\Lambda$. 
In the superconformal case, $\Lambda$ can be replaced by the ultraviolet coupling $\tau_{\mr{uv}}$ which is also a background field, and the path integral will depend on $\tau_{\mr{uv}}$. 

Now we choose our set of background fields to be 
\be \label{eq:CFbck}
\CF^{\mr{bck}} := \{\Lambda, m, (P^{\mr{bck}} \to \IX, \n^{\mr{bck}} ),b, \Phi^{\mr{grav}}  \} ~,
\ee
where 
\be \label{eq:Pbck-nabla-bck}
(P^{\mr{bck}} \to \IX, \n^{\mr{bck}}) := (p_1 \times p_2 \times p_3)_*(P^{\mr{phys}} \to \IX, \n^{\mr{phys}}) ~,
\ee
is a principal bundle with connection  over $\IX$ with  structure group
\be \label{eq:Gbck}
G^{\mr{bck}} :=  \big(\mr{Spin(4)} \times \mr{SU(2)}_{\mr{R}} \times G^{\mr{f}} \big)/C^{\mr{bck}} = (p_1 \times p_2 \times p_3)(G^{\mr{phys}}) ~.
\ee
Note that   $C^{\mr{bck}}$ is a subgroup of the center of $\mr{Spin(4)} \times \mr{SU(2)}_{\mr{R}} \times G^{\mr{f}}$ 
and note that the analog of \eqref{eq:metric-restriction} also holds for $\CF^{\mr{bck}}$. 
We might or might not wish to consider 
some subset of the gerbes $b$ to be dynamical.\footnote{If one adopts the quiche picture \cite{Freed:2022qnc,Freed:2022iao} then the \underline{five}-dimensional field $b$ is always dynamical. Moreover, the isomorphism to the gauge theory gerbe is likewise dynamical. The distinction we discuss in this paragraph is then phrased in terms of choices one makes on the topological boundary.}
The choice of which subgroup of gerbes 
is dynamical leads to different gauge theories for different gauge groups $G^{\mr{gauge}} \cong \wt{G}^{\mr{gauge}}/C^{\mr{grb}}$ with the 
same simply connected cover $\wt{G}^{\mr{gauge}}$. For simplicity, we will consider $b$ to be nondynamical. 
 In general, the fibration $\pi$ is not a direct product because the fiber of dynamical fields depends on the background fields. For example, the set of dynamical gauge fields depends on the 
background gerbe $b$. Moreover, there are consistency conditions on the background fields. See, for example, \eqref{eq:BackgroundGerbeCondition} above.   See also Remark \ref{rmk:invonpfvsfunctor}.  
 
It follows from the above discussion that the set of dynamical fields includes    $\Phi^{\mr{phys}}$ 
along with the connection 
on the $\wt{G}^{\mr{gauge}}/C^{\mr{grb}}$ gauge bundle with connection obtained by transfer of structure group 
associated with the projection homomorphism $p_1$.
%
%
%
%

 In section \ref{sec:gen4dn=2} we will 
explicitly describe one   topological twisting of an arbitrary Lagrangian $d = 4$ $\CN=2$ theory (with generic masses) using a transfer of structure group based on a homomorphism $\varphi^{\mr{tw}}: G^{\mr{tw}} \to G^{\mr{phys}}$, where 
\be 
G^{\mr{tw}} :=  \big( \mr{Spin(4)} \times \IT^{\mr{f}} \times \wt{G}^{\mr{gauge}} \big)/C^{\mr{tw}} ~, \label{eq:Gtw}
\ee
where $C^{\mr{tw}}$ is a subgroup of the center of $\mr{Spin(4)} \times \IT^{\mr{f}} \times \wt{G}^{\mr{gauge}}$. 
The key data that must be specified are the Abelian group $C^{\mr{tw}}$ and 
the homomorphism $\varphi^{\mr{tw}}$. We will take  $C^{\mr{tw}}$ to be a finite group. Moreover, we will require that the projection of 
$C^{\mr{tw}}$ to $Z(\mr{Spin(4)})$ surjects to $\langle (-1,-1)\rangle \cong \IZ_2$  so that we can have equations \eqref{eq:redstgrpPtw} and 
\eqref{eq:gen_spinc_structure} below. For simplicity, we restrict the discussion to the case where $\wt{G}^{\mr{gauge}}$ is a simple group.

The fields of the twisted theory are in part determined by the bundle with connection $(P^{\mr{tw}} \to \IX,\n^{\mr{tw}})$
for structure group $G^{\mr{tw}}$, (which must satisfy a condition analogous to \eqref{eq:metric-restriction}).  
We can write the fields of the twisted theory informally as: 
\be \label{eq:CFtw}
\CF^{\mr{tw}}:= \{ \tau, m, (P^{\mr{tw}} \to \IX, \n^{\mr{tw}} ), \Phi^{\mr{tw}} , b \} ~,
\ee
where $\Phi^{\mr{tw}} \in \Gamma(E^{\mr{tw}})$ and $E^{\mr{tw}} \to \IX$ is a suitable  bundle associated to $P^{\mr{tw}} \to \IX$, 
and 
\be \label{eq:redstgrpPtw}
\varphi^{\mr{tw}}_* (P^{\mr{tw}} \to \IX, \n^{\mr{tw}},\Phi^{\mr{tw}} ) = (P^{\mr{phys}} \to \IX, \n^{\mr{phys}},\Phi^{\mr{phys}} ) ~.
\ee
The choice of $G^{\mr{tw},\mr{bck}}$ and $\varphi^{\mr{tw},\mr{bck}}$ is such that there is a commutative diagram 
%
%
\be\label{eq:twpathint}
\begin{tikzcd}
    \CF^{\mr{tw}} \ar[r, "\varphi_*^{\mr{tw}}"] \ar[d, "\pi_{\Lambda}", swap] & \CF^{\mr{phys}} \ar[r, "e^{-S}"] \ar[d, "\pi_{\Lambda}", swap] & \IC \\
    \CF^{\mr{tw},\mr{bck}} \ar[r, "\varphi_*^{\mr{tw},\mr{bck}}"] & \CF^{\mr{bck}} \ar[r, "\mr{Z}"] & \IC ~,
\end{tikzcd}
\ee
where 
%
\begin{align}
G^{\mr{tw},\mr{bck}} &:= (\mr{Spin(4)} \times \IT^{\mr{f}})/C^{\mr{tw},\mr{bck}} \label{eq:Gtwbck} ~,\\
\CF^{\mr{tw}, \mr{bck}} &:= \{   m, (P^{\mr{tw},\mr{bck}}, \n^{\mr{tw},\mr{bck}}),b \} \label{eq:Ftwbck}~,
\end{align}
has the property that $\mr{Z} \circ \varphi_*^{\mr{tw},\mr{bck}}$  is locally constant as a function of 
$(P^{\mr{tw},\mr{bck}}, \n^{\mr{tw},\mr{bck}})$.  Again, given the key choices of $C^{\mr{tw}}$ and 
  $\varphi^{\mr{tw}}$, the finite Abelian group $C^{\mr{tw},\mr{bck}}$ and the homomorphism 
  $\varphi_*^{\mr{tw},\mr{bck}}$ are determined by the integration process associated to $\pi_\Lambda$. 
Once again, we note that in the superconformal case, $\CF^{\mr{tw}, \mr{bck}}$ also includes the ultraviolet coupling, and the topologically twisted partition function will depend on it. 

Imposing once again equation \eqref{eq:metric-restriction} we see that the partition function of the topologically twisted theory again depends on the generalized spin$^\mr{c}$ structure, and no other background field data, aside from the diffeomorphism type of $\IX$ and the 't Hooft fluxes.

\section{Twisting General Renormalizable Lagrangian 4d $\CN=2$ Theories\label{sec:gen4dn=2}}

In this section, we will describe the choices $C^{\mr{tw}}$ and $\varphi^{\mr{tw}}$  
%
for a general $d=4$ $\CN=2$ gauge theory with a simple gauge group, which give us a topologically twisted version of the theory.  

In order to construct the structure group $G^{\mr{tw}}$ of a topologically twisted theory 
we begin by defining $H:= p_4(C^{\mr{phys}}) \times \IZ_2$ and we construct a homomorphism 
\be \label{eq:phi-rho-G}
\varphi_\rho^G: H \to \IT^{\mr{f}}~, 
\ee
where $\IT^{\mr{f}}$ is a maximal torus of the flavor group. To construct it,
we start with the isotypical decomposition of $\rho$, as in Section \ref{sec:cohcond}, 
\be 
\rho \cong  \bigoplus_{u} \rho_u ~,
\ee
which gives us an isomorphism 
\be
\IT^{\mr{f}} \cong \prod_u \mr{U(1)}_{u}~.
\ee
As in Section \ref{sec:cohcond}, for each irreducible representation $\rho_u$, 
\begin{equation}
\langle\rho_u(p_4(C^{\mr{phys}})),-1\rangle \subset {\rm Aut}(\rho_u) ~,
\end{equation}
is a finite cyclic group isomorphic to  $\Z/{n_{u}}\Z$ for some positive integer $n_u \in \IZ_{>0}$.  
We consider this cyclic group to be the standard cyclic group of roots of unity in $\mr{U(1)}$.  Then we define 
\be 
\varphi_\rho^G(g, \zeta) := \prod_u \left( \rho_u(g^{-1})\zeta \right) ~.
\ee
%
Now define 
\be \label{eq:Gtw-tilde}
\widetilde G^{\mr{tw}}:= \mr{Spin(4)} \times \IT^{\mr{f}}\times \widetilde G^{\mr{gauge}} ~. 
\ee
%
Writing a typical element of $\wt{G}^{\mr{tw}}$ as $((u_1, u_2), \mu,g)$, where $(u_1, u_2) \in \mr{Spin(4)} \cong \mr{SU(2)}\times\mr{SU(2)}$, $\mu \in \IT^{\mr{f}}$ and $g \in \wt{G}^{\mr{gauge}}$, there is a crucial homomorphism based on the Witten homomorphism: 
\begin{align}
\begin{split}
    \varphi: \widetilde G^{\mr{tw}} &\longrightarrow \widetilde G^{\mr{phys}} ~,\\
    \varphi((u_1, u_2), \mu,g) &:= ( (u_1, u_2), u_2, \mu, g) ~, \label{eq:twhomgen} 
\end{split}
\end{align}
%

Now, define an injective homomorphism 
%
%
\be
\begin{aligned} \label{eq:psi-tw}
\psi^{\mr{tw}}: H &\longrightarrow Z(\widetilde G^{\mr{tw}}) \\
 (g,\zeta) &\longmapsto \psi^{\mr{tw}}(g,\zeta) := (\zeta, \zeta , \varphi_\rho^G(g, \zeta), g) ~,
\end{aligned}
\ee
and observe that 
\be \label{eq:phipsitw(H)}
\varphi( \psi^{\mr{tw}}(H)) \subset C^{\mr{phys}} ~,
\ee
when $C^{\mr{phys}}$ is the maximal admissible subgroup.  
In  the notation of Section \ref{sec:NonSpin}, we can take 
$C^{\mr{tw}}= \psi^{\mr{tw}}(H)$ when $C^{\mr{phys}}$ is the maximal admissible subgroup. In general, we will take: 
\begin{equation}\label{eq:AtwdefAphys}
C^{\mr{tw}}=\left\{(\zeta, \zeta, \mu, g):(\zeta, \zeta, \zeta, \mu, g) \in C^{\mr {phys }}\right\}~.
\end{equation}
In any case, $\varphi$ descends to a homomorphism 
\be \label{eq:varphibar}
\ov{\varphi}:  G^{\mr{tw}}:= \widetilde G^{\mr{tw}}/\psi^{\mr{tw}}(H)  \longrightarrow G^{\mr{phys}} = \wt{G}^{\mr{phys}}/ C^{\mr{phys}} ~. 
\ee
and in the notation of Section \ref{sec:NonSpin},
\be
 \varphi^{\mr{tw}}=\ov{\varphi}~.
\ee

Using the condition \eqref{eq:BackgroundGerbeCondition}, one can show that the transition functions 
\begin{equation}\label{eq:tpq-tw}
    \begin{aligned}
 t_{pq}^{\mr{tw}}:=[((\wt{t}^{+}_{pq},\wt{t}^{-}_{pq}),\wt{t}^{\,\mr{f}}_{pq},\wt{t}^{\,g}_{pq})]:U_{pq} &\longrightarrow G^{\mr{tw}}=\frac{\mr{Spin(4)}\times\IT^{\mr{f}}\times \wt{G}^{\mr{gauge}}}{\psi^{\mr{tw}}(H)}~,
    \end{aligned}  
\end{equation}
where the various components are the components of $t^{\mr{phys}}_{pq}$ as in \eqref{eq:tphys}, satisfy the cocycle condition. Thus we can construct a principal $G^{\mr{tw}}$-bundle $P^{\mr{tw}}$ with transition functions $t_{pq}^{\mr{tw}}$. The physical bundle
$P^{\mr{phys}}$ with connection is then obtained by transfer of structure group from the $P^{\mr{tw}}$ bundle with connection under the map induced by the group homomorphism $\ov{\varphi}: G^{\mr{tw}} \to G^{\mr{phys}}$.

\subsection{Three Inequivalent Twistings of 4d $\CN=4$ Super Yang-Mills Theory}\label{sec:3twist_n=4}
In this section, we describe the three inequivalent twistings of 4d,  $\CN=4$ super Yang-Mills theory discussed in \cite{Yamron:1988qc,Vafa:1994tf,Marcus:1995mq,Dabholkar:2020fde}, using the transfer of structure group. The three twisting homomorphisms are based on the three homomorphisms $\mr{Spin(4)} \rightarrow \mr{S U(4)}$ discussed in \cite{Vafa:1994tf}:
\begin{enumerate}
    \item The first homomorphism $\mr{Spin(4)} \rightarrow \mr{S U(4)}$ is chosen such that the \textbf{4} of $\mr{SU(4)}$ decomposes to $(\textbf{2},\textbf{1}) \oplus(\textbf{1},\textbf{2})$ under pullback. This corresponds to the Geometric Langlands (\textsf{GL}) twist \cite{Kapustin:2006pk}, originally discussed in \cite{Yamron:1988qc,Marcus:1995mq}.
\item The second homomorphism $\mr{Spin(4)} \rightarrow \mr{S U(4)}$ is chosen such that the \textbf{4} of $\mr{SU(4)}$ decomposes to $(\textbf{1},\textbf{2}) \oplus(\textbf{1},\textbf{1}) \oplus(\textbf{1},\textbf{1})$ under pullback. This corresponds to the usual Donaldson-Witten (\textsf{DW}) twist, thinking of the $\CN=4$ theory as an $\CN=2$ theory \cite{Witten:1988ze}. 
\item The third homomorphism $\mr{Spin(4)} \rightarrow \mr{S U(4)}$ is chosen such that the \textbf{4} of $\mr{SU(4)}$ decomposes to $(\textbf{1},\textbf{2}) \oplus(\textbf{1},\textbf{2})$ under pullback. This corresponds to the Vafa-Witten (\textsf{VW}) twist \cite{Vafa:1994tf}.
\end{enumerate}
We now describe the twisting group $G^{\mr{t w}}$ and twisting homomorphisms $\varphi^{\mr{t w}}: G^{\mr{t w}} \to G^{\mr {phys }}$ in each case. The notation in this section follows Example \ref{ex:4dN=4} above.
\begin{enumerate}
\item Define
\begin{equation}\label{eq:G-and-C-GL-Twist}
\begin{aligned}
& \wt{G}_1^{\mr{tw}}=\mr{Spin(4)} \times \mr{U(1)_R}\times \wt{G}^{\mr{gauge}}  \\
& C_1^{\mr{t w}}=\{(\zeta, \zeta, 1, g):(\zeta, \zeta, \zeta, g) \in C^{\mr{p h y s}}\}~.
\end{aligned}
\end{equation}
The twisting group is $G_1^{\mr{t w}}=\wt{G}_1^{\mr{t w}} / C_1^{\mr{t w}}$. The twisting homomorphism is given by
\begin{equation}\label{eq:varphi-GL-Twist}
\begin{aligned}
\varphi_1^{\mr{GL}}: \wt{G}_1^{\mr {tw}} & \longrightarrow \wt{G}^{\mr {phys }} \\
\big(\left(u_1, u_2\right), e^{i \theta},g\big) & \longmapsto\big(\left(u_1, u_2\right), \mr{diag}\big(e^{i \theta} u_1, e^{-i \theta} u_2\big),g\big)~,
\end{aligned}
\end{equation}
where $\left(u_1, u_2\right) \in \mr{SU(2)} \times \mr{SU(2)} \cong \mr{Spin(4)}.$ Then $\varphi_1^{\mr{GL}}$ descends to a homomorphism
\begin{equation}\label{eq:varphi-GL-Twist-bar}
\ov{\varphi}_1^{\mr{GL}}: G_1^{\mr{t w}} \longrightarrow G^{\mr {phys }}~.
\end{equation}
The $\mr{U(1)_R}$ factor in $\wt{G}^{\mr{t w}}$ reflects the fact that a $\mr{U(1)}$ subgroup of $\mr{SU(4)}$ commutes with the image of the first homomorphism $\mr{Spin(4)} \longrightarrow \mr{S U(4)}$ and hence is a residual $\mr{R}$-symmetry of the twisted theory. 
\item Define
\begin{equation}\label{eq:G-and-C-DW-Twist}
\begin{aligned}
& \wt{G}_2^{\mr{tw}}= \mr{Spin(4)} \times \mr{U(2)_R}\times \wt{G}^{\mr{gauge}} ~,\\
& C_2^{\mr{t w}}=\left\{(a, a, 1, g):(a, a, a, g) \in C^{\mr {phys }}\right\}~.
\end{aligned}
\end{equation}
The twisting group is $G_2^{\mr{t w}}=\wt{G}_2^{\mr{t w}} / C_2^{\mr{t w}}$. The homomorphism is given by
\begin{equation}\label{eq:varphi-DW-Twist}
\begin{aligned}
\varphi_2^{\mr{DW}}: \wt{G}_2^{\mr {tw}} & \longrightarrow \wt{G}^{\mr {phys }} \\
\big(\big(u_1, u_2\big), u, g\big) & \longmapsto\big(\big(u_1, u_2\big), \mr{diag}\big(\tfrac{u_2}{\sqrt{\mr{det} u}}, u\big), g\big)~.
\end{aligned}
\end{equation}
Then $\varphi_2^{\mr{DW}}$ descends to a homomorphism
\begin{equation}\label{eq:varphi-DW-Twist-bar}
\ov{\varphi}_2^{\mr{DW}}: G_2^{\mr{t w}} \longrightarrow G^{\mr {phys }}~.
\end{equation}
The $\mr{U(2)_R}$ factor in $\wt{G}^{\mr {tw }}$ reflects residual $\mr{R}$-symmetry of the twisted theory.
\item Define
\begin{equation}\label{eq:G-and-C-VW-Twist}
\begin{aligned}
& \wt{G}_3^{\mr {tw }}=\mr{Spin(4)} \times \mr{SU(2)}_R\times \wt{G}^{\mr{gauge}} ~,\\
& C_3^{\mr{tw}}=\left\{(\zeta, \zeta, 1, g):(\zeta, \zeta, \zeta, g) \in C^{\mr {phys }}\right\}~.
\end{aligned}
\end{equation}
The twisting group is $G_3^{\mr{t w}}=\wt{G}_3^{\mr{t w}} / C_3^{\mr{t w}}$. To define the twisting homomorphism, introduce the map
\begin{align}
    \begin{split}
f: \mr{SU(2)} \times \mr{SU(2)_R} &\longrightarrow \mr{SU(4)_R} \\
 \left(u,\left(\begin{smallmatrix}
\alpha & \beta\\-\bar{\beta} &
\bar{\alpha}
\end{smallmatrix}\right)\right) &\longmapsto\left(\begin{smallmatrix}
\alpha u & \beta u \\
\bar{\beta} u & \bar{\alpha} u
\end{smallmatrix}\right)~.
\end{split}
\end{align}
It is easy to check that $f$ is a well-defined homomorphism. We define the twisting homomorphism
by
\begin{equation}\label{eq:varphi-VW-Twist}
\begin{aligned}
\varphi_3^{\mr {VW}}: \wt{G}_3^{\mr{tw}} & \longrightarrow \wt{G}^{\mr {phys }} \\
\left(\left(u_1, u_2\right), u, g\right) & \longmapsto\left(\left(u_1, u_2\right), f\left(u_2, u\right), g\right)~.
\end{aligned}
\end{equation}
Clearly, $\varphi_3^{\mr{VW}}$ descends to a homomorphism
\begin{equation}\label{eq:varphi-VW-Twist-bar}
\ov{\varphi}_3^{\mr{VW}}: G_3^{\mr{t w}} \longrightarrow G^{\mr {phys }}~.
\end{equation}
Again, the twisted theory has residual $\mr{S U(2)_R~ R}$-symmetry.
\end{enumerate}

\section{Independence Of Topological Correlators Under Continuous Deformation Of Background Connection}\label{sec:backind}

In equation \eqref{eq:WittenHomConstraint} et. seq. we gave a general argument that 
backgrounds defined via transfer of structure group satisfying \eqref{eq:WittenHomConstraint} will have partition functions that are invariant under 
continuous deformations of the background connections. In this section, we review the more traditional argument that applies to Lagrangian field theories.   We begin by spelling out the sublocus of topologically twisted fields and the twisted action in more detail. 

The rules of $d=4$ $\CN=2$ supersymmetry dictate that $\Phi^{\mr{phys }}$ be a section of the direct sum of $G^{\mr{phys }}$-vector bundles associated to the representations listed in Tables \ref{tab:physvecfield} and \ref{tab:physhypfield}.
\begin{small}
\begin{table}[H]
\centering
\begin{tabular}{|c|c|}
\hline Description & Representation under $G^{\mr{phys }}$ \\
\hline scalar & $(\bm{1}, \bm{1}, \bm{1}, \bm{1},\mr{adj})$ \\
scalar & $(\bm{1}, \bm{1}, \bm{1}, \bm{1},\mr{adj})$ \\
Weyl spinor & $(\bm{2}, \bm{1}, \bm{2}, \bm{1},\mr{adj})$ \\
Weyl spinor & $(\bm{1}, \bm{2}, \bm{2}, \bm{1},\mr{adj})$ \\
Auxiliary field & $(\bm{1}, \bm{1}, \bm{3}, \bm{1},\mr{adj})$ \\
\hline
\end{tabular}
\caption{Matter fields of an untwisted off-shell $\mr{G}^{\mr{gauge}}$ vectormultiplet. The scalars are not related by complex conjugation.}
\label{tab:physvecfield}
\end{table}
\begin{table}[H]
\centering
\begin{tabular}{|c|c|}
\hline Description & Representation under $G^{\mr{phys }}$ \\
\hline Scalar & $(\bm{1}, \bm{1}, \bm{2}, \mr{def},\rho)$ \\
Weyl spinor & $(\bm{2}, \bm{1}, \bm{1}, \mr{def},\rho)$ \\
Weyl spinor & $(\bm{1}, \bm{2}, \bm{1}, \mr{def},\rho)$ \\
\hline
\end{tabular}
\caption{Fields of an untwisted on-shell hypermultiplet. Note that $\rho$ is a quaternionic representation of $\wt{G}$ and $\mr{def}$ denotes the defining representation of $\IT^{\mr{f}}$. The condition that the tuple in the second column is a representation of $G^{\mr{phys}}$ fixes the charges of various hypermultiplet fields under the $\mr{U(1)}_u$ factor of $\IT^{\mr{f}}$ to be integral multiples of $\frac{1}{n_u}$.
}
\label{tab:physhypfield}
\end{table} 
\end{small}
The  $G^{\mr{phys }}$-connections $\n^{\mr{phys}}$ form a torsor for $\Omega^1\left(P^{\mr{ph y s}},\mr{ad}\,P^{\mr{phys }}\right)$.
Only the $G^{\mr{gauge}}$-connection is dynamical. The other components of the $G^{\mr{phys}}$-connection correspond to background fields. Topological twisting then picks out a sublocus of those background connections consistent with the transfer of structure group \eqref{eq:redstgrpPtw} as will be explained below. 

To determine the sublocus of topologically twisted matter fields, we pull back the representations in Tables \ref{tab:physvecfield} and \ref{tab:physhypfield} to representations of $G^{\mr{t w}}$ by precomposing with $\varphi^{\mr {tw }}$. By a simple group theory argument, we obtain the irreducible representations of $G^{\mr{t w}}$ listed in Tables \ref{tab:twsvecfield} and \ref{tab:twshypfield}. We choose a complex structure on the quaternionic representation $\rho$, so that 
\begin{equation}\label{eq:decomp_rho}
    \rho\cong \varrho\oplus\ov{\varrho}~,
\end{equation}
where $\ov{\varrho}$ is the complex conjugate representation of $\varrho$. Various matter fields in Table \ref{tab:physhypfield} can then be decomposed under the decomposition \eqref{eq:decomp_rho}. The fields $M_{\dt{\alpha}}$ and $\wt{M}_{\dt{\alpha}}$ are the pullbacks of the hypermultiplet scalars decomposed into two $\mf{su(2)}_{\mr{R}}$-doublet scalars.
In particular, $M_{\dt{\alpha}}$ and $\wt{M}_{\dt{\alpha}}$ are related by complex conjugation. 
\begin{small}
\begin{table}[H]
\centering
\begin{tabular}{|l|c|c|c|}
\hline Description & Local chart component & Grassmann parity & Rep. under $G^{\mr{tw }}$ \\
\hline Scalar & $\phi \in \Omega^{0}(\IX, \mathsf{ad\,}P^{\mr{phys}})$ & even & $(\bm{1}, \bm{1}, \bm{1},\mr{adj})$ \\
Scalar & $\wt{\phi} \in \Omega^{0}(\IX, \mathsf{ad\,}P^{\mr{phys}})$ & even & $( \bm{1}, \bm{1}, \bm{1},\mr{adj})$ \\
1-form gaugino & $\psi_\mu \in \Pi\,\Omega^{1}(\IX, \mathsf{ad\,}P^{\mr{phys}})$ & odd & $(\bm{2}, \bm{2}, \bm{1},\mr{adj})$ \\
0-form gaugino & $\eta \in \Pi\,\Omega^{0}(\IX, \mathsf{ad\,}P^{\mr{phys}})$ & odd & $(\bm{1}, \bm{1}, \bm{1},\mr{adj})$ \\
SD 2-form gaugino & $\chi_{\mu \nu} \in \Pi\,\Omega_{g}^{2,+}(\IX, \mathsf{ad\,}P^{\mr{phys}})$ & odd & $(\bm{1, 3, 1},\mr{adj})$ \\
SD 2-form auxiliary field & $D_{\mu \nu} \in \Omega_{g}^{2,+}(\IX, \mathsf{ad\,}P^{\mr{phys}})$ & even & $(\bm{1, 3, 1},\mr{adj})$ \\
\hline
\end{tabular}
    \caption{Matter fields of the twisted off-shell $\CN=2$ vectormultiplet.}
    \label{tab:twsvecfield}
\end{table}
\end{small}
\begin{small}
\begin{table}[H]
    \centering
   \begin{tabular}{|l|c|c|c|}
\hline Description & Local chart component & \begin{tabular}{@{}c@{}}Grassmann \\ parity \end{tabular} & Rep. under $G^{\mr{t w}}$ \\
\hline Monopole field & $M_{\dt{\alpha}} \in \Gamma( S^- \otimes E^{\mr{tw}})$ & even & $(\bm{1}, \bm{2},  \mr{def},\varrho)$ \\
Conjugate monopole field & $\wt{M}^{\dt{\alpha}} := \big(M_{\dt{\alpha}}\big)^{\dagger}  \in \Gamma( S^- \otimes \ov{E}^{\mr{tw}})$ & even & $(\bm{1}, \bm{2},  \overline{\mr{d e f}},\ov{\varrho})$ \\
Weyl fermion & $\mu_\alpha  \in \Pi\,\Gamma( S^+ \otimes E^{\mr{tw}})$ & odd & $(\bm{2}, \bm{1},  \mr{def},\varrho)$ \\
Weyl fermion & $\wt{\lambda}_{\dt{\alpha}}  \in \Pi\,\Gamma( S^- \otimes E^{\mr{tw}})$ & odd & $(\bm{1}, \bm{2},  \mr{d e f},\varrho)$ \\
Weyl fermion & $\wt{\mu}_{\dt{\alpha}}  \in \Pi\,\Gamma( S^- \otimes \ov{E}^{\mr{tw}})$ & odd & $(\bm{1}, \bm{2},  \overline{\mr{d e f}},\ov{\varrho})$ \\
Weyl fermion & $\lambda_\alpha  \in \Pi\,\Gamma(S^+ \otimes \ov{E}^{\mr{tw}})$ & odd & $(\bm{2}, \bm{1},  \ov{\mr{def}},\ov{\varrho})$ \\
Auxiliary Weyl spinor & $h_\alpha  \in \Gamma(S^+ \otimes E^{\mr{tw}})$ & even & $(\bm{2}, \bm{1},  \mr{def},\varrho)$ \\
Auxiliary Weyl spinor & $\wt{h}_\alpha  \in \Gamma( S^+ \otimes \ov{E}^{\mr{tw}})$ & even & $(\bm{2}, \bm{1},  \ov{\mr{def}},\ov{\varrho})$ \\
\hline
\end{tabular}
    \caption{Fields of the twisted off-shell $\CN = 2$ hypermultiplet. }
    \label{tab:twshypfield}
\end{table}
\end{small}
These fields are collectively denoted by $\Phi^{\mr{tw}}$. 
Finally there is a $G^{\mr{tw}}$-connection $\n^{\mr{tw}}$ on $P^{\mr{tw}}$ satisfying the requirement analogous to \eqref{eq:metric-restriction}.
The transfer of structure group from $P^{\mr{tw}}$ to $P^{\mr {phys }}$ along the homomorphism $\varphi^{\mr{tw}}:G^{\mr {tw }}\to G^{\mr {phys }}$ 
then determines the connection $\varphi^{\mr{tw}}_*(\n^{\mr{tw}})$ on $P^{\mr {phys }}$.  
The background fields in the physical theory are determined by the pushforward \eqref{eq:redstgrpPtw}. The connections that enter the action are given by pushforwards of $\varphi^{\mr{tw}}_*(\n^{\mr{tw}})$ via the transfer of structure group of $P^{\mr{phys}}$ to $P^{\mr{gauge}}$, $\mr{Fr}(\IX)$, $P^{\mr{R}}$ and $P^{\mr{f},u}$. These connections are listed in Table \ref{tab:twconnGXRF}. 
\begin{table}[H]
    \centering
\begin{tabular}{|c|c|c|c|}
\hline Connection & Bundle & Local chart component & Dynamical/Background \\
\hline
$\left(p_1\right)_*\left(\varphi_*^{\mr{t w}}\left(\n^{\mr{t w}}\right)\right)$ & $\mr{Fr}(\IX)$ & $\omega_\mu$ & Background \\
$\left(p_2\right)_*\left(\varphi_*^{\mr{t w}}\left(\n^{\mr{t w}}\right)\right)$ & $P^{\mr{R}}$ & $\left(\omega_{+}\right)_\mu$ & Background \\
$\left(p_{3, u}\right)_*\left(\varphi_*^{\mr{t w}}\left(\n^{\mr{t w}}\right)\right)$ & $P^{\mr{f},u}$ & $\CA_\mu^u$ & Background \\$\left(p_4\right)_*\left(\varphi_*^{\mr{t w}}\left(\n^{\mr{t w}}\right)\right)$ & $P^{\mr {gauge }}$ & $A_\mu$ & Dynamical \\
\hline
\end{tabular}
    \caption{Various dynamical and background connections in the physical theory obtained by pushforward of $\n^{\mr{tw}}$, see Section \ref{sec:cohcond} for the definition of various bundles. Here $\omega_{+}$ denotes the self-dual part of the Levi-Civita connection.}
    \label{tab:twconnGXRF}
\end{table}
Note that the requirement analogous to \eqref{eq:metric-restriction} for $(P^{\mr{tw}},\n^{\mr{tw}})$
forces the connection $\omega_\mu$ on $\mr{Fr}(\IX)$ to be the LC connection $\omega^{\mr{LC}}$ and the twisting homomorphism $\varphi^{\mr{tw}}$ forces the connection on the $\mr{R}$-symmetry bundle $P^{\mr{R}}$ to be the SD part $\omega^{\mr{LC}}_{+}$ of the LC connection.\par
The supercharges are in the representation $(\bm{1}, \bm{2}, \bm{1}, \bm{2}, \bm{1})\oplus (\bm{1}, \bm{1}, \bm{2}, \bm{2}, \bm{1})$ of $G^{\mr {phys}}$. They can be pulled back to representations of $G^{\mr{t w}}$. Again, simple group theory arguments show that there is a supercharge $\s{Q}$ singlet under $G^{\mr{t w}}$. The supersymmetry algebra implies that
\begin{equation}
\CQ^2=-\frac{1}{2} Z ~,
\end{equation}
where $Z$ is the generator of the central charge in the physical theory (the action of which on fields is given in Table \ref{tab:centcharge}; fields of the vectormultiplet are neutral under $Z$). Using the action of supercharges in the physical theory, we obtain the action of $\CQ$ on $\Phi^{\mr{t w}}$ and on the connections in Table \ref{tab:twconnGXRF}.

One should think of $\CQ$ as an odd vectorfield on the supermanifold of all fields, background, and dynamical. There is a sub-supermanifold where the background fermions are set to zero, and we will write down the action of $\CQ$ on this submanifold. 
There is no off-shell formulation of an untwisted $\CN=2$ hypermultiplet with a finite number of auxiliary fields, but there exists one for the twisted hypermultiplet \cite{Karlhede:1988ax}.\footnote{The fact that the untwisted off-shell $\CN=2$ hypermultiplet requires infinitely many auxiliary fields is perhaps most elegantly seen in the formalism of projective superspace \cite{Lindstrom:1989ne,Lindstrom:2009afn}: here, the hypermultiplet superfield is an infinite series in a twistorial coordinate $\zeta \in \IC\IP^{1}$, with infinitely many coefficients corresponding to the infinitely many auxiliary fields.
%
The off-shell twisted $\CN=2$ hypermultiplet with a \underline{finite} number of auxiliary fields was given in \cite{Karlhede:1988ax}, the authors having arrived at it by explicit construction. It would be interesting to understand this from a projective superspace viewpoint, but we leave this for future work. We thank M. Ro\v{c}ek for a discussion about this point. Finally, we note that among the two presentations of the off-shell twisted $\CN=2$ hypermultiplet transformations in the book \cite{Labastida:2005zz}, their equations (5.51) do not close off-shell: in their language, the supersymmetry variation $\delta$ satisfies $\delta^{2}\varphi \not\sim \varphi$ and, in fact, $\delta^{4}\varphi \sim \varphi$ for a typical hypermultiplet field $\varphi$. Their second presentation, namely equations (5.56), does close off-shell, and that is what we use.} One introduces spinorial, Grassmann-even auxiliary fields $h^{\alpha}$ and $\wt{h}^{\alpha}$ \cite{Karlhede:1988ax,Labastida:2005zz,Hyun:1995mb}. 
The $\CQ$-action on fields is:
\begin{align}
\begin{aligned}
\label{eq:sQonfields}
[\CQ, \phi] &=0 ~, & [\CQ, \wt{\phi}] &=\eta ~, & \{\CQ, \eta\} &=[\phi, \wt{\phi}] ~, \\
\left[\CQ, A_\mu\right] &=\psi_\mu ~, & \left\{\CQ, \psi_\mu\right\} &=-D_\mu \phi~,&\left[\CQ, \CA^u_\mu\right] &= 0~,\\
\left\{\CQ, \chi_{\mu\nu}\right\} &= F_{\mu\nu}^{+}-D_{\mu\nu} ~, & \left[\CQ, D_{\mu\nu}\right] &=2\big(D_{[\mu} \psi_{\nu]}\big)^+ -\left[\phi, \chi_{\mu\nu}\right] ~,&\left[\CQ, \omega_\mu\right] &=0~,\\
\big[\CQ, M_{\dt{\alpha}}\big] &=\wt{\mu}_{\dt{\alpha}} ~, & \big[\CQ, \wt{M}_{\dt{\alpha}}\big] &=\wt{\lambda}_{\dt{\alpha}} ~, \\
\left\{\CQ, \wt{\mu}_{\dt{\alpha}}\right\} & =m M_{\dt{\alpha}}-\phi M_{\dt{\alpha}} ~, & \big\{\CQ, \wt{\lambda}_{\dt{\alpha}}\big\} &= -m \wt{M}_{\dt{\alpha}}+\wt{M}_{\dt{\alpha}}\phi ~, \\
\left\{\CQ, \lambda_\alpha\right\} &= h_\alpha, & \left\{\CQ, \mu_\alpha\right\} & =\wt{h}_\alpha ~,\\
[\CQ, h_\alpha] &= m\lambda_\alpha-\phi\lambda_\alpha ~, & [\CQ,\wt{h}_\alpha] &= -m\mu_\alpha+\mu_\alpha\phi~.
\end{aligned}
\end{align}
The central charge acts as multiplication by $\pm m$ on the hypermultiplet fields. 
The $\pm 1$ factors in the action of $Z,\ov{Z}$ on various hypermultiplet fields are shown in Table \ref{tab:centcharge}.


\begin{small}
\begin{table}[]
    \centering
\begin{tabular}{|c|c|c|c|c|c|c|c|c|} \hline
Field & $M$\Tstrut & $\wt{M}$\Tstrut & $\lambda$\Tstrut & $\wt{\lambda}$\Tstrut & $\mu$\Tstrut & $\wt{\mu}$\Tstrut & $h$\Tstrut & $\wt{h}$\Tstrut \\ \hline
\begin{tabular}{@{}c@{}}
Central charge\\
$(Z, \ov{Z})$
\end{tabular}
& $(-1,1)$ & $(1,-1)$ & $(-1, 1)$ & $(1,-1)$ & $(1,-1)$ & $(-1,1)$ & $(-1,1)$ & $(1,-1)$ \\ \hline
\end{tabular}
    \caption{The action of $(Z,\ov{Z})$ on various fields is given by the listed signs multiplying $m$.}
    \label{tab:centcharge}
\end{table}
\end{small}

From \eqref{eq:sQonfields},  $\mathcal{Q}$ squares to a gauge transformation plus a central charge transformation:
\begin{equation}
    \CQ^2=-\frac{1}{2}\delta_{m}^{\mr{central-charge}}+\delta_\phi^{\mr{gauge}}~,
\end{equation}
where the subscript $m$ is the parameter (a mass) for the central charge transformation. 

To formulate an action for the twisted theory, we begin with the action for the physical theory with fields $\Phi^{\mr{phys }}$, $\tau$, $m$ and $\n^{\mr {phys }}$ and then specialize the connection $\n^{\mr {phys }}$ to the pushforward of $\n^{\mr{t w}}$ and express $\Phi^{\mr {phys }}$ in terms of $\Phi^{\mr{t w}}$. 
Here $A_\mu$ and $\CA_\mu^u$ are YM and $\mr{U(1)}$ connections respectively, while $\omega_\mu$, $\left(\omega_{+}\right)_\mu$ are gravitational connections. A sensible way to formulate the action is in terms of a $\left(G^{\mr{gauge}} \times \IT^{\mr{f}}\right)$-gauge theory with matter coupled to background $\mathcal{N}=2$ conformal supergravity (where the $\IT^{\mr{f}}$ vectormultiplet is treated as a non-dynamial background vectormultiplet), with fermions in the supergravity multiplet and $\IT^{\mr{f}}$-vectormultiplet set to zero (to ensure a $\CQ$-exact energy-momentum tensor).
Note that $\CA_\mu^u$ is a $\mr{U(1)}/ p_{3,u}(C^{\mr{phys}})$-connection, and $p_{3,u}(C^{\mr{phys}}) \cong \mathbb{Z} / n_u \mathbb{Z}$. Thus,
\begin{equation}\label{eq:U1connrel}
\CA_\mu^u=\frac{1}{n_u} B_\mu^u~,
\end{equation}
where $B_\mu^u$  is a $\mr{U(1)}$-connection in the  $\mr{U(1)}_u$-factor of the $\IT^{\mr{f}}$-vectormultiplet. 

Furthermore, the scalars in the $\IT^{\mr{f}}$-vectormultiplets are set equal to the bare mass parameters: since $m \in \mr{Lie}\left(G^{\mr{f}}\right) \otimes \IC$ and $G^{\mr{f}}$ breaks to $\IT^{\mr{f}}$, $m$ can be identified with $\mr{rank}(\IT^{\mr{f}})$ complex parameters $m_u$, for $u=1, \ldots, \mr{rank}\big(\IT^{\mr{f}}\big)$. Let $\phi_u$, $\wt{\phi}_u$ for $u=1, \ldots, \mr{rank}\big(\IT^{\mr{f}}\big)$ denote the scalars in the $\IT^{\mr{f}}$-vectormultiplets. Then we set\footnote{The rescalings are chosen to make contact with the vectormultiplet action of \cite{Cushing:2023rha}.} 
\begin{equation}\label{eq:scalarmass}
    \phi_u=-m_u,\quad \wt{\phi}_u=\frac{m_u}{2}~,\quad u=1, \ldots, \mr{rank}\big(\IT^{\mr{f}}\big)~.  
\end{equation}
Finally choose generators $\{T^a\}_{a=1}^{\mr{dim}(G^{\mr{gauge}})}$ for $\mathsf{Lie}(G^{\mr{gauge}})$ and $\{T^u\}_{u=1}^{\mr{rank}\left(\IT^{\mr{f}}\right)}$ and denote the components of the bilinear form $\tau$ in this basis by\footnote{Note that there are no cross terms because of $\mr{Ad}(G^{\mr{gauge}}\times G^{\mr{f}})$-invariance of $\tau$.} $\tau_{ab},\tau_{uv}$. Also, let 
\begin{equation} \label{eq:Trho}
T^{a}_\rho=T^{a}_\varrho\oplus T^{a}_{\ov{\varrho}}~,    
\end{equation}
be the representation matrix of the generators $\{T^{a}\}_{a=1}^{\mr{dim}(G^{\mr{gauge}})}$ of $\mathsf{Lie}(G^{\mr{gauge}})$ under $\rho$. The action can be written using the results of \cite{deWit:1984rvr,Karlhede:1988ax,Cushing:2023rha}. It is given by\footnote{Our conventions for differential forms are the same as \cite{Cushing:2023rha}, but for spinor index contractions and self-duality differ mildly: we contract spinor indices using the North-West-South-West convention, and we write a self-dual tensor $\chi_{\mu\nu}$ in dotted spinor indices as the symmetric object $\chi_{(\dt{\alpha}\dt{\beta})}$.}
\begin{equation} \label{eq:Stwisted-N2-plus-bck}
\s{S}^{\mr{twisted}}_{\CN=2+\mr{bck}}=\s{S}^{\mr{twisted}}_{\mr{vec}}+\s{S}^{\mr{twisted}}_{\mr{hyp}}+\s{S}^{\mr{top}}~,   
\end{equation}
where 
\begin{align}
\hspace{-0.2in}\s{S}^{\mr{twisted}}_{\mr{vec}}&=\frac{1}{4 \pi} \int \mr{vol}(g)~ \mr{Im\,} \tau_{a b}\left[\frac{1}{4} F_{\mu \nu}^a F^{\mu \nu,b}-\frac{1}{2} D_{\mu \nu}^a D^{\mu \nu,b}-2\chi_{\mu \nu}^a D^{\mu} \psi^{\nu,b}+\frac{1}{2}\chi_{\mu \nu}^a\left[\phi, \chi^{\mu \nu}\right]^b\right.\nonumber \\
\hspace{-0.2in}&\qquad\quad -\left.\eta^a D_\mu \psi^{\mu,b}+\wt{\phi}^a D_\mu D^\mu \phi^b-\frac{1}{2}\phi^a[\eta, \eta]^b-\frac{1}{2}[\phi, \wt{\phi}]^a[\phi, \wt{\phi}]^b-\wt{\phi}^a\left[\psi_\mu, \psi^\mu\right]^b\right] ~, \label{eq:Stwisted-vec}\\
\hspace{-0.2in}\s{S}^{\mr{twisted}}_{\mr{hyp}}&=\int \mr{vol}(g)\left[-\wt{h}^\alpha\sigma^\rho_{\alpha\dt{\alpha}} D_\rho M^{\dt{\alpha}}+D_\rho \wt{M}^{\dt{\alpha}}\wt{\sigma}^\rho_{\dt{\alpha}\alpha}h^\alpha-\wt{h}^\alpha h_\alpha-m^2 \wt{M}_{\dt{\alpha}} M^{\dt{\alpha}}+m \wt{M}_{\dt{\alpha}} \phi M^{\dt{\alpha}}\right. \nonumber\\
\hspace{-0.2in}&\qquad\qquad\qquad\quad\quad -2m \wt{M}_{\dt{\alpha}} \wt{\phi} M^{\dt{\alpha}}+\wt{\lambda}_{\dt{\alpha}}\left(\wt{\sigma}^{\rho\sigma}\right)^{\dt{\alpha} \dt{\beta}} M_{\dt{\beta}} \chi_{\rho \sigma}+\eta \wt{\lambda}_{\dt{\alpha}} M^{\dt{\alpha}}-\wt{M}^{\dt{\alpha}} \wt{\sigma}_{\dt{\alpha} \dt{\beta}}^{\rho \sigma} \wt{\mu}^{\dt{\beta}} \chi_{\rho\sigma}\nonumber \\
\hspace{-0.2in}&\qquad\qquad\qquad\quad\quad -\eta \wt{M}_{\dt{\alpha}} \wt{\mu}^{\dt{\alpha}}+\lambda^\alpha \sigma^\mu_{\alpha\dt\beta} D_\mu \wt{\lambda}^{\dt\beta}-\wt{M}_{\dt{\alpha}}\left(\wt{\sigma}^\rho\right)^{\dt{\alpha} \alpha} \lambda_\alpha \psi_\rho + \mu^\alpha \sigma_{\alpha \dt{\beta}}^\rho M^{\dt{\beta}} \psi_\rho \nonumber\\
\hspace{-0.2in}&\qquad\qquad\qquad\quad\quad + \mu^\alpha \sigma^\rho_{\alpha\dt\beta} D_\rho \wt{\mu}^{\dt{\beta}}+\mu^\alpha \phi \lambda_\alpha -2\wt{\lambda}_{\dt{\alpha}} \wt{\phi} \wt{\mu}^{\dt{\alpha}} -m \mu^\alpha \lambda_\alpha-m \wt{\lambda}_{\dt\alpha} \wt{\mu}^{\dt\alpha}\nonumber \\
\hspace{-0.2in}&\qquad\qquad\qquad\quad\quad+\wt{M}_{\dt{\alpha}}\{\phi, \wt{\phi}\} M^{\dt{\alpha}} \left.-\wt{M}_{(\dt{\alpha}} F^{\dt{\alpha}\dt{\beta}} M_{\dt{\beta})}+\wt{M}_{(\dt{\alpha}} D^{\dt{\alpha}\dt{\beta}} M_{\dt{\beta})}\right] ~, \label{eq:Stwisted-hyp}\\
\hspace{-0.2in}\s{S}^{\mr{top}}&:=\frac{\im}{32\pi} \int_{\IX}~ \tau_{ab}(F^a \wedge F^b)+\frac{\im}{32\pi} \int_{\IX}~\tau_{uv}( F^u \wedge F^v) ~. \label{eq:Stwisted-top}
\end{align}
Here 
$F$ and $F^u$ are the field strengths of the dynamical gauge field $A_\mu$ and the background $\mr{U(1)}_u$-connection $B_\mu^u$, and the covariant derivatives are given by\footnote{There is no sum over the repeated index $u$ in the covariant derivatives.} 
\begin{align}
D_\mu \wt{\phi} &=\partial_\mu \wt{\phi}+[A_\mu,\wt{\phi}]~,
\\ D_\mu \phi &=\partial_\mu \phi+[A_\mu,\phi]~,
\\ D_{\mu} \psi_\nu &= \partial_\mu\psi_\nu-\Gamma_{\mu\nu}{}^{\rho}\psi_\rho+[A_\mu,\psi_{\nu}] ~, 
\\ D_{\mu} M^{\dt{\alpha}}_u&=\partial_\mu M^{\dt{\alpha}}_u-\frac{1}{2}\omega_\mu{}^{\ua\,\ub}(\wt{\sigma}_{\ua\,\ub})^{\dt{\alpha}}_{~\dt{\beta}}M^{\dt{\beta}}_u+A^a_\mu T^a_{\varrho}M^{\dt{\alpha}}_u+\frac{q_u}{n_u}B_\mu^u M^{\dt{\alpha}}_u~,
\\D_{\mu} \wt{M}^{\dt{\alpha}u} &=\partial_\mu\wt{M}^{\dt{\alpha}u}-\frac{1}{2}\omega_\mu{}^{\ua\,\ub}(\wt{\sigma}_{\ua\,\ub})^{\dt{\alpha}}_{~\dt{\beta}}\wt{M}^{\dt{\beta}u}+A^a_\mu T^a_{\ov{\varrho}}\wt{M}^{\dt{\alpha}u}-\frac{q_u}{n_u}B_\mu^u \wt{M}^{\dt{\alpha}u}~,
\\ D_\mu \wt{\lambda}^{\dt{\alpha}u} &=\partial_\mu \wt{\lambda}^{\dt{\alpha}u}-\frac{1}{2}\omega_\mu{}^{\ua\,\ub}(\wt{\sigma}_{\ua\,\ub})^{\dt{\alpha}}_{~\dt{\beta}}\wt{\lambda}^{\dt{\beta}u}+A^a_\mu T^a_{\varrho}\wt{\lambda}^{\dt{\alpha}u}+\frac{q_u}{n_u}B_\mu^u \wt{\lambda}^{\dt{\alpha}u}~,
\\ 
D_{\rho} \wt{\mu}^{\dt{\alpha}}_u &=\partial_\rho \wt{\mu}^{\dt{\alpha}}_u-\frac{1}{2}\omega_\rho{}^{\ua\,\ub}(\wt{\sigma}_{\ua\,\ub})^{\dt{\alpha}}_{~\dt{\beta}}\wt{\mu}^{\dt{\beta}}_u+A^a_\rho T^a_{\ov{\varrho}}\wt{\mu}^{\dt{\alpha}}_u-\frac{q_u}{n_u}B_\rho^u \wt{\mu}^{\dt{\alpha}}_u~,
\end{align} 
where $\omega_{\mu}{}^{\ua\,\ub}$ is the spin connection. The factor of $\frac{q_u}{n_u}$ in the covariant derivatives above comes from the charge $q_u\in\IZ$ of hypermultiplet fields under $\mr{U(1)}_u$. Using the relation \eqref{eq:U1connrel} we see that these covariant derivatives are local expressions for $\n^{\mr{tw}}$ on suitable associated $G^{\mr{tw}}$-vector bundles.
Moreover, since $F^u$ is the curvature for the background $\mr{U(1)}_u$-connection $B^{u}_\mu$, we have
\begin{equation}\label{eq:curv-u}
    \left[\frac{F^u}{2\pi}\right]=c_1(u)~,\quad i=1, \ldots, \mr{rank}(\IT^{\mr{f}}) ~,
\end{equation}
where $c_1(u)$ is as in \eqref{eq:BackgroundGerbeCondition}. 
It turns out that the action is $\CQ$-exact up to topological terms:
\begin{equation}
\s{S}^{\mr{twisted}}_{\CN=2+\mr{bck}}=\{\s{Q},V\}+\s{S}^{\mr{top}} ~,
\end{equation}
where
\begin{align}
\mr{V}&:= \mr{V}_{\mr{vec}}+ \mr{V}_{\mr{hyp}} ~, \label{eq:Vee}\\
\mr{V}_{\mr{vec}}&:=\frac{1}{4 \pi} \int_{\IX} \mr{vol}(g)~\mr{Im\,} \tau_{a b}\left(\frac{1}{4} \chi^{\mu \nu, a}\left(F_{\mu \nu}^{+}+D_{\mu \nu}\right)^b-\frac{1}{4} \eta^a[\phi, \wt{\phi}]^b+\frac{1}{2} \psi^{\mu, a} D_\mu \wt{\phi}^b\right) ~, \label{eq:Vee-vec}\\
%
%
\mr{V}_{\mr{hyp}}&:=\int_{\IX} \mr{vol}(g)\left(D_\mu \wt{M}^{\dt{\alpha}} \wt{\sigma}_{\dt{\alpha} \alpha}^\mu \lambda^\alpha+D_\mu M^{\dt{\alpha}} \wt{\sigma}_{\dt{\alpha} \alpha}^\mu \mu^\alpha+\frac{1}{2} m \wt{\lambda}_{\dt{\alpha}} M^{\dt{\alpha}}-\frac{1}{2} \mu_\alpha h^\alpha-\frac{1}{2} m \wt{M}^{\dt{\alpha}} \wt{\mu}_{\dt{\alpha}}\right.\nonumber\\
&\qquad\qquad\qquad\left.+\frac{1}{2} \wt{h}_\alpha \lambda^\alpha+\frac{1}{2}\left(\wt{M}^{\dt{\alpha}} \wt{\phi} \wt{\mu}_{\dt{\alpha}}+\wt{\lambda}^{\dt{\alpha}} \wt{\phi} M_{\dt{\alpha}}\right)- \wt{M}^{(\dt{\alpha}} \chi_{\dt{\alpha} \dt{\beta}} M^{\dt{\beta})}\right) ~. \label{eq:Vee-hyp}
%
%
\end{align}
Since $\mr{V}$ and $\CS^{\mr{top}}$ are gauge invariant and invariant under the action of central charge (see Table \ref{tab:centcharge}), it follows that on a closed $4$-manifold,
\begin{equation} \label{eq:QonStwisted}
\left\{\CQ,\s{S}^{\mr{twisted}}_{\CN=2+\mr{bck}}\right\}=0~.
\end{equation}
The response currents to variations of the background metric and background $\mr{U(1)}$ connections are $\CQ$-exact as a result of $\CQ$-exactness of the action. Thus the twisted partition function and topological correlators are formally locally constant as a function of the background connections.  

\subsection{Non-Abelian Monopole Equations}
Given that the action is $\CQ$-exact up to a topological term, one can show that the path integral localizes to supersymmetric backgrounds i.e.,  solutions of $\CQ$-fixed equations with fermions set to zero. From \eqref{eq:sQonfields}, we see that supersymmetric configurations correspond to the solutions of the following $\CQ$-fixed equations: 
\begin{align}
 \left. \begin{array}{l}D_{\dt{\alpha} \dt{\beta}}^{a} =F_{\dt{\alpha} \dt{\beta}}^{a} ~,\\ 
 h_{\alpha u} =0 ~, \\
 \wt{h}^u_\alpha=0 ~, 
 \end{array} \right\}\quad \begin{array}{l}
a=1,\ldots,\mr{dim}(G^{\mr{gauge}}) ~, \\
 u=1,\ldots,\mr{rank}(\IT^{\mr{f}}) ~.\\
 \end{array}
\end{align}
Setting the auxiliary fields of the dynamical vectormultiplets and hypermultiplets to their equations of motion
\begin{align}
 \left. \begin{array}{l}D_{\dt{\alpha}\dt{\beta}}^{a} =2 \wt{M}^u_{(\dt{\alpha}} T_\varrho^{a} M_{\dt{\beta})u} ~,\\ 
 h_{\alpha v} =\sigma_{\alpha \dt{\alpha}}^\mu D_\mu M^{\dt{\alpha}}_v ~,\\
 \wt{h}^v_\alpha = -D_\mu \wt{M}^{\dt{\alpha}v} \wt{\sigma}_{\dt{\alpha} \alpha}^\mu~,
 \end{array} \right\}\quad \begin{array}{l}
\text{$u$ summed over} ~, \\
 v=1,\ldots,\mr{rank}(\IT^{\mr{f}}) ~,\\
 \end{array}
\end{align}
yields
\begin{equation}\label{eq:nonabmoneq}
\begin{aligned}
\left.
\begin{array}{l}
F_{\dt{\alpha} \dt{\beta}}^{a}-2 \wt{M}^u_{(\dt{\alpha}} T_\varrho^{a} M_{\dt{\beta})u} =0 ~,\\
 \sigma_{\alpha \dt{\alpha}}^\mu D_\mu M^{\dt{\alpha}}_v =0 ~, \\
 \sigma_{\alpha \dt{\alpha}}^\mu D_\mu \wt{M}^{\dt{\alpha}v} = 0 ~,
\end{array} 
\right\} \quad \begin{array}{l}
\text{$u$ summed over} ~, \\
 v=1,\ldots,\mr{rank}(\IT^{\mr{f}}) \\
 \end{array}~.
\end{aligned}
\end{equation}
These $\CQ$-fixed equations are called \textit{non-Abelian monopole equations} \cite{Labastida:1995zj,Hyun:1995mb,Losev:1997tp,pidstrigach1995localisation,Witten:1999QFT2} 
and are generalizations of both the usual ASD instanton equations and the ``monopole'' or ``Seiberg-Witten'' equations \cite{Witten:1994cg}.  We can write these equations in coordinate-free notation as follows: the monopole field $M$ is a section of the associated $G^{\mr{tw}}$-vector bundle associated with the representation $(\bm{1}, \bm{2}, \mr{def},\varrho)$. Denote this vector bundle by $E^{\mr{mon}}$. The conjugate monopole field $\wt{M}$ is a section of the associated $G^{\mr{tw}}$-vector bundle associated to the representation $(\bm{1}, \bm{2}, \overline{\mr{def}},\overline{\varrho})$. Denote this vector bundle by $\ov{E}^{\mr{mon}}$. On every local chart $(U,x^\mu)$ on $\IX$, there is a natural bilinear map: 
\begin{equation}
\begin{split}
    \mu:\Gamma(\ov{E}^{\mr{mon}})\times \Gamma(E^{\mr{mon}})&\longrightarrow \Omega^{2,+}(U,\mr{ad}\,P^{\mr{gauge}}) \\
    (\wt{M},M)&\longmapsto \frac{1}{2!}\left(\frac{1}{2}e^{~\ua}_\mu e^{~\ub}_\nu\wt{\sigma}^{\dt{\alpha}\dt{\beta}}_{\ua\,\ub}\wt{M}^u_{(\dt{\alpha}} T_\varrho^{a} M_{\dt{\beta})u}T^{a}\right) dx^\mu\wedge dx^\nu~,
\end{split}    
\end{equation}
where we have chosen a local section $s:U\to P^{\mr{tw}}$ to write 
\begin{equation}
    \wt{M}=[(s,\wt{M}^u_{\dt{\alpha}m})]~,\quad M=[(s,M_{\dt{\alpha}u}^{\ov{n}})]~,
\end{equation}
where $m$ and $\ov{n}$ are indices for the representations $\varrho$ and $\ov{\varrho}$ respectively. We have chosen a basis for the orthonormal frame bundle with vielbein $e_{\mu}{}^{\ua}$ in the local chart. Finally as noted before, $D_\mu$ acting on $\wt{M}$ and $M$ is the Dirac operator coupled to $\ov{E}^{\mr{mon}}$ and $E^{\mr{mon}}$ respectively. The monopole equations can then be written as 
\begin{equation}
    F^+\left((p_4)_*(\n^{\mr{tw}})\right)=2\mu(\wt{M},M)~,\quad \slashed{D}_{E^{\mr{mon}}}M=0 ~,\quad \slashed{D}_{\ov{E}^{\mr{mon}}}\wt{M}=0~,
\end{equation}
where $F^+\left((p_4)_*(\n^{\mr{tw}})\right)$ is the SD part of the curvature of the $G^{\mr{gauge}}$-connection $(p_4)_*(\n^{\mr{tw}})$ on $P^{\mr{gauge}}$ and $\slashed{D}=\wt{\sigma}^\mu D_\mu$.
This allows us to generalize the non-Abelian monopole equations to any compact, simple gauge group.

The path integral for topological correlators and the partition function formally localizes to the moduli space of non-Abelian monopole equations. After performing the path integral, we obtain topological invariants which depend on the first Chern classes $c_1(u)$ and the background fields $\rho_u(\mu(b))$ for 1-form symmetries constrained according to \eqref{eq:BackgroundGerbeCondition}. This topological dependence on the background fields arises from the topological terms $\CS^{\mr{top}}$ in the action. 

In conclusion, 
after integrating out the dynamical fields we obtain a twisted partition function depending only on the data \ref{it:top_data1}--\ref{it:top_data3}.

\section{Topological Twisting Of General Class $\CS$ Theories}\label{sec:GenClassS-Twisting}

We now turn to class $\CS$ theories. In general, we cannot apply the general discussion of Sections \ref{sec:NonSpin}-\ref{sec:backind} because in general class $\CS$ theories are non-Lagrangian.
(Theories of type $A_1$ are Lagrangian, and this is a notable and important exception.)

\subsection{Pants Decompositions}

As noted in \cite{Gaiotto:2009hg,Gaiotto:2009we} weak coupling regimes are determined by a pants decomposition of the ultraviolet curve $C_{g,n}$ used to define the theory. Here $g$ is the genus and 
$n$ the number of punctures. 

Recall that a pants decomposition of a stable Riemann surface $C_{g,n}$ \cite{HatcherExpos} is a maximal collection of pairwise disjoint simple closed curves (also called \textit{cutting curves} or \textit{cuffs}).\footnote{A simple closed curve on $\IX$ is a map $c: [0,1] \to \IX$ that is closed, i.e., $c(0) = c(1)$, and also injective as a map $[0,1) \to \IX$. Following \cite{Manschot:2021qqe}, we will refer to cutting curves surrounding the punctures as \textit{external cutting curves}, and cutting curves $c_i$ associated with gauge groups as \textit{internal cutting curves}. In the mathematical literature, only the internal cutting curves are counted as \textit{cuffs}.} 
For a given connected surface $C_{g,n}$, any pants decomposition will consist of $(2g-2+n)$ pairs of pants.\footnote{The Euler characteristic $\chi(C_{g,n}) = 2-2g-n$ of a stable Riemann surface is negative, this excludes $C_{0,0}$, $C_{0,1}$, $C_{0,2}$, and $C_{1,0}$. 
Strictly speaking, one should distinguish between boundary components and punctures (the latter being marked points in the interior of the closure of the surface). Here the $n$ ``punctures'' are obtained by removing $n$ small open disks from the surface.}
Moreover, any pants decomposition consists of $(3g-3+n)$ internal cutting curves and $n$ external cutting curves.
We denote a typical pants decomposition by a list of cutting curves:
\begin{equation}\label{eq:pants-decom-list}
    \mf{p} = \{c_1,\ldots,c_{3g-3+2n}\}~.
\end{equation} 
The complement of these curves in $C_{g,n}$ is a collection of $(2g-2+n)$ pairs of pants\footnote{A pair of pants (hereafter abbreviated simply to ``pants'') is a surface homeomorphic to $C_{0,3}$.}  (also called trinions).

In these regimes we can express the theory as a gauge theory, coupled to possibly non-Lagrangian theories associated with the trinions. The key point is that we need only study general properties of the partition functions of these (possibly non-Lagrangian) trinion theories in the presence of background fields. This is all that is needed  to discuss the analogues of the cohomological conditions 
\eqref{eq:BackgroundGerbeCondition} and the twisting procedure.

\subsection{Trinion Theories}

The trinion theory\footnote{\label{foot:dataS}As noted in Section \ref{sec:a1classS}, the trinion theory is a class $\CS$ theory on a three-punctured sphere. The notation $T[\mathfrak{g}, D_{p_1}, D_{p_2}, D_{p_3}]$ specifies the gauge algebra of the 6d theory and the puncture data $D_{p_1},D_{p_2}, D_{p_3}$ at the three punctures. The data at a regular puncture $p$ includes a reductive Lie algebra $\mathfrak{g}_p$ and a homomorphism $\rho: \mf{su(2)} \to \mathfrak{g}_p$, among other things. The flavor symmetry algebra $\mathfrak{f}_{p}^{\mr{fl}}$ at puncture $p$ is given by the centralizer of the image of $\rho$ in $\mathfrak{g}_{p}$. See \cite{Chacaltana:2012zy,Tachikawa:2013kta,Tachikawa:2015bga}, for a more complete discussion of the punctures in class $\CS$ theories.} 
$T[\mathfrak{g}, D_{p_1}, D_{p_2}, D_{p_3}]$ is a relative field theory.  We would like to gauge a collection of trinion theories to produce a class $\CS$ theory, and we would like the latter to be an absolute 4d theory that can be defined on all standard four-manifolds. We will make the conservative assumption that this requires each component trinion theory to be an absolute theory definable on all standard four-manifolds.\footnote{It would be interesting if one could relax this conservative assumption by gauging the relative trinion theories to produce an absolute theory.}

It is thought that the trinion theory can be defined as an absolute 4d theory once some extra data is specified 
\cite{Tachikawa:2013hya}. For the trinion theory, it is argued in \cite{Tachikawa:2013hya}
that the only extra data is a choice of global form of the symmetry group. (Equivalently, a choice of global form of the group is used to define the Hitchin system.) 
We simply denote this extra data as $\CL$ so the absolute theory is denoted $T[\mathfrak{g}, D_{p_1}, D_{p_2}, D_{p_3};\CL]$.

For our considerations, we need to know the global symmetry \underline{group} of the theory, not just its Lie algebra. The group form of the symmetry group of 
class $\CS$ theories in general, and trinion theories, in particular, has been studied in \cite{Tachikawa:2013hya,Albertini:2020mdx,Bhardwaj:2021ojs,Bhardwaj:2021pfz,Bhardwaj:2021mzl,Etxebarria:2021lmq,DelZotto:2022ras,Heckman:2022suy,Garding:2023unh}.\footnote{See also \cite{Bashmakov:2022jtl,Bashmakov:2022uek} for some recent work on generalized global symmetries of class $\CS$ theories.}
The viewpoint of this paper, based on remarks from section \ref{sec:GeneralSetup}, 
differs slightly from the above references. We regard the global symmetry as a \underline{choice} based on the choice of domain category in the functorial formulation of field theory.

To each puncture $p$ we associate a global symmetry algebra $\mathfrak{g}_p^{\mr{fl}}$ and we denote by  $\wt{G_{p}}$ a Lie group with this algebra which is a product of a simply connected semi-simple Lie group and a torus. (Specification of the torus requires further discussion, not found here.) We can then define a theory $T[\mathfrak{g}, D_{p_1}, D_{p_2}, D_{p_3};\CL]$ with 
symmetry group of the form 
\be \label{eq:trinion-global-sym}
\wt{G}_T = 
  \mr{Spin(4)} \times \mr{SU(2)_R} \times  \wt{G_{p_1}} \times \wt{G_{p_1}}\times  \wt{G_{p_1}} ~.
\ee
Since, as we have mentioned, the theories $T[\mathfrak{g}, D_{p_1}, D_{p_2}, D_{p_3};\CL]$ are in general non-Lagrangian it is very useful to adopt the functorial viewpoint of what is meant by a ``theory'' as in section \ref{sec:GeneralSetup}.   Thus we take a ``theory'' to be   a functor

\be \label{eq:trinion-monoidal-functor-tilde}
\wt{\mr{Z}}_T: \mathsf{Bord}_{\leq 4}^{\wt{\CF}} \to \mathsf{VECT} ~,
\ee
where $\wt{\CF}$ is a set of background fields that includes the groupoid of principal $\wt{G}_T$ bundles with connection. 
We identify the $\mr{Spin(4)}$ factor with the covering group of the tangent bundle of $\IX$, and hence the bordism category $\mr{Bord}_{\leq 4}^{\wt{\CF}}$ only includes spin manifolds. The background fields $\wt{\CF}$ should include the full set of vectormultiplets for the flavor symmetry so the partition function 
on $\IX$ in the presence of background fields for $\wt{G}_T$ can be written 
informally as: 
\be 
\mr{Z}_T[\IX, \omega_{\mu}{}^{\ua\,\ub}, A^{\mr{R}}_\mu , \mr{VM}_{p_1}, \mr{VM}_{p_2}, \mr{VM}_{p_3} ] ~,
\ee
where $\mr{VM}_p$ is the \underline{full set} of fields of the standard $\CN=2$ vectormultiplet.

It has been noted in \cite{Tachikawa:2013hya,Bhardwaj:2021ojs}
 that the symmetry group of the trinion theory can be taken to be a quotient 
\be\label{eq:Trin-Glob-Sym}
G_T:=\wt{G}_T/C_T ~,
\ee
where $C_T$ is a subgroup of the center of the numerator. 
As in \eqref{eq:FactorFunctor} above we require that the functor $\wt{\mr{Z}}$ 
factors through 
\be \label{eq:trinion-monoidal-functor}
\mr{Z}_T: \mathsf{Bord}_4^{\CF} \to \mathsf{VECT} ~.
\ee

For purposes of topological twisting, we would like this bordism category to include non-spin manifolds. Therefore, the subgroup $C_T$ should  
satisfy \eqref{eq:p1p2-C-condition}. 
%
%
Applying the reasoning of section \ref{sec:GeneralSetup} we arrive at: 
\be\label{eq:Trin-C-condition2}
C_T \subset C_T^{\mr{max}}:= \{ (\zeta,\zeta,\zeta,\mu): (\zeta,\zeta,\zeta,\mu)\cdot \wt{\mr{Z}}_T = \wt{\mr{Z}}_T \}  \subset Z(\wt{G}_T) ~,
\ee
where $\mu\in Z(\wt{G_{p_1}} \times \wt{G_{p_1}}\times  \wt{G_{p_1}})$.

Very little is known about the trinion theories. Therefore we will restrict attention to the top level of the functor, i.e., the partition function.
%
%
Fortunately, one example of a partition function  is known, it is the superconformal index\footnote{See Appendix \ref{app:SupSymIndex} for a brief review.}
\cite{Romelsberger:2005eg,Kinney:2005ej,Gadde:2009kb,Gadde:2011ik,Gadde:2011uv,Tachikawa:2015bga} which is $\wt{\mr{Z}}_T$ evaluated on  $\IS^3 \times \IS^1$ with nonbounding spin structure on $\IS^1$, coupled to a $\wt{G}_T$ bundle with connections specialized so as to give an index. 
%
%
We can work out the action of $C_T$ on these. Given our present ignorance, we will simply 
\underline{postulate} that if $C_T$ acts trivially on the superconformal index it acts trivially on $\wt{\mr{Z}}$ and therefore $\wt{\mr{Z}}$ factorizes.\footnote{\label{foot:centralextension}In fact, although this assumption seems to be standard in the literature, it has a fundamental flaw. It is well known that the action of a symmetry group on operators $O \to U O U^\dagger$ differs slightly from that on states $\vert \psi \rangle \to U \vert \psi \rangle$. Put more formally, the group that acts faithfully on the Hilbert space is in general a central extension of the group that acts faithfully on the operator algebra. This happens quite frequently in quantum theory. Sure enough,  in the case of class $\CS$ theories, L. Hollands and A. Neitzke \cite{Hollands:2016kgm} discovered that the $\IZ_3$ central subgroup of $\mr{E_6}$ acts \underline{nontrivially} on the spaces of BPS states on $\IR^3$ in the Minahan-Nemeschansky theory. They show that the Hilbert space of BPS states has a nontrivial isotypical decomposition for the $\IZ_3$ action where the degeneracy spaces for all three irreps are infinite-dimensional, and has nontrivial BPS index. (The IR description of the degeneracy spaces has a $\IZ\oplus \IZ$ grading from an IR $\mr{U(1)}\times \mr{U(1)}$ global symmetry.)
It is conceivable that, when working on compact manifolds only the adjoint group of $\mr{E_6}$ acts faithfully on the Hilbert space, but we are not aware of any evidence for or against this. We will nevertheless proceed to determine $C_T^{\mr{max}}$ from the superconformal index, \emph{faute de mieux}, bearing in mind that our results might need to be refined by a central extension. } 
\par  
The full superconformal index is not known in closed analytic form for generic trinion theories but a simplification known as the Schur index is known (see Appendix \ref{app:SupSymIndex}). The Schur index for a $d=4$ $\CN=2$ theory (on $\IS^3 \times \IS^1$) has the trace formula 
\begin{equation}\label{eq:schur_index1}
    \CI_{\mr{Schur}}(q,\bm{g}):=\mr{Tr}_{\CH(\IS^3)}(-1)^{\mr{F}}q^{\Delta-\frac{R}{2}} \bm{g}~,
\end{equation}
where $\CH(\IS^3)$ is the Hilbert space of the theory on $\IS^3$, $\mr{F}$ is the fermion number, $\Delta$ is the dilatation generator, $R$ is the Cartan generator of $\mr{SU(2)_R}$, and $\bm{g} \in G^{\mr{f}}$ is the fugacity for the flavor symmetry (understood as an operator via its action on $\CH(\IS^3)$).  Only the operators with quantum numbers satisfying
\begin{equation}\label{eq:schurop}
\Delta=R+\frac{j_1+j_2}{2}~,
\end{equation}
contribute to the Schur index (these are the so-called Schur operators \cite{Gadde:2011ik,Gadde:2011uv,Rastelli:2014jja,Rastelli:2016tbz}), where $(j_1, j_2)\in \IZ_{\geq 0}\times \IZ_{\geq 0}$ in \eqref{eq:schurop} are the eigenvalues of the Cartan generators of the $\mr{Spin(4)} \cong \mr{SU(2)} \times \mr{SU(2)}$ isometry group of $\IS^3$.\footnote{We use the same symbols for the operators and their eigenvalues.}
Thus the Schur index \eqref{eq:schur_index1} reduces to
\begin{equation}\label{eq:schur_index2}
    \CI_{\mr{Schur}}(q, \bm{g})=\mr{Tr}_{\CH(\IS^3)} (-1)^{\mr{F}} q^{\frac{R+j_1+j_2}{2}} \bm{g} ~.
\end{equation}
    The fugacities are holonomies for background gauge fields. Thus an element $(\zeta, \zeta, \zeta, \mu) \in Z(\wt{G}_T)$ acts on the holonomies $q^{1/2}$ and $\bm{g}$ for the background gauge fields:
the holonomies simply get multiplied by $a$ and $\mu$. From the exponent of $q^{1/2}$ in \eqref{eq:schur_index2}, it is clear that the center acts on the fugacities as
\begin{align}\label{eq:center_action_fugacity}
    (q^{\frac{1}{2}}, \bm{g}) &\longmapsto ( \zeta q^{\frac{1}{2}}, \mu \bm{g} ) ~.
\end{align}
In this way, we can use the action of the center on the Schur index to determine $C_T^{\mr{max}}$. We now work out two explicit examples.  
\begin{ex}
The simplest example is that of the trinion theory of class $\CS$ of type $A_{N-1}$ with two ``full'' punctures and one ``simple'' puncture, so the global symmetry Lie algebra is $\fs\fu(N)\oplus \fs\fu(N) \oplus \fu(1)$. 
In this case 
\be 
\wt{G}_T = 
  \mr{Spin(4)} \times \mr{SU(2)_R} \times  \mr{SU(N)}  \times \mr{SU(N)} \times  \mr{U(1)} ~.
\ee

This is a Lagrangian theory expressible in terms of free hypermultiplets and background vectormultiplets.   We refer the reader to Appendix \ref{app:Lag-ffs-Trinion-Analysis} for our notation, and relevant details of the action and the partition function. Here we just make use of the superconformal 
index.   The Schur index of the trinion theory with two ``full'' punctures and one ``simple'' puncture is given by \cite{Gadde:2011ik,Tachikawa:2015bga}
\begin{align}\label{eq:schur_index_free_hyper}
\CI^{\mr{ffs}}_{\mr{Schur}}(q;\bm{a},\bm{b}, \alpha) 
&= \prod_{u,i=1}^N\prod_{n\geq 0}\left[1-q^{n+\frac{1}{2}}\left(\frac{\alpha a_i}{b_u}\right)\right]^{-1}\cdot\left[1-q^{n+\frac{1}{2}}\left(\frac{b_u}{\alpha a_i}\right)\right]^{-1}~,
\end{align}
where $\bm{a},\bm{b}\in\IT(\mr{SU(N)})$ are the fugacities for the Cartan of the two $\mr{SU(N)}$ flavor symmetries, parametrized as $\bm{a}=\mr{diag}(a_1,\ldots,a_N)$ and $\bm{b}=\mr{diag}(b_1,\ldots,b_N)$, and $\alpha\in \mr{U(1)}$ is the flavor fugacity for $\mr{U(1)}$.
Under the action of the central element 
\begin{equation}
(\zeta,\zeta,\zeta,\omega^{k_1}\textbf{1}_N,\omega^{k_2}\textbf{1}_N,\xi)\in \IZ_2\times\IZ_2\times\IZ_2\times\IZ_N\times\IZ_N\times\mr{U(1)}~,   
\end{equation}
where $\omega := e^{\frac{2\pi \im }{N}}$ and $k_{1}, k_{2} \in \{0, \ldots, N-1\}$, the fugacities transform as in \eqref{eq:center_action_fugacity}:
\begin{equation}\label{eq:fugtransf}
q^{\frac{1}{2}} \mapsto \zeta q^{\frac{1}{2}}~,\quad (\bm{a},\bm{b},\alpha) \mapsto (\omega^{k_1}\bm{a},\omega^{k_2}\bm{b},\alpha\xi)~.    
\end{equation}
From the explicit character expansions of \eqref{eq:schur_index_free_hyper}, we find that the Schur index is invariant under \eqref{eq:fugtransf} if and only if 
\begin{equation}
\zeta\omega^{k_1}\omega^{-k_2}\xi=1 ~,    
\end{equation}
and hence:  
\begin{equation}\label{eq:CTmax_freehyper}
C_T^{\mr{max}}=\{(\zeta,\zeta,\zeta,\omega^{k_1}\mathbf{1}_N,\omega^{k_2}\mathbf{1}_N,\zeta\omega^{-k_1}\omega^{k_2})\in Z(\wt{G}_T)\}~.    
\end{equation}
Clearly,
\be 
C_T^{\mr{max}} = \langle 
(-1,-1,-1,   \textbf{1}_N, \textbf{1}_N, -1   ) , 
(1,1,1,\omega \textbf{1}_N, \textbf{1}_N, \omega^{-1}   ),
(1,1,1,  \textbf{1}_N, \omega \textbf{1}_N, \omega)  \rangle ~.
\ee
There are different subgroups of $C_T^{\mr{max}}$ satisfying \eqref{eq:Trin-C-condition2}. They define 
different theories.

\paragraph{Topological Twisting Of The $\mr{ffs}$ Theory.}
To define a topological twisting we can take $\widetilde G^{\mr{tw}} := \mr{Spin(4)} \times \IT^{\mr{f}}$ with 
a homomorphism $\widetilde G^{\mr{tw}} \to \wt{G}_{T}$ given by $(u_1, u_2, \mu ) \mapsto ( (u_1, u_2), u_2, \mu) $ which will descend to a 
homomorphism $G^{\mr{tw}} \to G_T$ for groups  $C_T$ described above, where $G^{\mr{tw}} \cong \wt{G}^{\mr{tw}}/C^{\mr{tw}}$ and we can take $C^{\mr{tw}}$ to be the kernel of the homomorphism from $\wt{G}^{\mr{tw}}$. 
Via pullback under the homomorphism $G^{\mr{tw}} \to G_T$, the $\mr{SU(2)_R}$-connection is identified with the self-dual part of the spin connection. 

As explained in Section \ref{sec:backind}, the bosonic scalars of the hypermultiplet become chiral spinors upon twisting. For details of the untwisted and twisted actions of the $\mr{ffs}$ theory and our notation, see Appendix \ref{app:Lag-ffs-Trinion-Analysis}. The Gaussian path integral over the twisted bosonic spinor fields yields the product of a topological invariant of the background fields and an inverse of the determinant of the square of the coupled Dirac operator. In particular, due to $\CQ$-symmetry, the determinants cancel:
\begin{equation}\label{eq:gaussian-twisted}
\begin{split}
    \int [\CD\mr{HM}] e^{-S^{\mr{twisted}}_{\mr{hyp}}}=\left(\mr{Det}\left(\slashed{D}_{\CV\otimes\ov{\CV}}^+\slashed{D}_{\CV\otimes\ov{\CV}}^-\right)\right)^{-1}\mr{Det}\big(\im\slashed{D}_{\CV\otimes\ov{\CV}}^{+}\big) \mr{Det}\big(-\im\slashed{D}_{\CV\otimes \ov{\CV}}^{-}\big) = 1 ~,
\end{split}    
\end{equation}
and the partition function is simply given by a topological invariant of the background $\mr{SU(N)} \times \mr{SU(N)} \times \mr{U(1)}$ bundle with connection:
\be\label{eq:partition_func_free_hyper_twisted_final} 
\mr{Z}_T^{\mr{twisted}}  = \exp\left(-\frac{\mr{i}}{16\pi}\int_{\IX}\tau_{ab}~(F^a\wedge F^b)\right) ~.
\ee

\end{ex}

\bigskip
\begin{ex}
We also consider the $A_{N-1}$ type trinion theory with 3 ``full'' punctures ($\mr{fff}$) for which\footnote{There are some issues about whether one should use $\mr{SU(N)}$ or $\mr{U(N)}$. We will not attempt to address them here. }
\be 
\wt{G}_T = 
  \mr{Spin(4)} \times \mr{SU(2)_R} \times  \mr{SU(N)}  \times \mr{SU(N)} \times  \mr{SU(N)} ~.
\ee
The only further information known about the operator content comes from the superconformal index \cite{Romelsberger:2005eg,Kinney:2005ej} (see Appendix \ref{app:SupSymIndex} for a review). The Schur index of the $T_N$ theory has the following form \cite{Gadde:2011ik,Tachikawa:2015bga}: 
\begin{equation} \label{eq:SchurTN}
    \mathcal{I}_{\mr{Schur}}^{T_N}(q;\bm{a},\bm{b},\bm{c})=\frac{K(\bm{a})K(\bm{b})K(\bm{c})}{K_0}\sum_{\lambda}\frac{\chi_\lambda(\bm{a})\chi_\lambda(\bm{b})\chi_\lambda(\bm{c})}{\chi_\lambda(\bm{q^\rho})}~.
\end{equation}
Here, the sum is over representations of the global flavor symmetries, for which the flavor symmetry elements are
\begin{equation} \label{eq:SCabc}
    (\bm{a},\bm{b},\bm{c})\in \IT(\mr{SU(N)}\times\mr{SU(N)}\times\mr{SU(N)})~,
\end{equation}
where $\IT$ denotes the standard maximal torus
and $\chi_\lambda$ is the character of the representation $\lambda$.
Also,
\begin{equation}\label{eq:q-to-the-rho}
 \bm{q^\rho} :=\mr{diag}(q^{(N-1)/2},q^{(N-3)/2},\ldots,q^{(1-N)/2})\in\mr{SU(N)}~.   
\end{equation}
The prefactor in \eqref{eq:SchurTN} 
is the contribution of the flavor multiplets to the index for $\mr{SU(N)}\times \mr{SU(N)}\times \mr{SU(N)}$, where for any $\bm{\mr{a}} \in \{\bm{a}, \bm{b}, \bm{c}\}$,
\begin{align}
   \label{eq:Kay-a} K(\bm{\mr{a}})&:=\prod_{n\geq 0}\bigg[(1-q^{n+1})^{1-N}\prod_{i\neq j}\left(1-q^{n+1}\frac{\mr{a}_i}{\mr{a}_j}\right)^{-1}\bigg]~, \\ 
\label{eq:Kay-zero} K_0 &:=\prod_{d=2}^N\prod_{n\geq 0}(1-q^{d+n})~.
\end{align}
and $\bm{\mr{a}} :=\mr{diag}(\mr{a}_1,\ldots,\mr{a}_N)\in \IT(\mr{SU(N)})$. 

The element
\begin{equation}
 (\zeta,\zeta,\zeta,\omega^{k_1}\mathbf{1}_N,\omega^{k_2}\mathbf{1}_N,\omega^{k_3}\mathbf{1}_N)\in \IZ_2\times\IZ_2\times\IZ_2\times\IZ_N\times\IZ_N\times\IZ_N ~, 
\end{equation}
where $k_1, k_2, k_3 \in \{0, \ldots, N-1\}$, acts on the fugacities as in \eqref{eq:center_action_fugacity}, which implies
\begin{equation}
    \bm{q^\rho} \mapsto \zeta^{N-1} \bm{q^\rho}~, \quad (\bm{a},\bm{b},\bm{c})\mapsto (\omega^{k_1}\bm{a},\omega^{k_2}\bm{b},\omega^{k_3}\bm{c})~.
\end{equation}
Consider the term in the Schur index corresponding to a representation $\lambda$ of $N$-ality $N(\lambda)$. From the expression of the character $\chi_\lambda$ in terms of Schur polynomials, we infer that 
\begin{equation}
  \frac{\chi_\lambda(\bm{a})\chi_\lambda(\bm{b})\chi_\lambda(\bm{c})}{\chi_\lambda(\bm{q^\rho})}\longmapsto \big(\zeta^{N-1}\omega^{k_1} \omega^{k_2} \omega^{k_3}\big)^{N(\lambda)} ~ \frac{\chi_\lambda(\bm{a})\chi_\lambda(\bm{b})\chi_\lambda(\bm{c})}{\chi_\lambda(\bm{q^\rho})}~. 
\end{equation}
Note that the prefactor $K(\bm{a})K(\bm{b})K(\bm{c})/K_0$ is invariant under translation by the center. 
Since representations of all $N$-alities appear in the sum, we conclude that 
\begin{align}\label{eq:CTmax_TN_theory}
   \hspace{-0.5cm} C_T^{\mr{max}} &= \left\{
       \begin{array}{ll}
               \{(\zeta,\zeta,\zeta,\omega^{k_1}\mathbf{1}_N,\omega^{k_2}\mathbf{1}_N,\omega^{k_3}\mathbf{1}_N)\in Z(\wt{G}_T):\zeta\omega^{k_1} \omega^{k_2} \omega^{k_3}=1\} & \text{for } N \text{ even} ~, ~\\
               \{(\zeta,\zeta,\zeta,\omega^{k_1}\mathbf{1}_N,\omega^{k_2}\mathbf{1}_N,\omega^{k_3}\mathbf{1}_N)\in Z(\wt{G}_T):\omega^{k_1} \omega^{k_2} \omega^{k_3}=1\}  & \text{for } N \text{ odd} ~.
       \end{array}
    \right.
\end{align}
\end{ex}
\bigskip

We now return to the general trinion theory. Since the theory is supersymmetric, its partition function $\mr{Z}_T$ should satisfy a set of supersymmetric Ward identities 
analogous to those of the partition function of a hypermultiplet in the presence of external fields.  In particular,  upon choosing a homomorphism 
\be 
\varphi: G^{\mr{tw,bck}} \to  \wt{G}_T/C_T ~,
\ee
such that the pullback of the supersymmetry operators contains a singlet we obtain a function of the topologically twisted \tsf{VM}'s and the metric which is -- moreover -- $\CQ$-closed, and therefore metric independent. 
We take 
\be 
G^{\mr{tw,bck}} = ( \mr{Spin(4)} \times \IT^{\mr{f}} )/C^{\mr{tw,bck}} ~,
\ee
and as in equation \eqref{eq:WittenHomConstraint} we require: 
\be
    \begin{tikzcd}
        G^{\mr{tw,bck}} \ar[r, "\varphi^{\mr{tw,bck}}"] \ar[d, "p_1", swap]   &   \wt{G}_T/C_T \ar[d,"p_1\times p_2"] \\
       \mr{Spin(4)}/\langle (-1,-1)\rangle    \ar[r, "\varphi^{\mr{W}}"] &  (\mr{Spin(4)} \times \mr{SU(2)_R})/\langle (-1,-1,-1)\rangle  \\ 
    \end{tikzcd}
\ee

The description of $G^{\mr{tw,bck}}$ and $\varphi^{\mr{tw,bck}}$ is now very similar to that of the Lagrangian case. For example, for the $A_{N-1}$ $\mr{fff}$ trinion theory, we can take 
\be 
G^{\mr{tw,bck}} =(  \mr{Spin(4)} \times \IT( \mr{SU(N)} \times \mr{SU(N)} \times \mr{SU(N)} ) )/C ~,
\ee
\be
C = \{ (\zeta,\zeta,\omega^{k_1} \mathbf{1}_N, \omega^{k_2} \mathbf{1}_N, \omega^{k_3} \mathbf{1}_N ) ~|~ \zeta^{N-1} \omega^{k_1+k_2+k_3}=1 \}~,
\ee
and 
\be 
\varphi^{\mr{tw,bck}}( [(u_1, u_2), \mu]) = [ (u_1, u_2), u_2, \mu ]~. 
\ee

\subsection{Gaiotto Gluing}
We are now ready to discuss the general class $\CS$ theory. In a weak-coupling domain, these can be assembled from trinion theories and vectormultiplets by the Gaiotto gluing conjecture \cite{Gaiotto:2009we}. Recall that this states the following: 

Consider two class $\CS$ theories 
$T[\mathfrak{g}, C^{(1)}, D^{(1)}]$ and $T[\mathfrak{g}, C^{(2)}, D^{(2)}]$ on (stable) Riemann surfaces $C^{(1)}$ and $C^{(2)}$, with full punctures $D_{p_1}$ in theory one,  $D_{p_2}$ in theory two (see footnote \ref{foot:dataS}). Then, since $\wt{G}_{p_1}  = \wt{G}_{p_2}$ we can gauge the diagonal subgroup
$\wt{G}$ of $\wt{G}_{p_1} \times  \wt{G}_{p_2}$ with coupling constant $\tau$. (When gluing full punctures the beta function vanishes so the coupling is scale-independent.) Denote the new gauged theory by
\be\label{eq:GaiottoGluing1}
T[\mathfrak{g}, C^{(1)}, D^{(1)}]\times_{\tau} T[\mathfrak{g}, C^{(2)}, D^{(2)}] ~.
\ee
On the other hand, we can glue the Riemann surfaces $C^{(1)}$ to $C^{(2)}$ with disks around $p_1$ and $p_2$ using the plumbing fixture $z_1 z_2 = q$. Denote the resulting Riemann surface $C^{(1)}\times_{p_1,p_2,q} C^{(2)}$. The Gaiotto gluing conjecture states that 
\be\label{eq:GaiottoGluing2}
T[\mathfrak{g}, C^{(1)}, D^{(1)}]\times_{\tau} T[\mathfrak{g}, C^{(2)}, D^{(2)}]
= 
T[\mathfrak{g}, C^{(1)}\times_{p_1,p_2,q} C^{(2)}, D^{(1,2)} ] ~,
\ee
where $q=e^{2\pi \im \tau}$ and $D^{(1,2)}$ is the disjoint union of the punctures that were not glued.

\bigskip
\begin{remark}\label{rem:gluing}
One may wonder if one could gauge a quotient of $\wt{G}_p$ in the gluing procedure.\footnote{We thank Y. Tachikawa for some interesting remarks on this point.}
Moreover, one may ask how the data making the theory associated with the glued surface absolute, is related to such data for the two factors. These are interesting and nontrivial questions we must leave for another time.
\end{remark}

%
%

\subsection{Expressing The Theory As A Gauge Theory }

Now consider a pants decomposition of $C_{g,n}$ into a set $\CT$ of trinions. We let $p_i$ denote the external punctures, where $i$ runs over the set of external punctures. The internal punctures come in unordered pairs 
$\{ p_{\alpha,1}, p_{\alpha,2} \}$ with $\wt{G}_{p_{\alpha,1}} \cong  \wt{G}_{p_{\alpha,2}}$
where $\alpha$ runs over the set of internal cutting curves.  
Consider first the product of all the trinion theories associated to $T\in \CT$. This will have a global symmetry group based on 
\be \label{eq:GT-tilde-trinion}
\wt{G}_{\CT}  = \mr{Spin(4)} \times \mr{SU(2)_R} \times \prod_i \wt{G}_{p_i} \times \prod_{\alpha} \left( \wt{G}_{p_{\alpha,1}} \times \wt{G}_{p_{\alpha,2}}\right) ~.
\ee
The possible theories we can construct from this product of trinion theories will have global symmetry 
$\wt{G}_{\CT} /C_\CT$. Again, if we wish our domain category to include non-spin manifolds we use the considerations of section 
\ref{sec:GeneralSetup} to conclude that  
\be\label{eq:ClassS-condition2}
(p_1\times p_2)(C_\CT ) = \langle (-1,-1,-1) \rangle \subset Z(\mr{Spin(4)} \times \mr{SU(2)_R}) ~. 
\ee
as well as 
\be\label{eq:ClassS-condition1}
C_{\CT}  \subset C_{\CT}^{\mr{max}}:=  \{ (\zeta,\zeta,\zeta, z) ~\vert~  (\zeta,\zeta,\zeta,z)\cdot \wt{\mr{Z}}_T = \wt{\mr{Z}}_T \quad \forall\,\,T\in \CT \}  ~. 
\ee
where $z$ summarizes the component in the factors other than $\mr{Spin(4)} \times \mr{SU(2)_R}$. 

\bigskip
\begin{ex}
Suppose, for simplicity, that all the punctures are ``full'' punctures. 
For a given trinion $T\in \CT$ let $R_T(z)$ be the $N^{th}$ root of unity obtained by taking 
the product of the $\mr{SU(N)}$ factors in $z$ pertaining to $T$. Then we require that $\zeta R_T(z)=1$ 
for all $T\in \CT $ when $N$ is even and $R_T(z)=1$ for all $T\in \CT$ when $N$ is odd. 
\end{ex}
\bigskip

Now we wish to gauge the diagonal flavor symmetry associated with the cutting curves. 
For each $\alpha$ consider the homomorphism\footnote{Other choices of gauge group are possible \cite{Chacaltana:2010ks}.  We restrict to the simplest case. }
\be \label{eq:psi_glue_1}
\psi_{\mr{glue}}: \prod_{\alpha} \mr{SU(N)}_{\alpha}  \to \wt{G}_{\CT} ~,
\ee
defined by 
%
\be \label{eq:psi_glue_2}
\psi_{\mr{glue}} :\prod_{\alpha} g_{\alpha} \longmapsto 
\bigg(  \mathds{1}_{\mr{Spin(4)}}, \mathds{1}_{\mr{SU(2)_R}},  \prod_i \mathds{1}_{\wt{G}_i}, \prod_{\alpha} (g_\alpha,g_\alpha)    \bigg) ~.
\ee
Let $\pi: \wt{G}_{\CT} \rightarrow \wt{G}_{\CT} /C'_{\CT}$ be the projection where 
\be\label{eq:CT'_general}
C'_{\CT} = \biggl[ \mr{Spin(4)} \times \mr{SU(2)_R} \times \prod_i \wt{G}_{p_i} \times \prod_{\alpha} \left( \Delta(\wt{G}_{p_{\alpha}} )  \right)\biggr]  \cap C_\CT ~,
\ee
where $\Delta(G) \subset G \times G$ is the diagonal subgroup.   Then $\pi\circ \psi_{\mr{glue}} $ will have a   nontrivial kernel $K \subset Z\big(\prod_\alpha \mr{SU(N)}_\alpha\big)$. The gauge group associated with the cutting curves is 
\be\label{eq:Class-S-cutcurvegaugegroup}
\bigg( \prod_\alpha \mr{SU(N)}_\alpha \bigg)\bigg/K ~ . 
\ee
This determines what 't Hooft fluxes can be turned on, and what 
't Hooft fluxes must be summed over when computing physical quantities. 

The analog of $G^{\mr{phys}}$ from section \ref{sec:NonSpin} is now 
\be\label{eq:Gphys-Class-S}
G_{\mr{S}} = \bigg( \mr{Spin(4)} \times \mr{SU(2)_R} \times \prod_i \wt{G}_{p_i} \times \prod_{\alpha} \mr{SU(N)}_\alpha  \bigg)\bigg/C''_{\CT} ~.
\ee
(where $C''_{\CT}$ is the preimage of $C'_{\CT}$ under the product over $\alpha$ of the map $\Delta: \mr{SU(N)}\to \Delta(\mr{SU(N)})$). 

\noindent
\begin{remark}
\text{}
\begin{enumerate}
\item As a simple example consider the gluing of two $A_{N-1}$ trinion theories with $\mr{fff}$ punctures to form a genus two surface. Choosing $C= C^{\mr{max}}_T$ the gauge group of the theory is 
\be 
\left( \mr{SU(N)} \times \mr{SU(N)} \times \mr{SU(N)} \right) \big/\{ 
( \omega^{k_1}\mathbf{1}_N,\omega^{k_2}\mathbf{1}_N,\omega^{k_3}\mathbf{1}_N) \vert 
\omega^{k_1+k_2+k_3} = 1 \}~. 
\ee
Note that $K$ is not a direct product of the form $\prod_{\alpha} K_\alpha$.

\item The existence of a principal $G_{\mr{S}}$ bundle  $P_{\mr{S}} \to \IX$ in terms of cohomology classes of the bundles associated with transfer of structure group for the projections on the factors will be based on conditions analogous to the conditions \eqref{eq:BackgroundGerbeCondition} we found above. Some explicit examples are 
given in Appendix \ref{app:Class-S-Coho-Examples}.

\item It is an interesting question whether all the theories constructed above for different choices of $C$ have six-dimensional descriptions, and what the extra data are in those descriptions. This question becomes highly relevant when considering the action of $S$-duality. In section 
\ref{sec:a1classS} we will assume that all the theories have a 6d origin, and therefore the mapping class groups permute them as an $S$-duality group. 
\end{enumerate}
\end{remark}

\bigskip
\bigskip

Now we turn to the partition function of the theory $\CS$ on a manifold $\IX$. 
It will have the general form:  
\be \label{eq:Z-Class-S}
\hspace{-0.2in}\mr{Z}_{\CS}[\omega_\mu{}^{\ua\,\ub}, A_\mu^{\mr{R}}, \mr{VM}_i] = 
\int \prod_{c_\alpha} [d\mr{VM}_{c_\alpha}]   e^{- S_{\mr{VM}_{c_\alpha}} } \prod_{T} 
Z_{T}[\omega_\mu{}^{\ua\,\ub}, A_\mu^{\mr{R}}, \mr{VM}_{p_{T,1}}, 
\mr{VM}_{p_{T,2}},\mr{VM}_{p_{T,3}}]~, 
\ee
where the product over $c_{\alpha}$ is over the internal cutting curves and the product over $T$ is a product over the trinions, and the vectormultiplet path integrals are for the gauge group \eqref{eq:Class-S-cutcurvegaugegroup}.  
We choose the external $\mr{VM}_i$ so that the mass terms are generic, the external fermions are zero, and the structure group is reduced to the maximal torus. However, for the internal cutting curves $c_{\alpha}$ one must maintain the full $\mr{VM}$ in writing $\mr{Z}_{T}$ in the integrand of the path integral. 

\subsection{Topological Twisting }

Having framed the discussion as above, topological twisting proceeds in a completely analogous fashion to what we have described in previous sections.
We choose a homomorphism $\varphi^{\mr{tw,bck}}$ so that 
\be\label{eq:TopTwistForGenClassS}
    \begin{tikzcd}
        G^{\mr{tw}} \ar[r, "\varphi^{\mr{tw,bck}}"] \ar[d, "p_1",swap]   &   G_{\mr{S}} \ar[d,"p_1\times p_2"] \\
       \mr{Spin(4)}/\langle (-1,-1)\rangle    \ar[r, "\varphi^{\mr{W}}"] &  (\mr{Spin(4)} \times \mr{SU(2)_R})/\langle (-1,-1,-1)\rangle ~. \\ 
    \end{tikzcd} 
\ee
Much as in the previous discussion we can take 
\be
G^{\mr{tw}} = \bigg(\mr{Spin(4)} \times \IT\bigg( \prod_i \widetilde G_i\bigg)  \times \bigg(\prod_\alpha \mr{SU(N)}_\alpha\bigg)\bigg)\bigg/C^{\mr{tw}}_{\mr{S}}  ~.
\ee
To find $C^{\mr{tw}}_{\mr{S}} $ we begin with 
\be 
\widetilde G^{\mr{tw}} =\mr{Spin(4)} \times \IT\bigg( \prod_i \widetilde G_i\bigg)  \times \bigg(\prod_\alpha \mr{SU(N)}_\alpha\bigg) ~,
\ee
and consider $\widetilde \varphi^{\mr{tw}}:\widetilde G^{\mr{tw}} \rightarrow  G_{\mr{S}}$ defined by
\be \label{eq:varphi-tilde-class-S}
\widetilde \varphi^{\mr{tw}}( (u_1, u_2), \mu, g) ) := [ (u_1,u_2),u_2,\mu, g]  ~.
\ee
The homomorphism has a kernel, which we can take to be  $C^{\mr{tw}}_{\mr{S}}$.

Upon topological twisting the actions $S_{\mr{VM}_{c_\alpha}}$ will be precisely of the form 
equation \eqref{eq:Stwisted-vec}. After doing the path integral over the internal vectormultiplets the path integral will depend on background fields. From this, we derive  
$ G^{\mr{tw,bck}}$. The path integral will be invariant under continuous deformation 
of connection on the $ G^{\mr{tw,bck}}$-bundle and therefore will only depend on the 
topological data outlined in items \ref{it:top_data1}--\ref{it:top_data3}.

\section{Twisting $\mf{a}_1$ Class $\s{S}$ Theories And S-duality}\label{sec:a1classS}

The ideas presented so far can be used to twist $\mf{a}_1$ class $\s{S}$ theories.  The cohomological conditions 
for the existence of $P^{\mr{phys}}$ were presented in \cite{Manschot:2021qqe}. Since these theories are Lagrangian theories we can rederive the conditions of \cite{Manschot:2021qqe}
using  the discussion in section \ref{sec:cohcond}.\footnote{In \cite{Manschot:2021qqe} these conditions were cited as conditions for 
topological twisting. One of the points of this paper is that the cohomological conditions for the existence of $P^{\mr{phys}}$ should be distinguished from topological twisting. } 
Then, in section \ref{sec:Coh_cond_S-duality}  we will study how these conditions change under $S$-duality, using the fact that these are also class $\CS$ theories.

\subsection{Cohomological Conditions For The $\mf{a}_1$ Class $\CS$ Theory}\label{sec:a1-class-S-cohcond}

The weak coupling descriptions of  $T[\mf{a}_1,C_{g,n}]$   corresponding to different pants decompositions of $C_{g,n}$ were studied in \cite{Gaiotto:2009we,Gaiotto:2010be,Drukker:2009tz}. 
The gauge algebra of the theory is
$\mf{g}^{\mr{gauge}}= \mr{Lie}(G^{\mr{gauge}}) = \mf{su}(2)^{3g-3+n}$, and as before, $\wt{G}^{\mr {phys }}= \mr{Spin(4)} \times \mr{SU(2)_R} \times G^{\mr{f}}\times \wt{G}^{\mr{gauge}}$, 
where
\begin{equation}
\wt{G}^{\mr{gauge}} \cong \mr{S U(2)}^{3 g-3+n}, \quad n = \mr{rank}(G^{\mr{f}}) ~.
\end{equation}
We will assume masses are generic and there is a reduction of flavor structure group to its maximal torus so we choose the domain category so that $G^{\mr{f}} \cong \mr{U(1)}^n$. 
%
%
Note that
\begin{equation}
Z\big(\wt{G}^{\mr {phys }}\big) \cong  \mathbb{Z}_2 \times \mathbb{Z}_2 \times \mathbb{Z}_2 \times \mr{U(1)}^n\times \mathbb{Z}_2^{3 g-3+n}  ~.
\end{equation}
The different global forms of the physical theory are then characterized by choices of $C^{\mr {phys }} \subset Z(\wt{G}^{\mr {phys }})$, subject to the requirements discussed in section \ref{sec:NonSpin}. There are several choices of $C^{\mr{phys }}$ (see, for example, \cite[Section 2.3.1]{Gaiotto:2010be} for a related discussion).

We then identify
\begin{align}\label{eq:cS-Gtilde}
\wt{G}^{\mr{gauge}} \quad &= \prod_{\substack{ c_\alpha \\ \mr { internal }}} \mr{S U(2)}_{c_\alpha} ~, \quad 
G^{\mr{f}} \quad =\prod_{\substack{c_i \\\mr { external }}} \mr{U(1)}_{c_i} ~.
\end{align}
We define the projection maps,
\begin{equation}\label{eq:cS-proj-p1calpha}
p_{4,c_\alpha}:\wt{G}^{\mr{p h ys}}\longrightarrow\mr{SU(2)}_{c_\alpha} ~,\quad 
p_{3,c_i}:\wt{G}^{\mr{p h ys}}\longrightarrow \mr{U(1)}_{c_i}~,    
\end{equation}
to the $c_{\alpha}^{\mathrm{th}}$ and $c_{i}^{\mathrm{th}}$ factors respectively. The preliminary discussion so far has been for the physical $\mf{a}_1$ class $\CS$ theory on $\IR^{4}$ (after a Wick rotation from $\IR^{1,3}$). Next, we seek cohomological conditions -- guided by the discussion of Section \ref{sec:cohcond} -- from the existence of the bundle $P^{\mr{phys}} \to \IX$, for the 
$\mf{a}_1$ class $\CS$ theory on the 4-manifold $\IX$.  

As discussed above, we choose $C^{\mr{phys }}$ so that the field theory functor factors through 
a functor from a bordism category that includes non-spin manifolds. As before, this leads to 
\begin{equation}
    w_2(\IX)=w_2(P^{\mr{R}})~.
\end{equation}

To derive additional conditions, we first note that $p_4(C^{\mr {phys }}) \cong C^{\mr{grb}}\subset\IZ_2^{3 g-3+n}$, 
so the characteristic class $\mu(b) \in H^2(\IX, C^{\mr{grb}})$ of the $C^{\mr{grb}}$-gerbe $b$ can then be projected to a tuple of characteristic classes $\mu(b)_{c_\alpha} \in H^2\left(\IX, \mathbb{Z}_2\right)$, one for each internal cutting curve $c_\alpha$, by projecting $C^{\mr {phys }}$ onto the $c_\alpha^{th}$ factor using \eqref{eq:cS-proj-p1calpha}.\footnote{Explicitly, let $\varpi$ be the map $\varpi: H^2(\IX, C^{\mr{grb}}) \to H^{2}(\IX, \IZ_2^{3g-3+n})$ induced by the inclusion $C^{\mr{grb}} \hookrightarrow \IZ_{2}^{3g-3+n}$. Therefore $\varpi(\mu(b))$ is a $(3g-3+n)$-tuple of classes in $H^{2}(\IX, \IZ_2)$. Then $p_{4,c_{\alpha}}$ induces a natural projection onto the $c_{\alpha}^{th}$ factor of this tuple, which we denote by $\mu(b)_{c_\alpha} \in H^2(\IX, \IZ_2)$. If $\mu(b) \in \mathsf{ker\,}\varpi$, then the corresponding $\mu(b)_{c_\alpha} = 0$ for all $c_{\alpha}$.} 
Note that, depending on 
our choice of $C^{\mr {phys }}$, which determines $C^{\mr{grb}}$, it might happen that $\mu(b)_{c_\alpha}$ vanishes. We also recall remark \ref{item:gerbecond} above: one should include a topological boundary condition $\rho$ for the 5d gerbe $b$ which determines which 't Hooft fluxes are to be summed over.

The matter content of the theory is a collection of half-hypermultiplets $\Phi_{c_{1},c_2,c_3}$
labeled by unordered triples of cutting curves of the pants decomposition.\footnote{Therefore, the scalar fields in $\Phi_{c_{1},c_2,c_3}$ are in a real subspace 
of the $2_{c_1} \otimes 2_{c_2} \otimes 2_{c_3} \otimes 2_{\mr{SU(2)_R}}$ giving a total of 
16 independent real scalar fields. }
Let us begin by considering conditions associated with $\Phi_{c_i c_j c_\alpha}$, containing 
 one internal and two external punctures. 
Introduce the notation 
\begin{equation} \label{eq:P-f-ci}
P^{\mr{f}, c_i}=(p_{3, c_i})_*(P^{\mr {phys }})~.
\end{equation}
Then $P^{\mr{f}, c_i}$ is a $\mr{U(1)}/ p_{3, c_i}(C^{\mr {phys }})$-bundle. Let $c_1(c_i)$ denote the first Chern class of the $\mr{U(1)}$-bundle obtained by transfer of structure group of $P^{\mr{f}, c_i}$ along the isomorphism (see footnote \ref{foot:power_n_map}) $\mr{U(1)}/ p_{3, c_i}(C^{\mr {phys }})\xrightarrow{\cong} \mr{U}(1)$.
Now, item \ref{it:G-const4} of Section \ref{sec:cohcond} (or \eqref{eq:BackgroundGerbeCondition}) directly yields the conditions
\begin{equation}\label{eq:coh_cond_a1_1int}
\begin{aligned}
& c_1\left(c_i\right)+\mu(b)_{c_\alpha}+ w_2(\IX)=0 \bmod 2~, \\
& c_1(c_j)+\mu(b)_{c_\alpha} + w_2(\IX)=0 \bmod 2~.
\end{aligned}
\end{equation}
Next, we consider trinions $\Phi_{c_i,c_\alpha,c_\beta}$ bounded by one external cutting curve $c_i$ and two internal cutting curves $c_\alpha$, $c_\beta$. 
Following Section \ref{sec:cohcond}, we obtain
\begin{equation}\label{eq:coh_cond_a1_2int}
c_1(c_i)+\mu(b)_{ c_\alpha}+\mu(b)_{ c_\beta}+w_2(\IX)=0 \bmod 2 ~.
\end{equation}
Finally for trinions $\Phi_{c_\alpha c_\beta c_\gamma}$ bounded by three internal cutting curves $c_\alpha$, $c_\beta$, $c_\gamma$, the matter $\Phi_{c_\alpha c_\beta c_\gamma}$ is a half-hypermultiplet in the $\bm{2}_\alpha \otimes \bm{2}_\beta \otimes \bm{2}_\gamma$ of the $\mr{SU(2)}_{c_\alpha} \times \mr{S U(2)}_{c_\beta} \times \mr{S U(2)}_{c_\gamma}$ factor of $\wt{G}$. The discussion in Section \ref{sec:cohcond} leads to the cohomological condition
\begin{equation}\label{eq:coh_cond_a1_3int}
\mu(b)_{c_\alpha}+\mu(b)_{c_\beta}+\mu(b)_{c_\gamma}+w_2(\IX)=0 \bmod 2 ~.
\end{equation}
\subsection{Transformation Of Cohomological Conditions Under S-duality}\label{sec:Coh_cond_S-duality}
We next analyze the effect of S-duality transformations on these cohomological conditions. Recall that the S-duality group of the theory is the mapping class group $\mr{MCG}(C_{g,n})$ and acts on the space of gauge couplings of the theory (identified with the Teichm\"uller space of $C_{g,n}$, see \cite{Witten:1997sc,Gaiotto:2009hg,Gaiotto:2009we}). $\mr{MCG}(C_{g,n})$, being the group of isotopy classes of orientation-preserving diffeomorphisms of $C_{g,n}$, acts on the set of simple closed curves on $C_{g,n}$, and thus also acts on the (infinite) set of all pants decompositions of $C_{g,n}$. The set of equivalence classes of pants decompositions modulo the action of the mapping class group $\mr{MCG}(C_{g,n})$ is finite.\footnote{Note that $\mr{MCG}(C_{g,n})$ is isomorphic to the automorphism group of the \textit{pants graph} or \textit{pants complex} of $C_{g,n}$, see \cite{Margalit} for a proof. This is the pants complex of Hatcher and Thurston \cite{HatcherThurston1980} with the inclusion of 2-cells by Hatcher, Lochak, and Schneps \cite{hatcher2000teichmuller}. The resulting pants complex is connected and simply connected. $\mr{MCG}(C_{g,n})$ acts co-compactly on this complex \cite{wolf}.} 
Thus, there are finitely many S-duality orbits of theories based on $C_{g,n}$. We will show that there is a natural transformation of the cohomological conditions of theories in the same S-duality orbit. To do so, we introduce the \textit{graph} of a pants decomposition of $C_{g,n}$.

\begin{figure}
    \centering
    \includegraphics[scale=0.13]{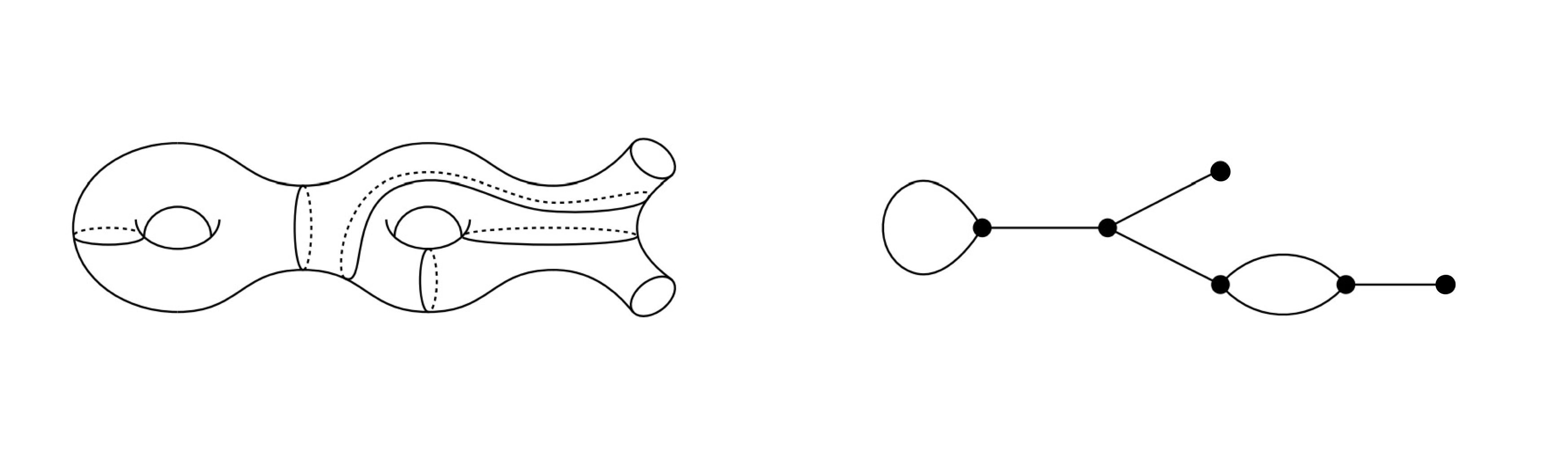}
    \caption{The graph of a pants decomposition of $C_{2,2}$. Figure adapted from \cite{wolf}.}
    \label{fig:ptoG(p)}
\end{figure}

Given a pants decomposition $\mf{p}$, we define an undirected graph $G(\mf{p})$ of $\mf{p}$ with vertices as the pairs of pants \textit{and} external cutting curves in $\mf{p}$, such that two pairs of pants are connected by an edge in $G(\mf{p})$ if they are bounded by a common internal cutting curve. Further, an external cutting curve (a vertex in $G(\mf{p})$) is connected by an edge in $G(\mf{p})$ to that pair of pants which it bounds (also a vertex in $G(\mf{p})$). Thus $G(\mf{p})$ has $2g-2+n$ trivalent vertices, $n$ univalent vertices and $3g-3+2n$ edges. An illustrative example is shown in Figure \ref{fig:ptoG(p)}.
Two pants decompositions $\mf{p}$, $\mf{p}'$ are in the same $\mr{MCG}(C_{g,n})$-orbit if and only if the corresponding graphs $G(\mf{p})$, $G(\mf{p}')$ are isomorphic \cite{wolf}.\footnote{Two graphs $G(\mf{p})$ and $G(\mf{p}')$ are said to be isomorphic if there exists a bijection between their vertex sets $f: V\big(G(\mf{p})) \to V\big(G(\mf{p}')\big)$ such that any two vertices $v_1, v_2 \in V\big(G(\mf{p})\big)$ of $G(\mf{p})$ are adjacent in $G(\mf{p})$ if and only if their images $f(v_1)$ and $f(v_2)$ are adjacent in $G(\mf{p}')$. When such an isomorphism exists between $G(\mf{p})$ and $G(\mf{p}')$, we will frequently write it as $f: G(\mf{p}) \to G(\mf{p}')$, with the understanding that it is an isomorphism between the corresponding vertex sets.}

We can now describe the cohomological conditions for the theory defined by the pants decomposition $\mf{p}$ in terms of the graph $G(\mf{p})$ as follows: we begin by labeling the vertices and edges in $G(\mf{p})$ as 
\begin{align}
   V\big(G(\mf{p})\big) :=\{t_1,\ldots,t_{2g-2+n},u_1,\ldots,u_n\}~ \label{eq:Vertices-pants-graph},
   \\ E\big(G(\mf{p})\big):=\{e_1,\ldots,e_{3g-3+n},\wt{e}_1,\ldots,\wt{e}_n\}~, \label{eq:Edges-pants-graph}
\end{align}
where $t$ and $u$ indicate trivalent and univalent vertices respectively, whereas the edge $e_\alpha$ connects two trivalent vertices and $\wt{e}_i$ connects a univalent vertex with a trivalent vertex. 

In section \ref{sec:a1-class-S-cohcond}, we encountered maps $c_i \mapsto c_1(c_i) \in H^{2}(\IX, \IZ)$ and $c_\alpha \mapsto \mu(b)_{c_\alpha} \in H^{2}(\IX, \IZ_2)$. By design, the set of external cutting curves $\{c_i\}_{i=1}^{n}$ is bijectively mapped to the subset $\{\wt{e}_i\}_{i=1}^{n}$ of edges of $G(\mf{p})$. Likewise, the set of internal cutting curves $\{c_\alpha\}_{\alpha=1}^{3g-3+n}$ also gets bijectively mapped to the subset $\{e_\alpha\}_{\alpha=1}^{3g-3+n}$. Note, however, that in each case, there might be several bijections due to different choices of indexing. We simply pick one, so that we have one-to-one map between elements of the sets $\{c_i\}_{i=1}^{n}$ and $\{\wt{e}_j\}_{j=1}^{n}$, and between elements of the sets $\{c_\alpha\}_{\alpha=1}^{3g-3+n}$ and $\{e_\beta\}_{\beta=1}^{3g-3+n}$.

Then, with this choice of map, we can associate fluxes to the edges in $G(\mf{p})$:
\begin{align}
 \label{eq:c1fi}  \wt{e}_{i} &\longmapsto  c_1(\wt{e}_i) \in H^{2}(\IX, \IZ) \quad \,\,\,\,\text{ for } i=1,\ldots,n~,\\ 
\label{eq:f2ea} e_{\alpha} &\longmapsto \mu(b)_{e_\alpha} \in H^{2}(\IX, \IZ_2) \quad \text{ for } \alpha=1,\ldots,3g-3+n~,
\end{align}
with the proviso that $\mu(b)_{e_\alpha}=0$ if $e_\alpha$ is a loop (i.e., an edge connecting one vertex to itself).
Then, the cohomological conditions described in the preceding section can be recast into equations at each trivalent vertex. If a trivalent vertex is connected by three edges $e_\alpha$, $e_{\beta}$, $e_\gamma$ to three other trivalent vertices, then we impose the condition 
\begin{equation}\label{eq:eabgf2}
\mu(b)_{e_\alpha}+ \mu(b)_{e_\beta}+\mu(b)_{e_\gamma}+w_2(\IX)=0~.   
\end{equation}
If a trivalent vertex is connected by edges $e_\alpha,e_\beta,\wt{e}_i$ to two trivalent and one univalent vertex, then we have the cohomological condition  
\begin{equation}\label{eq:eabf2c1fi}
c_1(\wt{e}_i)+\mu(b)_{e_\alpha}+\mu(b)_{e_\beta}+w_2(\IX)=0\bmod 2 ~.    
\end{equation}
Finally, if a trivalent vertex is connected by edges $e_\alpha$, $\wt{e}_i$, $\wt{e}_j$ to a trivalent and two univalent vertices, then 
\begin{align}
\label{eq:c1fiea} c_1(\wt{e}_i)+\mu(b)_{e_\alpha}+w_2(\IX)=0\bmod 2~,\\ 
\label{eq:c1fjea} c_1(\wt{e}_j)+\mu(b)_{e_\alpha}+w_2(\IX)=0\bmod 2~.   
\end{align}
%
Now consider two pants decompositions $\mf{p}$, $\mf{p}'$ with graphs $G(\mf{p})$, $G(\mf{p}')$ in the same $\mr{MCG}(\Sigma_{g,n})$-orbit. 
Let the vertices and edges of $G(\mf{p}')$ be labeled as 
\begin{align}
 V\big(G(\mf{p}')\big):=\{t'_1,\ldots,t'_{2g-2+n},u'_1,\ldots,u'_n\}~,
   \\ E\big(G(\mf{p}')\big):=\{e'_1,\ldots,e'_{3g-3+n},\wt{e}'_1,\ldots,\wt{e}'_n\}~.
\end{align}
We associate fluxes to the edges as in \eqref{eq:c1fi} and \eqref{eq:f2ea}.

Recall that for an undirected graph $G(\mf{p})$, its line graph $L(G(\mf{p}))$ is constructed in the following way: to each edge in $G(\mf{p})$, we assign a vertex in $L(G(\mf{p}))$\footnote{Therefore, $V\big(L(G(\mf{p}))\big) \cong E\big(G(\mf{p})\big)$.}
and for every two edges in $G(\mf{p})$ emanating from a common vertex, we assign an edge between their corresponding vertices in $L(G(\mf{p}))$. Since $G(\mf{p})\cong G(\mf{p}')$, by the Whitney graph isomorphism theorem \cite[Theorem 8.3]{harary1969graph}, their line graphs 
%
are also isomorphic:
%
%
\begin{equation}\label{eq:isolGpp'}
    L(G(\mf{p}))\cong L(G(\mf{p}'))~.
\end{equation}
Moreover, there is a one-to-one correspondence between graph isomorphisms $G(\mf{p}) \cong G(\mf{p}')$ and the corresponding line graph isomorphisms $L(G(\mf{p})) \cong L(G(\mf{p}'))$.\footnote{\label{foot:Whitney}The Whitney graph isomorphism theorem implies an equivalence between isomorphisms of graphs and isomorphisms of corresponding line graphs with the following exception: the graphs $K_3$ (the complete graph on three vertices, a.k.a. the three-vertex triangle) and $K_{1,3}$ (a complete bipartite graph, a.k.a. the four-vertex claw) are non-isomorphic but their line graphs are isomorphic. (See \cite[Ch. 8]{harary1969graph} for a discussion of this point.) However, this does not concern us for two reasons: first of all, we require only the forward implication of the theorem which says that \textit{if} two graphs are isomorphic, \textit{then} their corresponding line graphs are isomorphic, and second, the graphs $K_3$ and $K_{1,3}$ cannot arise from any pants decomposition.}
The vertices of the graphs $L(G(\mf{p}))$ and $L(G(\mf{p}'))$ are respectively, $E\big(G(\mf{p})\big)$ and $E\big(G(\mf{p}')\big)$. Choose an isomorphism $\bm{F}:E\big(G(\mf{p})\big)\to E\big(G(\mf{p}')\big)$ and let $f:G(\mf{p})\to G(\mf{p}')$ denote the graph isomorphism corresponding to $\bm{F}$. 

Let us denote an edge in $G(\mf{p})$ connecting two vertices $u$ and $v$ by $uv$. 
%
By arguments outlined in \cite[Theorem 8.3]{harary1969graph}, we have
\begin{equation}
    \bm{F}(uv)=f(u)f(v)~.
\end{equation}
In particular, if $e_\alpha$, $e_\beta$, and $e_\gamma$ are edges in $G(\mf{p})$ connecting a vertex $u$ to three other \textit{trivalent} vertices, then $\bm{F}(e_\alpha)$, $\bm{F}(e_\beta)$, $\bm{F}(e_\gamma)$ are three edges in $G(\mf{p}')$ connecting $f(u)$ to three other \textit{trivalent} vertices in $G(\mf{p}')$. Thus by \eqref{eq:eabgf2}, the cohomological condition at the trivalent vertex $f(u)$ in $G(\mf{p}')$ is  
\begin{equation}\label{eq:eabgf2pprime}
\mu(b')_{\bm{F}(e_\alpha)}+ \mu(b')_{\bm{F}(e_\beta)}+\mu(b')_{\bm{F}(e_\gamma)}+w_2(\IX)=0~,   
\end{equation}
where $b'$ is the gerbe for the theory defined by the pants decomposition $\mf{p}'$. 
Arguing along the same lines, we conclude that the cohomological conditions for the theory defined by $\mf{p}'$ are given by applying the map
\begin{equation}
    \wt{e}_i\longmapsto \bm{F}(\wt{e}_i)~,\quad e_\alpha\longmapsto \bm{F}(e_\alpha)~,
\end{equation}
to \eqref{eq:eabgf2}, \eqref{eq:eabf2c1fi}, \eqref{eq:c1fiea} and \eqref{eq:c1fjea}. 
Thus, we see that at least in this class of examples, $S$-dual theories with different Lagrangians require precisely the same cohomological conditions as needed for the existence of $P^{\mr{phys}}$.  

We emphasize that in general the cohomological conditions for two Lagrangian theories which are not in the same S-duality orbit are different and \textit{not} related by a nice transformation. Consider, for example, the $\mr{a}_1$ class $\CS$ theory on a 3 punctured torus $C_{1,3}$. Consider the two pants decompositions of $C_{1,3}$ shown in Figure \ref{fig:two_pants_3ptorus} below.
\begin{figure}[H]
    \centering
    \includegraphics[width=\linewidth]{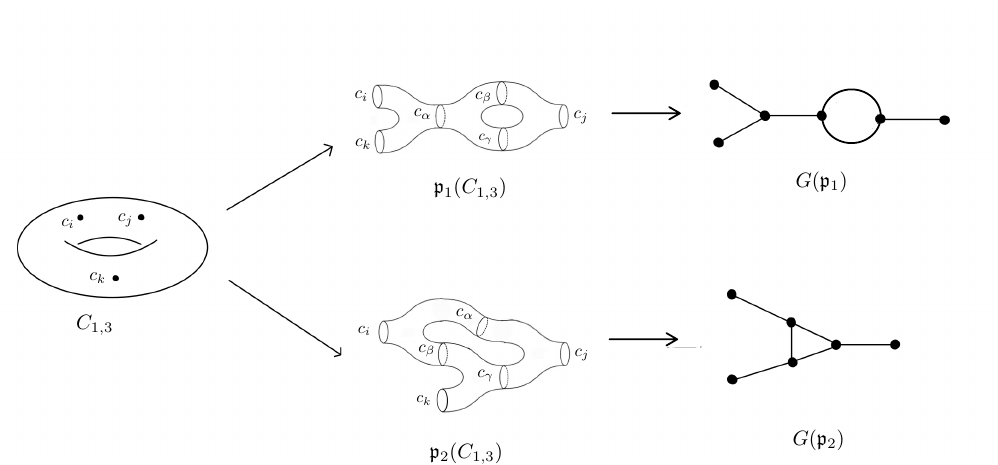}
    \caption{Two $\mr{MCG}(C_{1,3})$-inequivalent pants decompositions of $C_{1,3}$ and their associated pants graphs.}
    \label{fig:two_pants_3ptorus}
\end{figure}
The pants graphs are non-isomorphic as can be easily checked using their adjacency matrix and hence their corresponding Lagrangian theories are \textit{not} related by $S$-duality. Let us now look at the cohomological conditions we get for the two theories using the general prescription described above. There are three pairs of pants in both pants decompositions. With the labeling of cutting curves in Figure \ref{fig:two_pants_3ptorus}, we obtain the following cohomological conditions for the two theories:
\begin{enumerate}
\item  For the pants decomposition $\mf{p}_1(C_{1,3})$ we get
\begin{equation}
\begin{aligned}
c_1(c_i)+\mu(b)_{c_\alpha}+w_2(\IX) & =0 \bmod 2~, \\
c_1(c_k)+\mu(b)_{c_\alpha}+w_2(\IX) & =0 \bmod 2~, \\
\mu(b)_{c_\alpha}+\mu(b)_{c_\beta}+\mu(b)_{c_\gamma}+w_2(\IX) & =0 \bmod 2~, \\
c_1(c_j)+\mu(b)_{c_\beta}+\mu(b)_{c_\gamma}+w_2(\IX) & =0 \bmod 2~.
\end{aligned}
\end{equation}
\item For the pants decomposition $\mf{p}_2(C_{1,3})$ we get
\begin{equation}
\begin{aligned}
& c_1(c_i)+\mu(b)_{c_\alpha}+\mu(b)_{c_\beta}+w_2(\IX)=0 \bmod 2~, \\
& c_1(c_k)+\mu(b)_{c_\beta}+\mu(b)_{c_\gamma}+w_2(\IX)=0 \bmod 2~, \\
& c_1(c_j)+\mu(b)_{c_\alpha}+\mu(b)_{c_\gamma}+w_2(\IX)=0 \bmod 2~.
\end{aligned}
\end{equation}
\end{enumerate}
We see that in the second case, we have no restriction on background gerbes since we can always choose 
a suitable $c_1(c_{i,j,k})$ while in the first case, the background gerbes are restricted. Thus the cohomological conditions for the two theories are different.

To twist these theories, we need to provide $C^{\mr{tw}}$ and a homomorphism $\varphi^{\mr{tw}}:G^{\mr{tw}}\to G^{\mr{phys}}$ satisfying \eqref{eq:WittenHomConstraint}.
Since these theories are Lagrangian, we simply 
choose $C^{\mr{tw}}$ as in
\eqref{eq:AtwdefAphys} above, 
and define
\begin{equation}\label{eq:twhoma1classS}
\begin{aligned}
\varphi^{\mr {tw }}: G^{\mr{tw}} & \longrightarrow G^{\mr {phys }} \\
{\left[\left(\left(u_1, u_2\right), \mu, g\right)\right] } & \longmapsto\left[\left(\left(u_1, u_2\right), u_2, \mu, g\right)\right] .
\end{aligned}
\end{equation}

The theories in different $S$-duality orbits correspond to different topological field theories.

\appendix
\setcounter{table}{0}
\setcounter{figure}{0}
\renewcommand{\thetable}{A.\arabic{table}}
\renewcommand{\thefigure}{A.\arabic{figure}}

\section{Symbol Lists}\label{app:Symbols}

\begin{small}
\begin{table}[H]
\centering 

\caption{\label{tbl:index conventions}Conventions for indices.}
\end{table}
\renewcommand*{\arraystretch}{1.18}
\begin{longtable}[c]{|c|l|c|}
\hline
\textbf{Symbol} & \textbf{Description} & \begin{tabular}{@{}l@{}} \textbf{Reference} \end{tabular} \\
\hline\hline
 \endhead
 \caption{\label{tbl:symbol list}List of symbols used in the paper.}
 \endfirstfoot
 \caption[]{(Continued) List of symbols used in the paper.}
 \endfoot
 $\alpha$ & an element of $C^{\mr{phys}}$ & Sec. \ref{sec:cohcond} \\ \hline
 $\bm{a}$ & an element of $\IT(\mr{SU(N)})$ & Sec. \ref{sec:GenClassS-Twisting} \\ \hline
 $a_i$ & the $i^{th}$ element of $\bm{a}$ & Sec. \ref{sec:GenClassS-Twisting} \\ \hline
 $\mathsf{adj}$ & adjoint representation of a Lie group or algebra & Sec. \ref{sec:backind} \\ \hline
 $A_{\mu}$ & connection on $\mathsf{ad\,}P$ in a local chart & Sec. \ref{sec:GeneralSetup} \\ \hline
 $\CA_{\mu}^{u}$ & \begin{tabular}{@{}l@{}}connection on $\mathsf{ad\,}P^{\mr{f},u}$ in a local chart \\ a $\mr{U(1)}/p_{3,u}(C^{\mr{phys}})$ connection \end{tabular} & Sec. \ref{sec:backind} \\ \hline
 $A_{\mu}^{\mr{R}}$ & a background $\mr{SU(2)_R}$ $\mr{R}$-symmetry connection & Sec. \ref{sec:GenClassS-Twisting} \\ \hline
 $A_{\mu}^{\mr{SU(N)}}$ & a background $\mr{SU(N)}$ (flavor) connection & App. \ref{app:Lag-ffs-Trinion-Analysis} \\
\end{longtable}
 \end{small}

\setcounter{table}{0}
\setcounter{figure}{0}
\renewcommand{\thetable}{B.\arabic{table}}
\renewcommand{\thefigure}{B.\arabic{figure}}
 \section{Transfer Of Structure Group \label{app:transredstructuregroup}}

In this appendix, we recall the notion of transfer of structure group.
%
%
  For simplicity, we will work in the category of smooth manifolds (where the objects are smooth manifolds and morphisms are smooth maps). Let $G$ be a Lie group and $\IX$ a smooth manifold. Let $\CC(G, \IX)$ be the groupoid of principal $G$-bundles over $\IX$. Let $G_1, G_2$ be two Lie groups. Given a homomorphism $\phi: G_1 \longrightarrow G_2$, we have a functor $\phi_*: \CC\left(G_1, \IX\right) \longrightarrow \CC\left(G_2, \IX\right)$ that maps a principal $G_1$-bundle $P_1 \stackrel{\pi_1}{\longrightarrow} \IX$ to $P_1 \times_\phi G_2 \stackrel{\pi'_1}{\longrightarrow} \IX$ where
\begin{equation}
P_1 \times_\phi G_2 := \left(P_1 \times G_2\right) / \sim_\phi ~,
\end{equation}
with   equivalence relation is defined by
\begin{equation}
(p_1 \cdot h, g) \sim_\phi\left(p_1, \phi(h) g\right), \quad(p_1, g) \in P_1 \times G_2, h \in G_1 ~.
\end{equation}
The projection $\pi'_1: P_1 \times_\phi G_2 \longrightarrow \IX$ is defined as $\pi'_1([(p_1, g)])=\pi_1(p_1)$ and the right $G_2$ action is given by
\begin{equation}
[(p_1, g_1)] \cdot g_2=\left[\left(p_1,g_1\cdot g_2\right)\right] ~.
\end{equation}

Let $P_1 \stackrel{\pi_1}{\longrightarrow} \IX$ be a principal $G_1$-bundle. The \emph{ transfer of structure group} of $P_1 \stackrel{\pi_1}{\longrightarrow} \IX$ from $G_1$ to $G_2$ using the homomorphism 
$\phi: G_1 \longrightarrow G_2$ is, by definition,  the image $\phi_*(P_1 \stackrel{\pi_1}{\longrightarrow} \IX)$ under the above functor. One can show that in terms of patches and transition functions, this means that 
if the local trivialization of $P_1 \xrightarrow{\pi_1} \IX$ is 
given by $\{U_p,t_{pq}\}$, then the local trivialization of $\phi_*(P_1 \stackrel{\pi_1}{\longrightarrow} \IX)$ is given by  $\{U_p,\wt{t}_{pq}\}$  with 
$\wt{t}_{pq} = \phi\circ t_{pq} $.
\par A related notion is a \textit{reduction of structure group} defined as follows: let $P_2 \stackrel{\pi_2}{\longrightarrow} \IX$ be a principal $G_2$-bundle. A reduction of structure group of $P_2 \stackrel{\pi_2}{\longrightarrow} \IX$ from $G_2$ to $G_1$ using the homomorphism 
$\phi: G_1 \longrightarrow G_2$ is a principal $G_1$-bundle 
$P_1 \xrightarrow{\pi_1}\IX$ together with an isomorphism of $\phi_*(P_1 \xrightarrow{\pi_1}\IX) $ with $P_2 \xrightarrow{\pi_2} \IX$. One can show that in terms of patches and transition functions, this means that 
there is an atlas and local trivialization of $P_2 \xrightarrow{\pi_2} \IX$
given by $\{U_p,\wt{t}_{pq}\}$ such that there exists a $G_1$ bundle 
$P_1\to \IX$ with local trivializations   $\{U_p,t_{pq}\}$  such that 
$\wt{t}_{pq} = \phi\circ t_{pq}$.

It is worth noting that there is never an obstruction to transferring the structure group of a principal bundle, but there are, in general, nontrivial obstructions to the existence of a reduction of structure group.  
Here are some examples of reductions of structure groups: 

Fix a positive integer $n \in \IZ_{> 0}$. Given an $n$-manifold $\IX$ and a structure group $G$, a $G$-structure \cite{ChernG} on $\IX$ is a principal $G$-subbundle of the frame bundle $F\IX$. Cases where the $G$-structure corresponds to standard tensor fields are familiar. For example, an $\mr{O(n)}$ structure defines a Riemannian metric on $\IX$ and can be viewed as a principal $\mr{O(n)}$ subbundle of the $\mr{GL(n,\IR)}$ frame bundle, or equivalently, a reduction of structure group from $\mr{GL(n,\IR)}$ to $\mr{O(n)}$ via the natural inclusion. A further reduction to $\mr{SO(n)}$ structure defines an orientation\footnote{An orientation \textit{without} a choice of Riemannian metric corresponds to a reduction from $\mr{GL(n,\IR)}$ to $\mr{GL^{+}(n,\IR)}$, the subgroup of matrices of positive determinant. A volume form corresponds to a reduction from $\mr{GL(n, \IR)}$ to $\mr{SL(n,\IR)}$, the subgroup of matrices of determinant $+1$.} 
on $\IX$, and a reduction to $\mr{Spin(n)}$ via the standard homomorphism $\mr{Spin(n)} \to\mr{O(n)}$ defines a spin structure. Indeed, some of these reductions might not exist -- obstructions are quantified by certain characteristic classes. The aforementioned examples are of tangential structures, i.e., structure group reductions for the frame bundle. However, the idea of the reduction of structure group applies more generally to principal bundles.

The notion of transfer of structure group can be extended to include the data of a connection. Let $\CC^{\mr{conn}}(G, \IX)$ be the groupoid of principal $G$-bundles with connection. If we view a connection as a rule for the coherent lifting of paths then a 
connection $\n_1$ on $P_1 \stackrel{\pi_1}{\longrightarrow} \IX$ lifts a path 
$\gamma: [0,1] \to \IX$ to a path $\wt{\gamma}:[0,1] \to P_1$, uniquely specified 
once we are given the initial point $\wt{\gamma}(0)$ in the fiber above $\gamma(0)$. 
The image connection $\phi_*(\n_1)$ is simply the connection on 
$P_1 \times_\phi G_2 \stackrel{\pi'_1}{\longrightarrow} \IX$ given by the rule to 
lift the path $\gamma(t)$ to  $[(\wt{\gamma}(t),1)]$ given the initial point 
$[(\wt{\gamma}(0), 1)]$. 

In terms of local patches and transition functions $\{U_p, t_{pq}\}$ for $P_1$ we have  
locally-defined $\mr{Lie}(G_1)$-valued one-forms   $A_p$,  then $\phi_*(P_1 \xrightarrow{\pi_1} \IX)$
has local trivializations $\left\{U_p, \phi\circ t_{pq}\right\}$ and the pushed-forward 
connection $\phi_*(\n_1)$ is represented by the locally-defined 
$\mr{Lie}(G_2)$-valued one-forms $\phi_*( A_p) $ where 
$\phi_*: \mr{Lie}(G_1) \to \mr{Lie}(G_2)$ is the induced morphism of Lie algebras.

\setcounter{table}{0}
\setcounter{figure}{0}
\renewcommand{\thetable}{C.\arabic{table}}
\renewcommand{\thefigure}{C.\arabic{figure}}
\section{Generalized Spin$^\mr{c}$ Structure}\label{app:generalized-spin-c}
In this section, we elaborate on the notion of generalized spin$^\mr{c}$ structures on smooth, oriented 4-manifolds and give some examples. Our conventions and notation for \v{C}ech cohomology are from Section \ref{sec:cohcond}. Define the generalized spin$^\mr{c}$ group
\begin{equation} \label{eq:generalized-spin-c-group}
\mr{Spin}^d_{\mr{C}}(4)=\frac{\mr{Spin(4)} \times \mr{U(1)}^d}{\mr{C}} ~,
\end{equation}
where $\mr{C}$ is a \underline{finite} subgroup of the center:
\begin{equation}
\mr{C} \subset Z(\mr{Spin(4)} \times \mr{U(1)}^d)\cong (\mathbb{Z}_2 \times \mathbb{Z}_2) \times \mr{U(1)}^d~,    
\end{equation}
such that
\begin{align} \label{eq:p1-mrc}
p_1(\mr{C}) &= \langle (-1, -1) \rangle \subset Z(\mr{SU(2)\times SU(2)}) = Z(\mr{Spin}(4)) ~.
\end{align}
This ensures that
\begin{align}\label{eq:p1-genspinc-so4}
p_1( \mr{Spin}^d_{\mr{C}}(4)) &\cong \mr{Spin(4)} / p_1(\mr{C}) \cong \mr{SO(4)}~.
\end{align}

We define a   principal $\mr{Spin}_{\mr{C}}^d(4)$-bundle   $P \to \IX$ to be 
 a generalized spin$^\mr{c}$ structure  if
\begin{equation} \label{eq:generalized-spin-structure-bundle-condition}
p_{1,*}(P) \cong \mr{OFr}(T\IX)~.
\end{equation}
A connection $\n$ on a principal $\mr{Spin}_{\mr{C}}^d(4)$-bundle $P$ is called a generalized spin$^\mr{c}$ connection if $p_{1,*}(P)(\n) \cong  \n^{\mr{LC}}$. A generalized spin$^\mr{c}$ structure is thus a particular case of a $G$- structure on a 4-manifold with $G=\mr{Spin}^d_{\mr{C}}(4)$.\footnote{Reference \cite{Brennan:2023vsa} discusses ``$Spin_G(4)$-structures,'' which can be given a similar interpretation. It is possible that these are relevant to special topological twistings with nongeneric masses. See also reference \cite{Albanese:2020bhe} for other generalizations of spin$^{\mr{c}}$ such as $\mr{Spin}^{\mr{h}}(n) := (\mr{Spin(n)}\times\mr{Sp(1)})/\IZ_2$ and $\mr{Spin}^{k}(n) := (\mr{Spin}(n) \times \mr{Spin}(k))/\IZ_2$ in $n$-dimensions (and any positive integer $k \in \IZ_{>0}$). Reference \cite{Avis:1979de} is an interesting early work on generalizations of spin structures, motivated by quantum gravity. 
\cite{Avis:1979de} considers a reduction of structure group of the tangent bundle to  $\mr{Spin}_{G}(4) := (\mr{Spin(4)}\times G)/\IZ_2$ where $G$ is a Lie group with center $\IZ_2$.}
Below, we present some examples to illustrate that a generalized spin$^\mr{c}$ structure is in general different from the standard spin$^\mr{c}$ structure. 
The latter exists on all oriented four-manifolds.\footnote{\label{foot:always-spinc}This follows from the important fact that on such manifolds, any class $w_2 \in H^{2}(\IX, \IZ_2)$ admits an integral lift, due to the existence of a characteristic element. This is equivalent to the vanishing of $W_3 = \mathsf{Bock}(w_2)$ where $\mathsf{Bock}: H^{2}(\IX, \IZ_2) \to H^{3}(\IX, \IZ)$ is the Bockstein map at degree $2$, associated with the short exact sequence $0 \longrightarrow \IZ \xrightarrow{\times 2} \IZ \xrightarrow{\bmod 2} \IZ_2 \longrightarrow 0$. See \cite{TeichnerVogt2000,GompfStipsicz1999,Scorpan2005} for details.} 
\begin{ex}\label{ex:Example-C1}
Consider the subgroup 
\begin{equation}\label{eq:Example-C1-group-C}
 \mr{C}=\langle (-1,-1,-1,\ldots,-1)\rangle=\mr{diag}(\IZ_2^{d+2})\subset  \IZ_2 \times\IZ_2\times\mr{U(1)}^d~.
 \end{equation}
We first show that for this choice of $\mr{C}$, a $\mr{Spin}_{\mr{C}}^d(4)$-structure exists on any orientable 4-manifold, and is equivalent to a choice of $d$ independent spin$^\mr{c}$ structures. 
\par
We start with transition functions
\begin{equation}
    t_{pq}^{\mr{SO(4)}}:U_{pq}\longrightarrow\mr{SO(4)}~,
\end{equation}
for $\mr{OFr}(\IX)$, and choose local lifts 
\begin{equation}
\left(\widetilde{t}_{p q}^{+}, \widetilde{t}_{p q}^{-}\right): U_{p q} \longrightarrow \mr{SU(2)\times\mr{SU(2)}}\cong\mr{Spin(4)} ~,    
\end{equation}
such that the diagram
\begin{equation}
\begin{tikzcd}
	{U_{pq}} && {\mathsf{Spin(4)}} \\
	& {\mathsf{SO(4)}}
	\arrow["{(\widetilde{t}^+_{pq},\widetilde{t}^-_{pq})}", from=1-1, to=1-3]
	\arrow["{t_{pq}^{\mathsf{SO(4)}}}",swap, from=1-1, to=2-2]
	\arrow["\pi", from=1-3, to=2-2]
\end{tikzcd}
\end{equation}
commutes, where $\pi:\mr{Spin(4)}\to\mr{SO(4)}$ is the projection map. The $\IZ_2$-valued cochain
\begin{align}
  \zeta_{p q r}^{\pm} &:=\wt{t}_{p q}^{\pm} \wt{t}_{q r}^{\pm} \wt{t}_{r p}^{\pm}~,  
\end{align}
is a \v{C}ech representative of the Stiefel-Whitney class of $T\IX$:
\begin{equation}
    \left[\zeta_{p q r}^{ \pm}\right]:=w_2(T\IX) \in H^2(\IX, \mathbb{Z}_2)~.
\end{equation}
Choose $d$ integral lifts of $w_2(T\IX)$, i.e., $d$ integer classes $\{c_1(u) \in H^{2}(\IX, \IZ)\}_{u=1}^d$ such that   
\begin{equation}\label{eq:example-c1-c1(u)}
    c_1(u)\equiv w_2(T\IX)\bmod 2~.
\end{equation}
(Such integral lifts always exist -- see footnote \ref{foot:always-spinc}.)
There exist complex line bundles $\{\CL_u\}_{u=1}^{d}$ with first Chern classes $\{c_1(u)\}_{u=1}^{d}$. Endowing each $\CL_u$ with a Hermitian metric on its fibers, we can regard $\CL_u$ as a $\mr{U(1)}$-bundle. Let us choose transition functions functions $(h_u)_{pq}:U_{pq}\to \mr{U(1)}$ on $\CL_u$. We further choose local lifts $(\wt{h}_u)_{pq}:U_{pq}\to\mr{U(1)}$ such that the diagram 
\begin{equation}
\begin{tikzcd}
	{U_{pq}} && {\mathsf{U(1)}} \\
	& {\mathsf{U(1)}}
	\arrow["{(\widetilde{h}_u)_{pq}}", from=1-1, to=1-3]
	\arrow["{(h_u)_{pq}}", from=1-1, to=2-2, swap]
	\arrow["P_{2}", from=1-3, to=2-2]
\end{tikzcd}
\end{equation}
commutes, where $P_{2} : \mr{U(1)} \to \mr{U(1)}$ is the squaring map, that acts as $z \mapsto z^{2}$ for $z \in \mr{U(1)}$. Therefore,
\begin{equation} \label{eq:example-c1-etau-intermsof-tildeh}
(\eta_u)_{p q r} :=(\widetilde{h}_u)_{p q}(\widetilde{h}_u)_{q r}(\widetilde{h}_u)_{r p}: U_{pqr} \longrightarrow \mr{\IZ}/2\IZ \subset \mr{U(1)} ~,\quad u=1,\ldots,d~, 
\end{equation}
represents the  $\bmod~2$ reduction of $c_1(u)$:
\begin{equation}
\left[(\eta_u)_{p q r}\right] = r_{2}(c_1(u)) \in H^{2}(\IX, \IZ/2\IZ) ~,\quad u=1,\ldots,d~, 
\end{equation}
or equivalently, the obstruction to finding a line bundle $\CW_{u}$ such that $\CW_{u}^{\otimes 2} \cong \CL_{u}$. Note that by definition, this $\bmod~2$ reduction is simply the second Stiefel-Whitney class $w_{2}(T\IX)$ due to \eqref{eq:example-c1-c1(u)}. So the representatives $\zeta^{\pm}_{pqr}$ and $(\eta_u)_{pqr}$ are cohomologous in $\IZ/2\IZ$. Therefore, there exist coboundaries $(\varepsilon_u)_{pqr}:U_{pqr}\to\IZ_2$ such that 
\begin{equation}\label{eq:cocycle_relation_dspin-c}
\zeta^{\pm}_{pqr}=(\varepsilon_u)_{p q r}(\eta_u)_{p q r}~.    
\end{equation}
Let $(\varepsilon_u)_{pq}:U_{pq}\to \mr{U(1)}$ denote the trivializing maps of these coboundaries, that is,
\begin{equation}
    (\varepsilon_u)_{p q r}=(\varepsilon_u)_{pq}(\varepsilon_u)_{qr}(\varepsilon_u)_{rp}~.
\end{equation}
We define transition functions
\begin{equation}
\wt{t}_{p q}: U_{p q} \longrightarrow \mr{Spin(4)\times U(1)}^d ~,
\end{equation}
via
\begin{equation}
\wt{t}_{p q}:=\left(\widetilde{t}_{p q}^{+}, \widetilde{t}_{p q}^{-}, 
(\varepsilon_1)_{p q}(\widetilde{h}_1)_{p q}, \cdots, (\varepsilon_d)_{p q}(\widetilde{h}_d)_{p q}\right)~.    
\end{equation}
Then \eqref{eq:cocycle_relation_dspin-c} implies that
\begin{equation}
\wt{t}_{p q}\wt{t}_{ qr}\wt{t}_{rp}=\left(\zeta_{p q r}^{+}, \zeta_{p q r}^{-},(\varepsilon_1)_{pqr}\left(\eta_1\right)_{p q r}, \cdots,(\varepsilon_d)_{pqr}\left(\eta_d\right)_{p q r}\right)\in\mr{C}~.   
\end{equation}
Thus the transition functions
\begin{equation}
    t_{pq}:=[\wt{t}_{pq}]:U_{pq}\longrightarrow\mr{Spin}^d_{\mr{C}}(4)~,
\end{equation}
(where $[\cdot]$ indicates an equivalence class due to the quotient by $\mr{C}$), satisfy the cocycle condition
\begin{equation}
t_{p q}(x) t_{q r}(x) t_{r p}(x)=\mathds{1}_{\mr{Spin}^d_{\mr{C}}(4)}~.
\end{equation}
Therefore, using these transition functions, we can construct a $\mr{Spin}^d_{\mr{C}}(4)$-bundle $P\to\IX$. Finally, because of \eqref{eq:p1-genspinc-so4}, the bundle $P$ satisfies
\begin{equation}
    (p_1)_*(P)\cong \mr{OFr}(\IX)~.
\end{equation}

The defining data of a $\mr{Spin}^d_{\mr{C}}(4)$ structure is a choice of $d$ integral lifts of $w_2(T\IX)$. This is the same data needed to choose $d$ spin${}^{\mr{c}}$-structures. The groups  $\mr{Spin}^d_{\mr{C}}(4)$ and $(\mr{Spin}^{\mr{c}}(4))^{d}$ are not isomorphic: they have distinct third homotopy groups:
\begin{align}
\begin{split}
  \pi_{3}\big( (\mr{Spin}^{\mr{c}}(4))^{d} \big) &\cong (\IZ \oplus \IZ)^{\otimes d} ~,\\
  \pi_{3}\big( \mr{Spin}^d_{\mr{C}}(4) ) &\cong \IZ \oplus \IZ ~.
  \end{split}
\end{align}
The groupoids of principal bundles based on these two structure groups are not isomorphic. However, once the latter bundles are subject to constraint \eqref{eq:generalized-spin-structure-bundle-condition}, with a similar constraint 
for each of the factors in $(\mr{Spin}^{\mr{c}}(4))^{d}$ there will be a 1-1 correspondence. Let us elaborate on this. 

An element of the set of $d$ spin$^{\mr{c}}$-structures is a collection of equivalence classes of pairs of maps $\{ [\wt{t}^{\,i}_{pq}, h^{i}_{pq}] \}_{i=1}^{d}$ where $\wt{t}^{\,i}_{pq}: U_{pq} \to \mr{Spin(4)}$ and $h^{i}_{pq} : U_{pq} \to \mr{U(1)}$. For a fixed $i$, both elements of this pair fail by a common $\IZ_2$-valued cochain. However, since the \v{C}ech 2-cocycle constructed out of any choice\footnote{The transition functions are determined up to possible bundle automorphisms (a.k.a. gauge transformations or frame rotations).}
of $\wt{t}^{\,i}_{pq}$ is a representative of the second Stiefel-Whitney class of $T\IX$, the only choice one has is in picking the transition functions $h^{i}_{pq}$. With this \v{C}ech  data, we can define a map 
\begin{align}\label{eq:Example-C1-MAP1}\begin{split}
    \text{Set of $d$ spin$^{\mr{c}}$ structures} &\longrightarrow \text{Set of $\mr{Spin}^d_{\mr{C}}(4)$ structures)} \\
    \big\{[\wt{t}^{\,i}_{pq}, h^{i}_{pq}]\big\}_{i=1}^{d} &\longmapsto [\wt{t}^{\,i}_{pq}, h^{1}_{pq}, \ldots, h^{d}_{pq}] ~,
    \end{split}
\end{align}
where we have picked -- without loss of generality -- one of the transition functions to $\mr{Spin(4)}$, namely the $i^{th}$ one. We can also define another map
\begin{align}\label{eq:Example-C1-MAP2}\begin{split}
    \text{Set of $\mr{Spin}^d_{\mr{C}}(4)$ structures} &\longrightarrow \text{Set of $d$ spin$^{\mr{c}}$ structures} \\
    [\wt{t}^{\,i}_{pq}, h^{1}_{pq}, \ldots, h^{d}_{pq}] &\longmapsto \big\{[\wt{t}^{\,i}_{pq}, h^{j}_{pq}]\big\}_{j=1}^{d} ~,
    \end{split}
\end{align}
The composition of the maps \eqref{eq:Example-C1-MAP1} and \eqref{eq:Example-C1-MAP2} clearly defines a one-to-one correspondence of the set of $d$ spin$^{\mr{c}}$ structures to itself. This example illustrates that it is not sufficient to say that the groups are not isomorphic to claim that this structure is different from a choice of $d$ spin$^{\mr{c}}$ structures.

\end{ex}
\begin{ex}\label{ex:Example-C2}
Consider a \underline{finite} subgroup $\mr{C}$ of the following subgroup of the center of $\mr{Spin(4)}\times\mr{U(1)}^d$:
\begin{equation}\label{eq:C-example-C2}
\mr{C} \subset \bigg\{\left(\zeta, \zeta, \mu_1, \cdots, \mu_d\right) \in \mathbb{Z}_2 \times \mathbb{Z}_2 \times \mr{U(1)}^d ~|~ \zeta \prod_{k=1}^{d}\mu_k =1\bigg\}~.
\end{equation}
Let us begin with transition functions
\begin{equation}
    t_{pq}^{\mr{SO(4)}}:U_{pq}\longrightarrow\mr{SO(4)}~,
\end{equation}
for $\mr{OFr}(\IX)$, and choose local lifts 
\begin{equation}
\left(\widetilde{t}_{p q}^{+}, \widetilde{t}_{p q}^{-}\right): U_{p q} \longrightarrow \mr{SU(2)\times\mr{SU(2)}}\cong\mr{Spin(4)} ~,    
\end{equation}
such that the diagram
\begin{equation}\label{eq:spin4_lift}
\begin{tikzcd}
	{U_{pq}} && {\mathsf{Spin(4)}} \\
	& {\mathsf{SO(4)}}
	\arrow["{(\widetilde{t}^+_{pq},\widetilde{t}^-_{pq})}", from=1-1, to=1-3]
	\arrow["{t_{pq}^{\mathsf{SO(4)}}}",swap, from=1-1, to=2-2]
	\arrow["\pi", from=1-3, to=2-2]
\end{tikzcd}
\end{equation}
commutes, where $\pi:\mr{Spin(4)}\to\mr{SO(4)}$ is the projection map. The $\IZ_2$-valued cochain
\begin{align}
  \zeta_{p q r}^{\pm} &:=\wt{t}_{p q}^{\pm} \wt{t}_{q r}^{\pm} \wt{t}_{r p}^{\pm}~,  
\end{align}
is a \v{C}ech representative of the Stiefel-Whitney class of $T\IX$:
\begin{equation}
    \left[\zeta_{p q r}^{ \pm}\right]:=w_2(T\IX) \in H^2(\IX, \mathbb{Z}_2)~.
\end{equation}
Note that $p_{u+1}(\mr{C})$ (for any $u = 1, \ldots, d$) is a cyclic group, since all finite subgroups of $\mr{U(1)}$ are cyclic. Therefore, we define $d$ integers $\{n_u\}_{u=1}^{d}$ by
\begin{equation}
p_{u+1}(\mr{C}) \cong \mathbb{Z} / n_u \mathbb{Z} \subset \mr{U(1)} ~, \quad u=1, \ldots, d~ ,
\end{equation}
where $p_{u+1}(\mr{C})$ denotes the projection onto the $(u+1)^{th}$ factor of $\mr{C}$.

For each $\mr{U(1)}$ factor in $\mr{Spin(4)} \times \mr{U(1)}^{d}$, we introduce a complex line bundle $\CL_u$ with first Chern class $c_1(u) \in H^2(\IX, \IZ)$. At this point, this is a completely arbitrary choice, but we will momentarily present a cohomological constraint for the existence of a $\mr{Spin}_{\mr{C}}^{d}(4)$-structure which will involve these Chern classes as well as the second Stiefel-Whitney class of $T\IX$.

Let us endow all complex line bundles with Hermitian inner products on their fibers so that they become $\mr{U(1)}$ bundles. Let $(h_u)_{pq}:U_{pq}\to \mr{U(1)}$ be transition functions on the bundle $\CL_u$. Also choose local lifts $(\wt{h}_u)_{pq}:U_{pq}\to\mr{U(1)}$ such that the diagram 
\begin{equation}\label{eq:U(1)_lift}
\begin{tikzcd}
	{U_{pq}} && {\mathsf{U(1)}} \\
	& {\mathsf{U(1)}}
	\arrow["{(\widetilde{h}_u)_{pq}}", from=1-1, to=1-3]
	\arrow["{(h_u)_{pq}}", from=1-1, to=2-2, swap]
	\arrow["P_{n_u}", from=1-3, to=2-2]
\end{tikzcd}
\end{equation}
commutes, where $P_{n_u} : \mr{U(1)} \to \mr{U(1)}$ is the $n_u^{th}$ power map, that acts as $z \mapsto z^{n_u}$ for $z \in \mr{U(1)}$. Then  the cochain defined by 
\begin{equation} \label{eq:example-c2-etau-intermsof-tildeh}
(\eta_u)_{p q r} :=(\widetilde{h}_u)_{p q}(\widetilde{h}_u)_{q r}(\widetilde{h}_u)_{r p}: U_{pqr} \longrightarrow \mr{\IZ}/n_{u}\IZ \subset \mr{U(1)} ~,\quad u=1,\ldots,d~. 
\end{equation}
represents the  $\bmod~n_u$ reduction of $c_1(u)$:
\begin{equation}
\left[(\eta_u)_{p q r}\right]=r_{n_u}(c_1(u)) \in H^{2}(\IX, \IZ/n_{u}\IZ) ~,\quad u=1,\ldots,d~.    
\end{equation}
For any fixed $u$, the class $\left[(\eta_u)_{p q r}\right]$ (or equivalently, the $\IZ_{n_u}$-valued cochain $(\eta_u)_{p q r}$) measures the obstruction to the existence of a globally-defined line bundle $\CW_{u}$ such that $\CW_{u}^{\otimes n_{u}} \cong \CL_{u}$.

At this point, we have $\IZ_2$-valued cochains $\zeta^{\pm}_{pqr}$, and the $d$ cochains of the form $(\eta_{u})_{pqr}$ for $u = 1, \ldots, d$, where, for a fixed $u$, $(\eta_{u})_{pqr}$ is $\IZ_{n_u}$-valued.\footnote{It is instructive to contrast this with the \v{C}ech arguments that establish the existence of a standard spin$^\mr{c}$ structure. There, $d=1$, $u=1$, and $n_{u} = 2$. One has two $\IZ_2$-valued cochains, and due to the result of footnote \ref{foot:always-spinc}, they represent the \textit{same} $\IZ_2$ cohomology class, and must consequently differ by a coboundary.}
 
Let us define
\begin{align}
    N &:= \mr{lcm}(2, n_1, \ldots, n_d) ~.
\end{align}
Therefore, $N = 2p$ for some $p \in \IZ_{\geq 1}$. Hence, the maps
\begin{align}
    \rho_s: \mathbb{Z} / 2 \mathbb{Z} &\longrightarrow \mathbb{Z} /N\mathbb{Z} ~,\\ 
 \iota_u: \mathbb{Z} / n_u \mathbb{Z} &\longrightarrow \mathbb{Z} /N\mathbb{Z} ~, \quad u = 1, \ldots, d ~,
\end{align}
 are well-defined. They induce the following maps in cohomology
 \begin{align}
     \rho_{s} : H^{\bullet}(X, \IZ/2\IZ) &\longrightarrow H^{\bullet}(\IX, \IZ/N\IZ) ~,\\
     \iota_{u} : H^{\bullet}(X, \IZ/n_{u}\IZ) &\longrightarrow H^{\bullet}(\IX, \IZ/N\IZ)  ~, \quad u = 1, \ldots, d ~.
 \end{align}
 which, in general, are neither injective nor surjective.\footnote{For example, the induced map $\rho_{s}$ on cohomology has a kernel isomorphic to the $p$-torsion submodule of $H^{\bullet}(\IX, \IZ/2\IZ)$, whereas the induced map $\iota_{u}$ has a kernel isomorphic to the $(N/n_u)$-torsion submodule of $H^{\bullet}(\IX, \IZ/n_u\IZ)$.}

 We define transition functions
\begin{equation}
\wt{t}_{p q}: U_{p q} \longrightarrow \mr{Spin(4)\times U(1)}^d ~,
\end{equation}
by 
\begin{equation} 
\wt{t}_{p q}:=\left(\widetilde{t}_{p q}^{+}, \widetilde{t}_{p q}^{-}, 
(\widetilde{h}_1)_{p q}, \cdots, (\widetilde{h}_d)_{p q}\right)~.    
\end{equation}
With this definition,
\begin{equation}
\wt{t}_{p q}\wt{t}_{ qr}\wt{t}_{rp}=\left(\zeta_{p q r}^{+}, \zeta_{p q r}^{-},\left(\eta_1\right)_{p q r}, \cdots,\left(\eta_d\right)_{p q r}\right)~.   
\end{equation}
It follows from the definition \eqref{eq:C-example-C2} that if 
\begin{equation}\label{eq:cochain_product}
\zeta_{p q r}^{ \pm} \cdot \prod_{u=1}^{d}\left(\eta_u\right)_{p q r} = 1 ~,
\end{equation}
the transition functions $\wt{t}_{p q}$ satisfy
\begin{equation}
\wt{t}_{p q}\wt{t}_{ qr}\wt{t}_{rp}\in \mr{C}~,   
\end{equation}
as desired.
In particular, the transition functions
\begin{equation}
    t_{pq}:=[\wt{t}_{pq}]:U_{pq}\longrightarrow\mr{Spin}^d_{\mr{C}}(4)~,
\end{equation}
(where $[\cdot]$ indicates an equivalence class due to the quotient by $\mr{C}$), satisfy the cocycle condition
\begin{equation}
t_{p q}(x) t_{q r}(x) t_{r p}(x)=\mathds{1}_{\mr{Spin}^d_{\mr{C}}(4)}~.
\end{equation}

Then, following the same line of reasoning as in Section \ref{sec:cohcond}, \eqref{eq:cochain_product} implies the following cohomological condition:
\begin{equation}\label{eq:coh-cond-example-c2}
    \boxed{\rho_ s\left( w_2(\IX) \right) + \sum_{u=1}^{d}\iota_u \circ r_{n_u} (c_1(u)) = 0 \in H^{2}(\IX, \IZ/N\IZ) ~.}
\end{equation}
We now argue that this cohomological condition enables us to construct explicitly a $\mr{Spin}_{\mr{C}}^{d}(4)$ cocycle. For this, we assume that a collection of line bundles $\CL_u$ exists with its first Chern classes chosen to satisfy \eqref{eq:coh-cond-example-c2}.

The cohomological condition \eqref{eq:coh-cond-example-c2} implies that the product of the \v{C}ech cochains corresponding to the individual terms represents a coboundary:
\begin{align}
    \rho_{s}\big(\zeta^{\pm}_{pqr}\big) \cdot \left[\prod_{u=1}^{d}\iota_{u}\big((\eta_{u})_{pqr}\big)\right] &= \varepsilon_{pqr} : U_{pqr} \to \IZ/N\IZ \subset \mr{U(1)} ~.
\end{align}
Since $\varepsilon_{pqr}$ is a coboundary, there exist trivializing maps
\begin{align}
 \varepsilon_{pq} : U_{pq} \to \IZ/N\IZ \subset \mr{U(1)} ~,    
\end{align}
such that
\begin{align}
    \varepsilon_{pqr} = \varepsilon_{pq} \varepsilon_{qr} \varepsilon_{rq} ~.
\end{align}
Therefore,
\begin{align}\label{eq:example-c2-coboundary-condition}
     \rho_{s}\big(\zeta^{\pm}_{pqr}\big) \cdot \left[\prod_{u=1}^{d}\iota_{u}\big((\eta_{u})_{pqr}\big)\right] = \varepsilon_{pq}\varepsilon_{qr}\varepsilon_{rp} ~.
\end{align}
Now, we can define the modified transition functions\footnote{\label{foot:choice-of-root-of-phase}By the $d^{th}$ root of a generic phase $\varepsilon_{pq}$, we mean the following: we choose \underline{a} solution $y$ to the equation $y^{d} = \varepsilon_{pq}$ and use it in \eqref{eq:example-c1-htildeprime}--\eqref{eq:example-c1-stilde-cocycle}. The conclusion \eqref{eq:example-c1-stilde-cocycle-valued-in-C} remains valid because the $d^{th}$ power of \underline{any} solution by definition equals $\varepsilon_{pq}$. So \eqref{eq:example-c1-stilde-cocycle-valued-in-C} is valid and independent of the branch chosen to define the $d^{th}$ root. We could have avoided taking the $d^{th}$ root -- at the expense of an undemocratic treatment of the $\mr{U(1)}$ bundles -- by arbitrarily picking one of the $d$ $\mr{U(1)}$ bundles -- say the $u_0^{th}$ one -- and defining its modified transition functions as $(\wt{h}_{u_0})'_{pq} := (\wt{h}_{u_0})_{pq}\cdot (\varepsilon_{pq})^{-1}$, whereas $(\wt{h}_{u})'_{pq} := (\wt{h}_{u})_{pq}$ for $u = 1, \ldots, d$, $u \neq u_0$.}
\begin{align}\label{eq:example-c1-htildeprime}
    (\wt{h}_{u})'_{pq} := (\wt{h}_{u})_{pq} \big(\varepsilon_{pq}\big)^{-1/d} ~,
\end{align}
as well as maps $\wt{s}_{pq}: U_{pq} \to \mr{Spin(4)} \times \mr{U(1)}^{d}$ via
\begin{align}\label{eq:example-c1-stilde}
    \wt{s}_{pq} &:= \left(\widetilde{t}_{p q}^{+}, \widetilde{t}_{p q}^{-}, 
(\widetilde{h}_1)'_{p q}, \cdots, (\widetilde{h}_d)'_{p q}\right) \nonumber\\
&= \left(\widetilde{t}_{p q}^{+}, \widetilde{t}_{p q}^{-}, 
(\widetilde{h}_1)_{p q}\varepsilon_{pq}^{-1/d}, \cdots, (\widetilde{h}_d)_{p q} \varepsilon_{pq}^{-1/d} \right) ~.
\end{align}
Note that
\begin{align}\label{eq:example-c1-stilde-cocycle}
   \wt{s}_{pq}\wt{s}_{qr}\wt{s}_{rp} &= \big(\zeta^{+}_{pqr}, \zeta^{-}_{pqr}, (\eta_{1})_{pqr}\varepsilon_{pqr}^{-1/d}, \cdots, (\eta_{d})_{pqr}\varepsilon_{pqr}^{-1/d} \big) ~.
\end{align}
It follows directly from \eqref{eq:example-c2-coboundary-condition} that
\begin{align}\label{eq:example-c1-stilde-cocycle-valued-in-C}
   \wt{s}_{pq}\wt{s}_{qr}\wt{s}_{rp} &\in \mr{C} ~.
\end{align}
Therefore, the transition functions,
\begin{align}
    s_{pq} := [\wt{s}_{pq}] : U_{pq} \longrightarrow \big(\mr{Spin(4)} \times \mr{U(1)}^{d}\big)/\mr{C} ~,
\end{align}
satisfy the cocycle condition
\begin{align}
    s_{pq}s_{qr}s_{rp} &= \mathds{1}_{\mr{Spin}_{\mr{C}}^{d}(4)} ~.
\end{align}
Moreover, because of \eqref{eq:spin4_lift} the $\mr{Spin}^d_{\mr{C}}(4)$-bundle $P\to\IX$ with transition functions $t_{pq}$ satisfies 
\begin{equation}
\left(p_1\right)_*(P) \cong \mr{OFr}(T \IX)~.
\end{equation}
Thus we have constructed a $\mr{Spin}^d_{\mr{C}}(4)$-structure on $\IX$. For $d=1$, this is precisely the standard $\mr{Spin}^c(4)$-structure.

It follows from the above argument that the set of $\mr{Spin}^d_{\mr{C}}(4)$-structures on $\IX$ is a torsor for the subgroup of elements  
\be 
(c(1), \dots, c(d)) \in H^{2}(\IX, \IZ)^d ~.
\ee
that satisfy:  
\begin{equation}\label{eq:coh-cond-example-c2-new}
     \sum_{u=1}^{d}\iota_u \circ r_{n_u} (c(u))  = 0 \in H^{2}(\IX, \IZ/N\IZ) ~. 
\end{equation}

\end{ex}

\begin{ex}\label{ex:Example-C3}
Fix a positive even integer $n \in \mathbb{Z}_{> 0}$ and consider the finite group
\begin{equation}\label{eq:Example-C3-group-C}
\mr{C}_n=\left\{(\zeta, \zeta, \mu) \in \mathbb{Z}_2 \times \mathbb{Z}_2 \times \mr{U(1)} ~|~ \mu^{n / 2} \zeta=1\right\}~.
\end{equation}
Then $\mr{C}_n$ is generated by
\begin{equation}
(-1,-1, g) \in \mathbb{Z}_2 \times \mathbb{Z}_2 \times \mr{U(1)} ~,
\end{equation}
where $g \in \mr{U(1)}$ is any element of order $n$. Moreover, since $n$ is even,
\begin{equation}
\mr{C}_n \cong \mathbb{Z} / n \mathbb{Z}~.
\end{equation}
We denote the corresonding generalized spin$^\mr{c}$ group by
\begin{equation}
\mr{Spin}^1_{\mr{C}_n}(4)=\big(\mr{Spin(4)} \times \mr{U(1)}\big)/\mr{C}_n~.
\end{equation}
As in Example \ref{ex:Example-C2}, we introduce transition functions $t_{pq}^{\mr{SO(4)}}$ for $\mr{OFr}(\IX)$ and choose local lifts $(\wt{t}_{pq}^{+}, \wt{t}_{pq}^{-}): U_{pq} \to \mr{Spin(4)}$ which satisfy a commutative diagram of the form \eqref{eq:spin4_lift}. Therefore, as before, the $\IZ_2$-valued cochain
\begin{align}
    \zeta_{pqr}^{\pm} &:= \wt{t}^{\pm}_{pqr} \wt{t}^{\pm}_{pqr} \wt{t}^{\pm}_{pqr} ~,
\end{align}
is a \v{C}ech representative of the second Stiefel-Whitney class of $T\IX$:
\begin{align}
    [\zeta^{\pm}_{pqr}] := w_2(T\IX) \in H^{2}(\IX, \IZ_2) ~.
\end{align}
Let us introduce a complex line bundle $\CL$ with a first Chern class $c_1(\CL)$ which is at the moment an arbitrary integral class, but will be further constrained by a cohomological condition later. Endowing $\CL$ with a Hermitian metric on its fibers, we can regard $\CL$ as $\mr{U}(1)$-bundle. Let $\{h_{pq}\}$ denote transition functions on this bundle. As in \eqref{eq:U(1)_lift}, we choose local lifts $\wt{h}_{pq}:U_{pq}\to\mr{U(1)}$ such that the diagram 
\begin{equation}
\begin{tikzcd}
	{U_{pq}} && {\mathsf{U(1)}} \\
	& {\mathsf{U(1)}}
	\arrow["{\widetilde{h}_{pq}}", from=1-1, to=1-3]
	\arrow["{h_{pq}}", from=1-1, to=2-2, swap]
	\arrow["P_{n}", from=1-3, to=2-2]
\end{tikzcd}
\end{equation}
commutes, where $P_{n}$ is the $n^{th}$ power map, which acts as $z \mapsto z^{n}$ for $z \in \mr{U(1)}$. Note that the global existence of the lifts $\wt{h}_{pq}$ is equivalent to the existence of a complex line bundle $\CW$ such that $\CW^{\otimes n} = \CL$. Existence of $\CW$ is equivalent to $c_1(\CL)$ being $n$ times an integral class $y \in H^{2}(\IX, \IZ)$ (which is then the first Chern class of $\CW$, i.e., $y = c_1(\CW)$).

The cochain
\begin{align}
    \eta_{pqr} := \wt{h}_{pq} \wt{h}_{qr} \wt{h}_{rp} ~,
\end{align}
represents the $\bmod ~ n$ reduction of $c_1(\CL)$:
\begin{align}
    [\eta_{pqr}] := r_{n}\big(c_1(\CL)\big) ~,
\end{align}
where $r_{n}$ is reduction modulo $n$, and measures the obstruction to the existence of a line bundle $\CW$.

We define transition functions
\begin{equation}
\wt{t}_{p q}:=\left(\widetilde{t}_{p q}^{+}, \widetilde{t}_{p q}^{-}, 
(\widetilde{h})_{p q}\right)~.    
\end{equation}
With this definition,
\begin{equation}
\wt{t}_{p q}\wt{t}_{ qr}\wt{t}_{rp}=\left(\zeta_{p q r}^{+}, \zeta_{p q r}^{-},\eta_{p q r}\right)~.   
\end{equation}
It follows from the definition \eqref{eq:Example-C3-group-C} that if 
\begin{equation}\label{eq:cochain_product_exampleC3}
\zeta_{p q r}^{ \pm} \cdot \big(\eta_{p q r}\big)^{n/2} = 1 ~,
\end{equation}
the transition functions $\wt{t}_{p q}$ satisfy
\begin{equation}
\wt{t}_{p q}\wt{t}_{ qr}\wt{t}_{rp}\in \mr{C}_{n} ~,   
\end{equation}
as desired.
If there exist functions 
\begin{align}
    \varepsilon_{pq} : U_{pq} \to \IZ/n\IZ ~,
\end{align}
such that
\begin{align}\label{eq:example-c3-coboundary-condition}
   \iota(\zeta^{\pm}_{pqr}) \cdot \big(\eta_{pqr}\big)^{n/2} &= \varepsilon_{pq}\varepsilon_{qr}\varepsilon_{rp} \equiv \varepsilon_{pqr} ~,
\end{align}
this is equivalent to the following cohomological condition
\begin{equation}\label{eq:coh-cond-example-c3}
    \boxed{\iota\big(w_2(\IX)\big) + \frac{n}{2} r_{n}\big(c_1(\CL)\big) = 0 \in H^{2}(\IX, \IZ/n\IZ) ~,}
\end{equation}
which should be viewed as a constraint on $c_1(\CL)$.

We now show that the condition \eqref{eq:example-c3-coboundary-condition} (or equivalently \eqref{eq:coh-cond-example-c3}) enables us to satisfy the cocycle condition for $\mr{Spin}_{\mr{C}_n}^{1}(4)$. We simply define the modified transition functions\footnote{We choose \underline{a} solution $y$ to $y^{n/2} = \varepsilon_{pq}$, use it in \eqref{eq:example-c3-htildeprime}--\eqref{eq:example-c3-stilde-cocycle}. The conclusion \eqref{eq:example-c3-stilde-cocycle-valued-in-C} remains true independent of the choice of branch for the $(n/2)^{th}$ root because by definition, the $(n/2)^{th}$ power of \textit{any} solution is the same. The reasoning is similar to what was described in footnote \ref{foot:choice-of-root-of-phase} for Example \ref{ex:Example-C1}, except in this example, we do not have the option to be undemocratic with only one $\mr{U(1)}$ bundle at our disposal.}
\begin{align}\label{eq:example-c3-htildeprime}
    \wt{h}'_{pq} := \wt{h}_{pq}(\varepsilon_{pq})^{-2/n} ~,
\end{align}
and maps $\wt{s}_{pq}:U_{pq}\to \mr{Spin(4)} \times \mr{U(1)}$ via
\begin{align}\label{eq:example-c3-stilde}
    \wt{s}_{pq} &:= \left(\widetilde{t}_{p q}^{+}, \widetilde{t}_{p q}^{-}, 
\widetilde{h}'_{p q}\right) = \left(\widetilde{t}_{p q}^{+}, \widetilde{t}_{p q}^{-}, 
\widetilde{h}_{p q}\varepsilon_{pq}^{-2/n}\right) ~.
\end{align}
Note that
\begin{align}\label{eq:example-c3-stilde-cocycle}
   \wt{s}_{pq}\wt{s}_{qr}\wt{s}_{rp} &= \big(\zeta^{+}_{pqr}, \zeta^{-}_{pqr}, \eta_{pqr}\varepsilon_{pqr}^{-2/n}\big) ~.
\end{align}
It follows directly from \eqref{eq:example-c3-coboundary-condition} that
\begin{align}\label{eq:example-c3-stilde-cocycle-valued-in-C}
   \wt{s}_{pq}\wt{s}_{qr}\wt{s}_{rp} &\in \mr{C}_{n} ~.
\end{align}
Therefore, the transition functions,
\begin{align}
    s_{pq} := [\wt{s}_{pq}] : U_{pq} \longrightarrow \big(\mr{Spin(4)} \times \mr{U(1)}\big)/\mr{C}_n ~,
\end{align}
satisfy the cocycle condition
\begin{align}
    s_{pq}s_{qr}s_{rp} &= \mathds{1}_{\mr{Spin}_{\mr{C}_n}^{d}(4)} ~.
\end{align}
A few remarks are in order. First of all, if $n = 2$, the map $\iota$ reduces to the identity map, and the condition \eqref{eq:coh-cond-example-c3} reduces to the condition for a standard spin$^{\mr{c}}$ structure, namely $w_2(\IX) = c_1(\CL) \bmod~2$. In this case, one recovers the familiar fact that the evenness of $c_1(\CL)$ is equivalent to the manifold being a spin manifold.

For $n > 2$, if $c_1(\CL)$ is $n$ times an integral class, i.e., if there exists an integral class $y \in H^{2}(\IX, \IZ)$ such that $c_1(\CL) = n \cdot y$, then the cohomological condition \eqref{eq:coh-cond-example-c3} is equivalent to $\iota(w_2(\IX)) = 0$. In this case, one possibility is that $\IX$ be spin, i.e., $w_2(\IX) = 0$. The other possibility is that $w_2(\IX)$ be in the kernel of the map $\iota: H^{2}(\IX, \IZ_2) \to H^{2}(\IX, \IZ/n\IZ)$. This possibility merits further study.

\end{ex}

\setcounter{table}{0}
\setcounter{figure}{0}
\renewcommand{\thetable}{D.\arabic{table}}
\renewcommand{\thefigure}{D.\arabic{figure}}
\section{\label{app:SCIMultiplets}Superconformal Multiplets}
  In this section, we will briefly review superconformal multiplets for the $d = 4$ $\CN=2$ superconformal algebra $\mf{su}(2,2|2)$.\footnote{This is the real form appropriate to $d=4$ Lorentzian signature, which is implicit in standard treatments of radial quantization. The appropriate real form for $d=4$ Euclidean signature is $\mathfrak{su^{*}(4|2)}$, which would lead to a noncompact summand in the Euclidean $\mr{R}$-symmetry group. This distinction will be unimportant for us, however, see \cite{Cushing:2023rha} for a discussion in the context of topological twisting.} 
The reader is referred to \cite{Cordova:2016emh,Rastelli:2014jja,Dolan:2002zh,Eberhardt:2020cxo} for more details. 
The superconformal algebra admits a bosonic subalgebra,
\begin{align}
    \mf{su}(2,2) \oplus \mf{su}(2)_{\mr{R}} \oplus \mf{u(1)}_{\mr{R}} \subset \mf{su}(2,2|2) ~.
\end{align}
Note that $\mf{su}(2,2) \cong \mf{so}(4,2)$, and there is an $\mf{su}(2)_{\mr{R}}\oplus\mf{u}(1)_{\mr{R}}$ symmetry, we will denote the charge under the $\mr{R}$-symmetry by $(R;r)$ where $R\in\IZ_{\geq 0}$ is the Dynkin label for the $\mf{su}(2)_{\mr{R}}$ and $r\in\IR$ is the charge under $\mf{u}(1)_{\mr{R}}$. The scaling dimension will be denoted by $\Delta$ and the Dynkin labels for the representation under the (spin lift of) the 4d rotation group $\mr{Spin(4)}\cong\mr{SU(2)}\times\mr{SU(2)}$ will be denoted by $(j_1;j_2)\in \IZ_{\geq 0}\times\IZ_{\geq 0}$. An operator with these eigenvalues under the generators of the superconformal algebra will then be denoted by $[j_1;j_2]_{\Delta}^{(R;r)}$. In particular, the supercharges transform in the following representations:
\begin{equation}
    \CQ \in[1 ; 0]_{\frac{1}{2}}^{(1 ;-1)}, \quad \ov{\CQ} \in[0 ; 1]_{\frac{1}{2}}^{(1 ; 1)}~.
\end{equation}
The basic generating multiplets are the long multiplets and short multiplets satisfying certain unitarity bounds \cite{Minwalla:1997ka}. Short multiplets are obtained from the long multiplet by imposing certain shortening conditions. These multiplets are listed in Tables \ref{tab:Q-short} and \ref{tab:Q_bar-short}.
\begin{small}
\begin{table}[H]
\centering
\def\arraystretch{1.1}
\begin{tabular}{|c|l|l|}
\hline Name & \multicolumn{1}{|c|}{ Primary } & \multicolumn{1}{|c|}{ Unitarity Bound } \\
\hline \hline$L$ & {$[j_1 ; j_2]_{\Delta}^{(R ; r)}$} & $\Delta>2+j_1+R-\frac{1}{2} r$ \\
\hline \hline$A_1$ & {$[j_1 ; j_2]_{\Delta}^{(R ; r)}, \quad j_1 \geq 1$} & $\Delta=2+j_1+R-\frac{1}{2} r$ \\
\hline$A_2$ & {$[0 ; j_2]_{\Delta}^{(R ; r)}$} & $\Delta=2+R-\frac{1}{2} r$ \\
\hline \hline$B_1$ & {$[0 ; j_2]_{\Delta}^{(R ; r)}$} & $\Delta=R-\frac{1}{2} r$ \\
\hline
\end{tabular}
    \caption{Multiplets with $\CQ$-shortening conditions, adapted from \cite[Table 13]{Cordova:2016emh}.}
    \label{tab:Q-short}
\end{table}
\end{small}
\begin{small}
\begin{table}[H]
\centering
\def\arraystretch{1.1}
\begin{tabular}{|c|l|l|}
\hline Name & \multicolumn{1}{|c|}{ Primary } & \multicolumn{1}{|c|}{ Unitarity Bound } \\
\hline \hline $\bar{L}$ & {$[j_1 ; j_2]_{\Delta}^{(R ; r)}$} & $\Delta>2+j_2+R+\frac{1}{2} r$ \\
\hline \hline $\bar{A}_1$ & {$[j_1 ; j_2]_{\Delta}^{(R ; r)}, \quad j_2 \geq 1$} & $\Delta=2+j_2+R+\frac{1}{2} r$ \\
\hline $\bar{A}_2$ & {$[j_1 ; 0]_{\Delta}^{(R ; r)}$} & $\Delta=2+R+\frac{1}{2} r$ \\
\hline \hline $\bar{B}_1$ & {$[j_1 ; 0]_{\Delta}^{(R ; r)}$} & $\Delta=R+\frac{1}{2} r$ \\
\hline
\end{tabular}
    \caption{Multiplets with $\ov{\CQ}$-shortening conditions, adapted from \cite[Table 14]{Cordova:2016emh}.}
    \label{tab:Q_bar-short}
\end{table}
\end{small}
Full multiplets are two-sided, i.e., they satisfy both $\CQ$ and $\ov{\CQ}$-shortening conditions which leads to restriction on some of the quantum numbers. The consistent two-sided multiplets are listed below.\footnote{The notation on the LHS of \eqref{eq:2sided_multiplets} is borrowed from \cite{Dolan:2002zh}.} 
We refer the reader to \cite[Table 15]{Cordova:2016emh} for details on the restrictions on the quantum numbers.\footnote{Note that the scaling dimensions of the short multiplets are determined by shortening conditions; this is not indicated in the notation.} 
\be\label{eq:2sided_multiplets}
\begin{aligned}
    \mathcal{A}_{R, r(j_1, j_2)}^{\Delta} &= L \ov{L}[j_1 ; j_2]_{\Delta}^{(R ; r)} ~,\\
    \mathcal{C}_{R, r(j_1, j_2)} &= A_{\ell} \ov{L}[j_1 ; j_2]^{(R ; r)} ~, \quad &\mathcal{B}_{R, r(0, j_2)} &= B_1 \ov{L}[0 ; j_2]^{(R>0 ; r)} ~,\\
    \mathcal{E}_{r(0, j_2)} &=B_1 \ov{L}[0 ; j_2]^{(0 ; r)} ~, \quad &\wh{\mathcal{C}}_{R(j_1, j_2)} &=A_{\ell} \ov{A}_{\ov{\ell}}[j_1 ; j_2]^{(R ; j_1-j_2)} ~,\\
    \mathcal{D}_{R(0, j_2)} &=B_1 \ov{A}_{\ov{\ell}}[0 ; j_2]^{(R ;-j_2-2)} ~, \quad &\wh{\mathcal{B}}_R &= B_1 \ov{B}_1[0 ; 0]^{(R ; 0)} ~.
\end{aligned}
\ee
where $\ell,\ov{\ell}=1,2$. The conjugate multiplets $\overline{\mathcal{C}}_{R, r(j_1, j_1)}, \overline{\mathcal{B}}_{R, r(j_1, 0)}, \overline{\mathcal{E}}_{r(j_1, 0)}$, and $\overline{\mathcal{D}}_{R(j_1, 0)}$ can be obtained by complex conjugation. The full operator content of each of these multiplets is tabulated in \cite{Cordova:2016emh}. 
\par

Vectormultiplets realize the short multiplet $A_2 \ov{B}_1[0;0]^{(0 ; 1)}$  and hypermultiplets realize the short multiplet $B_1 \ov{B}_1[0;0]^{(1 ; 0)}$. 
Superconformal multiplets containing currents for symmetries are called conserved current multiplets. Here we are only interested in the flavor current multiplet and the stress tensor multiplet (also called the energy-momentum multiplet). See \cite{Cordova:2016emh} for other conserved current multiplets. 
\paragraph{Flavor current multiplet.}
The $R=2$ multiplet $B_1 \ov{B}_1[0;0]^{(2;0)}_2$ is a flavor current multiplet containing eight bosonic and eight fermionic operators. The full flavor multiplet \cite{Cordova:2016emh} has the structure shown in Fig. \ref{fig:flavor_multiplet}.
\begin{figure}[H]
\centering
\begin{tikzcd}[row sep=small]
	&& {\boxed{[0 ; 0]_2^{(2 ; 0)}}} \\
	& {\boxed{[1 ; 0]_{\frac{5}{2}}^{(1 ;-1)}}} && {\boxed{[0 ; 1]_{\frac{5}{2}}^{(1 ; 1)}}} \\
	{\boxed{[0 ; 0]_3^{(0 ;-2)}}} && {\boxed{[1 ; 1]_3^{(0 ; 0)}}} && {\boxed{[0 ; 0]_3^{(0 ; 2)}}}
	\arrow["{\mathcal{Q}}"', from=1-3, to=2-2]
	\arrow["{\widetilde{\CQ}}", from=1-3, to=2-4]
	\arrow["{\mathcal{Q}}"', from=2-2, to=3-1]
	\arrow["{\widetilde{\mathcal{Q}}}", from=2-2, to=3-3]
	\arrow["{\mathcal{Q}}"', from=2-4, to=3-3]
	\arrow["{\widetilde{\mathcal{Q}}}", from=2-4, to=3-5]
\end{tikzcd}
\caption{\label{fig:flavor_multiplet}The structure of the flavor multiplet, adapted from \cite[Fig. 5]{Eberhardt:2020cxo}.}
\end{figure}
Let us denote the bottom component by $\mu^{(\!(AB)\!)a}$, where $A=1,2$ is the index for the fundamental representation of $\mr{SU(2)_R}$ and $a$ is the index for the adjoint representation of the flavor group.\footnote{The notation $T^{(\!(AB)\!)}$ denotes a symmetric traceless rank-$2$ tensor transforming in the adjoint ($\underline{\bm{3}}$) of $\mr{SU(2)_R}$.}
Then the full multiplet can be constructed by acting on the bottom component by the supercharges:
\begin{align}
\left[\CQ_{A\alpha},\mu^{(\!(BC)\!)a}\right]&=\delta_A{}^{(\!(B}\psi_{\alpha}^{C)\!)a}\in [1 ; 0]_{\frac{5}{2}}^{(1 ;-1)}~,
\\
\left[\wt{\CQ}_{A\dt\alpha},\mu^{(\!(BC)\!)a}\right]&=\delta_A{}^{(\!(B}\wt{\psi}_{\dt\alpha}^{C)\!)a}\in [0 ; 1]_{\frac{5}{2}}^{(1 ; 1)}~,
\\
\label{eq:Q-action_flavor1}\left\{\wt{\CQ}_{A\dt\alpha},\psi_{\alpha}^{Ba}\right\}&=\delta_A{}^B \wt{\sigma}^\mu_{\dt{\alpha}\beta} J^a_\mu\in [1 ; 1]_3^{(0 ; 0)}~,
\\
\label{eq:Q-action_flavor2}\left\{\CQ_{A\alpha},\wt{\psi}_{\dt\beta}^{Ba}\right\}&=\delta_A{}^B \sigma^\mu_{{\alpha}\dt\beta} J^a_\mu\in [1 ; 1]_3^{(0 ; 0)}~.
\end{align}
The current $J^a_\mu$ is the conserved current for the flavor symmetry of the theory.
\paragraph{Stress tensor multiplet.} The multiplet $A_{2} \ov{A}_{2}[0 ; 0]_{2}^{(0 ; 0)}$ is the stress tensor multiplet whose top component is the stress tensor of the theory. The generators of the $d=4$ $\CN=2$ superconformal algebra must be integrals of local currents over a spatial slice. The bosonic generators are integrals of the stress tensor $T_{\mu\nu}$ and the $\mf{su}(2)_{\mr{R}}\oplus\mf{u}(1)_{\mr{R}}$  $\mr{R}$-symmetry currents $(j_\mu^{(\mr{R})})^A$, $j_\mu^{(\mr{R})}$. The fermionic generators are integrals of conserved fermionic supercurrents $G_{\mu\alpha}^A$ and $\wt{G}_{\mu\dt{\alpha}}^A$. The superconformal algebra dictates that the stress tensor, $\mr{R}$-symmetry current and supercurrent sit in the same multiplet, called the stress tensor multiplet. In particular,
\begin{equation}\label{eq:Q-tranformation_stress-tensor-multiplet}
\{ \CQ_{\alpha A} , \wt{G}^{B}_{\mu\dt \beta } \} = \sigma^{\nu}_{\alpha\dt\beta} \delta_A{}^{B} T_{\mu\nu} ~,\quad  \{ \wt{\CQ}_{\dt{\alpha} A} , G^{B}_{ \mu\beta} \} = \wt{\sigma}^{\nu}_{\dt\alpha\beta} \delta_A{}^{B} T_{\mu\nu} ~.   
\end{equation}
The full stress tensor multiplet has the form \cite{Cordova:2016emh,Eberhardt:2020cxo} shown in Fig. \ref{fig:stmultiplet}.
\begin{figure}[H]
\centering
\begin{tikzcd}[row sep=small,column sep=small]
	&& {\boxed{[0;0]_2^{(0;0)}}} \\
	& {\boxed{[1 ; 0]_{\frac{5}{2}}^{(1 ;-1)}}} && {\boxed{[0 ; 1]_{\frac{5}{2}}^{(1 ; 1)}}} \\
	{\boxed{[2 ; 0]_3^{(0 ;-2)}}} && {\boxed{[1 ; 1]_3^{(2 ; 0)} \oplus[1 ; 1]_3^{(0 ; 0)}}} && {\boxed{[0 ; 2]_3^{(0 ; 2)}}} \\
	& {\boxed{[2 ; 1]_{\frac{7}{2}}^{(1 ; 1)}}} && {\boxed{[1 ; 2]_{\frac{7}{2}}^{(1 ;-1)}}} \\
	&& {\boxed{[2 ; 2]_4^{(0 ; 0)}}}
	\ar["\CQ"', from=1-3, to=2-2]
	\ar["{\widetilde{\CQ}}", from=1-3, to=2-4]
	\ar["\CQ"', from=2-2, to=3-1]
	\ar["{\widetilde{\CQ}}", from=2-2, to=3-3]
	\ar["\CQ"', from=2-4, to=3-3]
	\ar["{\widetilde{\CQ}}", from=2-4, to=3-5]
	\ar["{\widetilde{\CQ}}", from=3-1, to=4-2]
	\ar["\CQ"', from=3-3, to=4-2]
	\ar["{\widetilde{\CQ}}", from=3-3, to=4-4]
	\ar["\CQ"', from=3-5, to=4-4]
	\ar["{\widetilde{\CQ}}", from=4-2, to=5-3]
	\ar["\CQ"', from=4-4, to=5-3]
\end{tikzcd}
\caption{\label{fig:stmultiplet}The structure of the stress tensor multiplet, adapted from \cite[Fig. 6]{Eberhardt:2020cxo}.}
\end{figure}
The top component of the supermultiplet is the stress tensor (which is, in fact, at the bottom of the diagram), and the middle components (with $\Delta = 3$) are the $\mr{R}$-symmetry currents. The operators with $\Delta=\frac{7}{2}$ are the supercurrents. 

\setcounter{table}{0}
\setcounter{figure}{0}
\renewcommand{\thetable}{E.\arabic{table}}
\renewcommand{\thefigure}{E.\arabic{figure}}
\section{The Superconformal Index\label{app:SupSymIndex}}
In this appendix, we review the superconformal index. For recent reviews, see \cite{Rastelli:2014jja,Rastelli:2016tbz,Gadde:2020yah}. First, let us review the supersymmetric index as a trace over the Hilbert space in the Hamiltonian quantization of a generic supersymmetric theory in $d$ spacetime dimensions, coupled to conserved currents of some global symmetries (i.e., to background flavor bundles). To compute it, we pick a supercharge\footnote{In terms of background supergravity, this is a choice of generalized Killing spinor, i.e., a solution to the differential constraints obtained by demanding the vanishing of the supersymmetry variations of the gravitino, in a bosonic background where the auxiliary fields of the supergravity multiplet may have nontrivial profiles \cite{Festuccia:2011ws,Klare:2012gn,Cassani:2012ri,Klare:2013dka,Karlhede:1988ax,Cushing:2023rha}.}
$\CQ$ with an adjoint $\CQ^+$ where $\{\CQ, \CQ^+\} = 2\mr{H}$, and $\CQ^2 = 0$. In a unitary theory, this implies that the spectrum of  $\mr{H}$ is positive semidefinite. The index is simply a weighted supertrace of states in $\Ker \,\mr{H}$, i.e., states in the cohomology of $\CQ$ and $\CQ^+$, or equivalently, the Hilbert space of a radially quantized theory on $\IS^{d-1}\times \IR$. Concretely, the index has a trace formula of the form
\begin{align}
    \CI(\{ \mu_i \}) &= \mr{Tr}_{\Ker\,\mr{H}}(-1)^{\mr{F}}e^{-\sum_i \mu_i T_i} e^{-2\beta\{\CQ, \CQ^+\}} ~, \label{eq:supIndex}
\end{align}
where $\{T_i\}$ is a complete set of generators that commute with $\CQ$ and with each other, $\{\mu_i\}$ is a set of ``fugacities,'' or ``chemical potentials'' for the associated global symmetries\footnote{At the moment, we are including superconformal fugacities (finitely many in number) as well as fugacities associated with flavor symmetries (which could be arbitrarily many in number, and which commute with all generators of the superconformal algebra). We will shortly refine this definition.} 
and $\mr{F}$ is the fermion number. As is the case for the original Witten index \cite{Witten:1982df}, only the states with $\mr{H} = 0$ contribute to the index, and consequently the index does not depend on $\beta$. The index so defined is a function of the fugacities that couple to the conserved charges of the corresponding global symmetries.\footnote{Sometimes, the factor $e^{-\sum_{i}\mu_i T_i}$ is replaced by $\prod_{i}\wt{\mu}_i^{C_i}$ with $\wt{\mu}_i$ referred to as the fugacities and $C_i$ the conserved charges.  }
There are various refinements of the index, and specializations that are useful in different contexts, especially for probing enhanced supersymmetry \cite{Gadde:2011uv}.
%

%

Expressed as a trace over the full Hilbert space $\CH$ of states on $\IS^3$ the $d=4$ $\CN=2$ superconformal index has the generic form
%
\begin{align}\label{eq:supcIndex}
   \hspace{-0.1in}& \CI(p, q, t; \{a_i\}) \nonumber\\
   \hspace{-0.1in} &:= \mathsf{Tr}_{\CH}(-1)^{\mr{F}} p^{\frac{1}{3}(\Delta + \frac{1}{2}j_1 - R - r) + \frac{1}{2}j_2} q^{\frac{1}{3}(\Delta + \frac{1}{2}j_1 - R - r) - \frac{1}{2}j_2} t^{\frac{1}{2}(R+r)} e^{-\beta(\Delta - j_1 - R + \frac{1}{2}r)}\prod_{i=1}^{\mr{rk\,}G^{\mr{f}}} a_{i}^{f_i} ~. 
\end{align}
Recall that the index is written with respect to a choice of supercharge\footnote{\label{foot:supconfgenerators}The supercharges of the $d=4$ $\CN=2$ superconformal algebra in radial quantization are $\CQ_{A\alpha}, \wt{Q}^{A}_{\dt{\alpha}}, \CS^{A\alpha} := \CQ^{\dagger A\alpha}, \wt{S}_{A}^{\dt{\alpha}} := \wt{Q}_{A}^{\dagger\dt{\alpha}}$. Here $A$ is an $\mr{SU(2)_R}$ symmetry index, and $\alpha$ and $\dt{\alpha}$ are spinor indices, which take values $+, -$ and $\dt{+}, \dt{-}$ respectively. The supercharge used to define the index is $\CQ := \CQ_{1-}$, for which $\{\CQ_{1-}, (\CQ_{1-})^{\dagger}  \} = \Delta - j_1 - 2 R + \frac{1}{2} r$. This is the coefficient of $-\beta$ in \eqref{eq:supcIndex}. This $\CQ$ should not be confused with the scalar supercharge in the bulk of the paper.}
$\CQ$, which determines an $\CN=1$ subalgebra of an $\CN=2$ algebra. The commutant of $\CQ$ in the superconformal algebra $\mathfrak{su}(2,2|2)$ is $\mathfrak{su}(2,1|1)$, which has rank $3$. Here $(p, q, t)$ are the three superconformal fugacities, and $\{a_i\}$ is a set of fugacities associated with global symmetries $G^{\mr{f}}$ that commute with each other and with the superconformal algebra (and $\{f_i\}$ are the conserved flavor charges), $\Delta$ is the dilatation generator (a.k.a. the conformal Hamiltonian in radial quantization), $j_1$ and $j_2$ are the Cartan generators of the $\mr{SU(2)}_{1} \times \mr{SU(2)}_{2}$ isometry group of the spatial $\IS^3$, and $R$ and $r$ are the Cartan generators of $\mr{SU(2)_R}$ and $\mr{U(1)_R}$ respectively. As previously stated, the trace formula makes a choice of supercharge. The formula \eqref{eq:supcIndex} is written for a supercharge with quantum numbers $\Delta = \frac{1}{2}R = -r = \frac{1}{2}$, $(j_1, j_2) = (0, -1)$.\footnote{As in the main text, we use the same symbols for the generators and their eigenvalues. The eigenvalues are normalized in accordance with Appendix \ref{app:SCIMultiplets}.}

The superconformal index counts short multiplets modulo an equivalence relation that sets to zero the contributions of short multiplets that can recombine to long ones \cite{Kinney:2005ej}.\footnote{A consequence of the definition \eqref{eq:supcIndex} is that the index is zero on long multiplets, and therefore, it \emph{cannot} be used to distinguish between two short multiplets for which it evaluates to the same result. See \cite{Gadde:2009dj} for a discussion of this issue.}
Superconformal multiplets and their shortening conditions are briefly reviewed in Appendix \ref{app:SCIMultiplets}. (See \cite[App. B]{Beem:2013sza} for further details). 
%
%

The background $\IS^3 \times \IS^1$ has been studied extensively in the context of supersymmetric localization (see \cite{Pestun:2016zxk} for a review). $\IS^3 \times \IS^1$ is a non-K\"{a}hler complex manifold that admits a Hermitian metric, and admits nontrivial holomorphic line bundles classified by a rank one Dolbeault cohomology group $\CH^{0,1}(\IS^3 \times \IS^1)$ (although complex line bundles on it are trivial since $H^{2}(\IS^3 \times \IS^1, \IZ) = 0$). The parameter dependence of the partition function on background fields is discussed in \cite{Closset:2013vra,Closset:2014uda}, where it is argued that the path integral of a supersymmetric theory on a Hermitian manifold depends only on the complex structure deformations of the manifold. 

While it is often remarked that the supersymmetric index is the partition function on $\IS^3 \times \IS^1$, a more precise statement is that it is proportional to the partition function up to a scheme-independent (and hence physical) multiplicative factor involving the supersymmetric Casimir energy \cite{Assel:2014paa,Lorenzen:2014pna,Assel:2015nca}. 
This multiplicative factor is dependent on the central charges $a_{4d}$ and $c_{4d}$ and reduces to $1$ for a conformal theory.

In general, the three-parameter superconformal index \eqref{eq:supcIndex} is not known in closed analytic form as a function of $(p, q, t)$. There is at least one technique for determining it in terms of a series expansion in these variables by exploiting generalized S-duality and recasting the problem to one of diagonalizing a set of commuting difference operators known as the elliptic Ruijsenaars-Schneider Hamiltonians \cite{Gaiotto:2012xa, Razamat:2013qfa}. The eigenfunctions of these operators are functions of $(p, q, t)$ and one can reconstruct the index as a sum of products of these ``wavefunctions,'' weighted with coefficient functions that encode data at the punctures in Class $\CS$ theory. In principle, this algorithm can be used to determine a series expansion of $\CI(p,q,t, \{a_i\})$ in all fugacities.\footnote{See also \cite{Pan:2021mrw} for some recent work on expressing the Schur index (with flavor fugacities) in closed form.}

In practice, however, it is preferable to have a specialization that is computable with less effort and nevertheless contains useful information. One such specialization is the so-called Schur limit of the superconformal index, with the resulting index known as the Schur index.\footnote{Its name stems from the existence of closed-form expressions for the ``wavefunctions,'' in terms of Schur polynomials.}
It can be obtained from \eqref{eq:supcIndex} by setting $q = t$ and noting that the $p$-dependence is $\CQ$-exact and drops out\footnote{One uses the fact that the index is independent of $\beta$, and in particular, counts states with 
$$
\Delta - j_1 - R + \frac{1}{2}r = 0 ~ . 
$$
This determines $j_1$ for a given $(\Delta, r, R)$. The exponents of $p$, $q$, and $t$ in \eqref{eq:supcIndex} are combinations of Cartan generators that commute with the supercharge $\CQ$ used to define the index in the first place. The power of $p$ in \eqref{eq:supcIndex} becomes $\frac{1}{2}\big\{ \wt{Q}_{1,\dt{+}}, \big(\wt{Q}_{1,\dt{+}}\big)^{\dagger}\big\} = \frac{1}{2}(\Delta + j_2 - R - \frac{1}{2}r)$ -- see footnote \ref{foot:supconfgenerators}. The $p$-dependence is therefore $\CQ$-exact, and the index counts states with $\Delta + j_2 - R - \frac{1}{2}r = 0$. This simplification leads to \eqref{eq:schur_index_appendix}}
leading to
\begin{align}\label{eq:schur_index_appendix}
  \CI_{\mr{Schur}}(q, \{a_i\}) &= \mathsf{Tr\,}_{\CH(\IS^3)} (-1)^{\mr{F}} q^{\Delta - \frac{R}{2}}  \prod_{i=1}^{\mr{rk}\,G^{\mr{f}}} a_{i}^{f_i} ~.
\end{align}
The states counted by this index -- known as Schur operators (or equivalently, Schur states) -- are harmonic representatives of cohomology classes of a nilpotent supercharge defined by the sum of a Poincar\'{e} supercharge $\CQ$ and a conformal supercharge $S$.\footnote{The Schur index is an intrinsically $\CN=2$ index.} 
The Schur index is, in fact, related to $q$-deformed 2d topological Yang-Mills theory \cite{Gadde:2009kb,Gadde:2011ik}. It is now known that Schur operators form the structure of a two-dimensional chiral algebra \cite{Beem:2013sza,Beem:2014rza,Lemos:2014lua}, and the Schur index is the supercharacter of the vacuum module of this chiral algebra. The Schur index also has interesting modularity properties related to conformal anomalies \cite{Razamat:2012uv}. Finally, the Schur index is also conjecturally related to a wall-crossing invariant \cite{Cordova:2015nma} defined in terms of the protected spin character of \cite{Gaiotto:2010be}. So, despite its apparent simplicity relative to the ``full'' superconformal index, it encodes a lot of useful information about the protected operator spectrum of the theory.\footnote{See also \cite{Gaiotto:2024kze} for a relation between pure $\mr{SU(2)}$ SYM and the SYK model, where Schur indices play an important role.}
For a superconformal $\CN=2$ theory, the index \textit{is} the partition function on $\IS^3 \times \IS^1$ as stated above. A natural question is whether there is a supergravity background on which the partition function computes the superconformal index, with couplings to background $\CN=2$ conformal supergravity fields.\footnote{For the $\CN=1$ index, a background based on new minimal supergravity was found in \cite[Sec. 5]{Festuccia:2011ws}.}
Such superconformal backgrounds have been discussed in \cite{Hama:2012bg,Pestun:2014mja}, with a detailed discussion in \cite{Pan:2019bor}, where explicit expressions may be found for the metric, the generalized Killing spinors, and profiles of background fields of $\CN=2$ conformal supergravity (in suitable parametrizations, bases of fields, and spinors).\footnote{\label{foot:toptwistschur}It would be interesting to apply the methods of \cite{Cushing:2023rha} to determine a background on which the partition function computes a ``topologically twisted Schur index,'' but as discussed in the main text, the interpretation of such an index would probably be appropriate only for a conformal theory. As is well known \cite{Witten:1988ze}, the trace of the energy-momentum tensor for topologically twisted $\CN=2$ SYM is a total derivative: the theory has global scale invariance, but not conformal invariance.}
(For a recent quick introduction to $\CN=2$ conformal supergravity, see \cite{Cushing:2023rha}.) The background is an $\IS^3$ fibration over $\IS^1$, where the fiber $\IS^3$ itself is a $\IT^2$ fibration over an interval.  In this formalism, one can turn on flat connections for background $\CN=2$ vectormultiplets for a maximal torus of the flavor symmetry group. These correspond to the flavor fugacities appearing in the index.

To our knowledge, a mass deformation of the Schur index is not known. On the one hand, it is quite natural --  from the standpoint of $\CN=2$ supergravity -- to consider couplings to background $\CN=2$ vectormultiplets with their scalars tuned to mass deformations (e.g., as we did in \eqref{eq:scalarmass}). However, for non-conformal theories, one cannot perform a Weyl rescaling from $\IS^3 \times \IR$ to $\IR^4$, and there is no state-operator correspondence. The interpretation of the resulting mass-deformed index is therefore less clear.\footnote{The authors of \cite{Cordova:2015nma} propose an extension of the Schur index to non-conformal Lagrangian $\CN=2$ theories (with $G^{\mr{gauge}}$-vectormultiplets and hypermultiplets in some representation of $G^{\mr{gauge}}$) in terms of a matrix integral over the maximal torus of $G^{\mr{gauge}}$. However, they note that it is a priori unclear why this extension will lead to a protected observable, and that further investigation is merited. Another non-conformal generalization of the Schur index which is based on isolating a subalgebra of the superconformal algebra that includes only the isometries of $\IS^3 \times \IR$ and $\mr{SU(2)_R}$ was proposed in \cite{Dumitrescu2016}, but remains an unpublished proposal at the time of writing this.}

\setcounter{table}{0}
\setcounter{figure}{0}
\renewcommand{\thetable}{F.\arabic{table}}
\renewcommand{\thefigure}{F.\arabic{figure}}
\section{Lagrangian Analysis Of The $\mr{ffs}$ Trinion Theory}\label{app:Lag-ffs-Trinion-Analysis}

In this appendix, we describe the $\mr{ffs}$ trinion theory as a Lagrangian theory of a free hypermultiplet and present the action. Let $V$ be the $N$-dimensional defining representation of $\mr{SU(N)}$.  The field content of the theory is a hypermultiplet valued in the quaternionic space $E\oplus \ov{E}$ where $E$ is the representation $V \otimes \ov{V}$ of $\mr{SU(N)} \times \mr{SU(N)}\times \mr{U(1)}$ with $\mr{U(1)}$ acting as scalars.\footnote{In fact we could have chosen global symmetry group $\mr{U(N)} \times \mr{U(N)} \times \mr{U(1)}$ but a subgroup of the center isomorphic to $\mr{U(1)}\times \mr{U(1)}$ acts trivially, so we choose to take $\mr{SU(N)} \times \mr{SU(N)} \times \mr{U(1)}$ for simplicity. Otherwise, we would need to discuss non-effective group actions and stacks. This is beyond the scope of this paper.}

To write the Lagrangian, we write the hypermultiplet in terms of a pair of $\mathcal{N}=1$ chiral multiplets. The field content is same as a chiral multiplet $Q^{m\ov{n}}$ and an anti-chiral multiplet $\wt{Q}'^{m\ov{n}}$ both in $V \otimes \ov{V}$.\footnote{An anti-chiral multiplet is distinguished from a chiral multiplet by having a Weyl spinor of the opposite chirality. In Euclidean signature, chiral and antichiral multiplets are independent and unrelated by complex conjugation: the complex conjugate of a chiral multiplet is another chiral multiplet.}   
Let us write their component fields as\footnote{Here $m,n=1,\ldots, N$ are indices for $V$, and $\ov{m},\ov{n}=1,\ldots,N$ are indices for $\ov{V}$. We use the convention that upper left (right) indices are for the first (second) factor and lower left (right) indices are for the dual space of the first (second) factor. We could equivalently have treated $(q^2)^{m\ov{n}}$ and $(\wt{q}^{1})^{\ov{n}m}$ as independent scalars.}
\begin{equation}\label{eq:ffs-Q-Qtildeprime}
Q^{m\ov{n}}=\left((q^2)^{m\ov{n}}, (\psi_{\alpha})^{m\ov{n}}\right) ~,\quad \wt{Q}'^{m\ov{n}}=\left((q^1)^{m\ov{n}}, (\wt{\chi}_{\dt{\alpha}})^{m\ov{n}}\right)~.    
\end{equation}
Define complex scalars $\wt{q}^{A,\ov{n}m}$ valued in $\ov{V}\otimes V$ via
\begin{align}\label{eq:realityscalars}
        \wt{q}^{A,\ov{n}m} = (q_{A,m\ov{n}})^{\dagger} =  \Omega^{m\ov{n},\ov{n}'m'}\varepsilon^{AB}q_{B,m'\ov{n}'} ~,
\end{align}
where $\Omega$ is a symplectic form\footnote{There exists a choice of basis exhibiting $V \otimes \ov{V} \oplus \ov{V} \otimes V$ as a real vector space with real dimension $4N^2$. In this basis, $\Omega$ is a $2N^2 \times 2N^2$ matrix with a block diagonal form, with $N^2$ blocks each of the form $\left(\begin{smallmatrix}
    0 & 1\\
    -1 & 0
\end{smallmatrix}\right)$.}
on $V \otimes \ov{V} \oplus \ov{V} \otimes V$, and $\varepsilon^{AB}$ is the completely antisymmetric invariant tensor of $\mathfrak{su}(2)_{\mr{R}}$, with $\varepsilon^{12}=1$. The matrix 
\begin{equation}
\left(\begin{array}{cc}
q^1 & q^2 \\
-\wt{q}_1 & \wt{q}_2
\end{array}\right) ~,
\end{equation}
(with $V\otimes \ov{V}$ indices suppressed) is then a quaternion in the standard quaternionic structure on $V\otimes \ov{V}\oplus \ov{V} \otimes V$ and $\mr{SU(2)_R}$ acts on the scalars by right multiplication by unit quaternions. 

For later use, we also define conjugate spinors
    \begin{align}\label{eq:realityferms}
    \begin{split}
        \chi^{\alpha,\ov{n}m} &:= \big( \psi_{\alpha, m \ov{n}}\big)^{\dagger} = \varepsilon^{\alpha\beta}\Omega^{m\ov{n},\ov{n}'m'}\psi_{\beta,m'\ov{n}'} ~,\\
        \wt{\psi}^{\dt{\alpha},\ov{n}m} &:= \big( \wt{\chi}_{\dt{\alpha},m\ov{n}} \big)^{\dagger} = \varepsilon^{\dt{\alpha}\dt{\beta}}\Omega^{m\ov{n},\ov{n}'m'}\wt{\chi}_{\dt{\beta},m'\ov{n}'} ~.
     \end{split}
    \end{align}
Let $\wt{P}_T$ denote a principal $\wt{G}_T$ bundle over $\IX$, where
\begin{align}
    \wt{G}_{T} := \mr{Spin(4)} \times \mr{SU(2)_R} \times \mr{SU(N)} \times \mr{SU(N)} \times \mr{U(1)} ~.
\end{align}
For this bundle to exist, $\IX$ must be a spin manifold. We assume this is so for the moment. In the main text, we have discussed how this can be relaxed by defining the theory in terms of a $\wt{G}_{T}/C_T$ bundle, for a suitable choice of subgroup $C_T$ of the center of $\wt{G}_T$.

To write the action in curved space, we follow the procedure outlined in section \ref{sec:backind} or \cite{Karlhede:1988ax, Cushing:2023rha}, coupling the hypermultiplet theory to a \underline{bosonic} background of off-shell $\CN=2$ conformal supergravity that preserves $\CN=2$ supersymmetry. Such a background can be defined by consistently setting the gravitinos and auxiliary spinor fields in the $\CN=2$ conformal supergravity multiplet to zero, as well as their supersymmetry variations to zero.\footnote{\label{foot:nodiffconst}For the untwisted theory, this leads to a differential constraint on the supersymmetry parameter. However, as is well known from \cite{Witten:1988ze,Karlhede:1988ax,Cushing:2023rha}, upon twisting, there is a globally-defined scalar supersymmetry parameter corresponding to $\CQ$ symmetry, and no differential constraints are needed. The bosonic auxiliary fields of the supergravity multiplet are allowed to have nontrivial profiles consistent with supersymmetry \cite{Festuccia:2011ws}. However, for simplicity we assume that the 2-form field in the $\CN=2$ conformal supergravity multiplet is set to zero: $T_{\mu\nu}^{\pm} = 0$, as this simplifies our partition function. For examples of more general backgrounds, see \cite{Hama:2012bg,Pestun:2014mja}.}
The supergravity background contains a connection for $\mr{SU(2)_R} \times \mr{U(1)_R}$ symmetry, but we treat the $\mr{U(1)_R}$ symmetry as rigid and gauge its connection to zero (see \cite{Cushing:2023rha} for details). The $\mr{SU(2)_R}$ connection plays an important role in the specialization of twisting, discussed later. We also couple to background vectormultiplets for the aforementioned $\mr{SU(N)} \times \mr{SU(N)} \times \mr{U(1)}$ flavor symmetry: to this end, we introduce connections $A_\mu^{\mr{SU(N)}^{(1)}}$, $A_\mu^{\mr{SU(N)}^{(2)}}$, $ A_\mu^{\mr{U(1)}}$. Although we write separate connections to clarify how they contribute to the covariant derivatives, it is useful to think of a composite background vectormultiplet for the $\mr{SU(N)} \times \mr{SU(N)} \times \mr{U(1)}$ flavor symmetry. We denote the composite curvature for this background vectormultiplet by $F_{\mu\nu}^{a}$ where $a, b = 1, \ldots, 2(\mr{N}^2-1) + 1$ run over the generators of the Lie algebra of the gauge group as usual. The background vectormultiplet also contains an adjoint-valued auxiliary field $D_{AB}^{a}$ which transforms as a triplet of $\mathsf{SU(2)_R}$. 

The action of the free hypermultiplet coupled to background fields is 
\begin{align}
\label{eq:hypaction}
\CS_{\mr {Hyper }} &=\int_{\IX} \mr{vol}(g)\bigg[\n^{\mr{f}}_\mu (\wt{q}_A)^{\ov{n} m} \n_{\mr{f}}^\mu (q^A)_{m \ov{n}}- \frac{1}{24}\big(\mathscr{R} + 3 \mathscr{D}\big) (\wt{q}_A)^{\ov{n}m}(q^A)_{m\ov{n}} \nonumber\\
&\qquad\qquad\quad\quad  + (\wt{q}_A)^{\ov{n}m}D^{AB}(q_B)_{m\ov{n}}  +\im \wt{\psi}^{\ov{n}m} \slashed{D}_{\CV\otimes \ov{\CV}}^{+} \psi_{m\ov{n}} 
- \im \chi^{\ov{n}m}\slashed{D}^{-}_{\CV \otimes \ov{\CV}}\wt{\chi}_{m\ov{n}} \bigg]\nonumber \\
&\,\,\,\,+\frac{1}{4 \pi} \int_{\IX} \mr{vol}(g)\left[\mr{Im\,} \tau_{ab}\left(\frac{1}{4} F_{\mu \nu}^a F^{\mu \nu,b}-\frac{1}{2} D_{AB}^a D^{AB,b}\right)+\frac{\im}{32\pi} \tau_{ab}(F^a \wedge F^b)\right]~.
\end{align}
Here, $\mathscr{D}$ is the auxiliary scalar in the $\CN=2$ conformal supergravity multiplet, $\mathscr{R}$ is the Ricci scalar curvature of $\IX$, $\n_{\mu}^{\mr{f}}$ (or $\n^{\mu}_{\mr{f}}$) are (metric + background gauge (flavor $\mr{f}$) + $\mr{SU(2)_R}$)-covariant derivatives, and $\slashed{D}^{+}$ denotes the chiral Dirac operator coupled to background gauge (flavor) and $\mr{SU(2)_R}$ connections -- we will momentarily write their explicit expressions.

We have made the following important specialization of the background vectormultiplet: we have set its scalars to zero (to preserve the full flavor symmetry) and we have set its fermions to zero as doing so eliminates coupling terms between spinors in the hypermultiplet and the spinors in the background vectormultiplet. However, this imposes an algebraic constraint on $D_{AB}^{a}$ in terms of the supersymmetry parameter defining the $\CN=2$ supergravity background and the curvature $F_{\mu\nu}^{a}$.\footnote{\label{foot:Dconstraint}Upon further specialization to the twisted theory, this constraint will simplify to \eqref{eq:susy_back_constraint_twisted}. The supersymmetry constraints on the background vectormultiplet can be easily deduced from the results of \cite{Cushing:2023rha}, but are not needed to follow the present discussion.} 

 The covariant derivatives appearing above are connections on associated bundles induced from the $\wt{G}_T$-connection $\n_T$ on $\wt{P}_T$, and $\slashed{D}_{\CV\otimes \ov{\CV}}^{+}$ denotes the Dirac operator coupled to the vector bundle $\CV\otimes \ov{\CV}$ associated to $\wt{P}_T$ by the representation $V\otimes \ov{V}$ of $\mr{SU(N)}\times\mr{SU(N)}\times\mr{U(1)}$.
 
 Let $\omega_\mu$, $A_\mu^{\mr{R}}$, $A_\mu^{\mr{S U}(N)^{(1)}}$, $A_\mu^{\mr{S U}(N)^{(2)}}$ and $A_\mu^{\mr{U(1)}}$ denote local coordinate representatives of the connections on vector bundles associated to $\left(p_1\right)_*\left(\n_T\right),\left(p_2\right)_*\left(\n_T\right),\left(p_3\right)_*\left(\n_T\right)$,
$\left(p_4\right)_*\left(\n_T\right)$, and $\left(p_5\right)_*\left(\n_T\right)$ respectively. Here $\omega_\mu$ is the spin connection. The covariant derivatives are
\begin{align}
\hspace{-0.15in} \n_{\mu}^{\mr{f}} q^A{}_{m\ov{n}} &=\partial_\mu q^A{}_{m\ov{n}}+ \big(A_\mu^{\mr{R}}\big)^A{}_B ~q^B{}_{m\ov{n}}+ \big(A_\mu^{\mr{S U}(N)^{(1)}}\big)^{m'}{}_{m} ~q^2{}_{m'\ov{n}}\delta^A_2 + \big(A_\mu^{\mr{S U}(N)^{(2)}}\big)^{\ov{n}'}{}_{\ov{n}} ~q^1{}_{m\ov{n}'}\delta^A_1 \nn
\hspace{-0.15in} &\quad+A_\mu^{\mr{U(1)}} q^A{ }_{m\ov{n}}~, \label{eq:nabla-f-q}\\
\hspace{-0.15in} \n_{\mu}^{\mr{f}} \wt{q}_A{}^{\ov{m}\,n} &=\partial_\mu \wt{q}_A{}^{\ov{m}\,n}+ \big(A_\mu^{\mr{R}}\big)^B{}_A~ \wt{q}_B{}^{\ov{m}\,n}+ \big(A_\mu^{\mr{S U}(N)^{(1)}}\big)^{\ov{m}}{}_{\ov{m}'}~\wt{q}_2{}^{\ov{m}'\,n}\delta_A^2+ \big(A_\mu^{\mr{S U}(N)^{(2)}}\big)^{n}{}_{n'}~\wt{q}_1{}^{\ov{m}\,n'}\delta_A^1 \nn
\hspace{-0.15in} &\quad + A_\mu^{\mr{U(1)}} \wt{q}_A{}^{\ov{m}\,n}~, \label{eq:nabla-f-q-tilde} \\
\hspace{-0.15in} (D_{\CV\otimes \ov{\CV}})_\mu \psi_{\alpha m\ov{n}}&=\partial_\mu \psi_{\alpha m\ov{n}}-\frac{1}{2} \omega_\mu{}^{\ua\,\ub}(\sigma_{\ua\, \ub})^{\beta}{ }_{\alpha} \psi_{\beta m\ov{n}}+ \big(A_\mu^{\mr{SU(N)}^{(1)}}\big)^{m'}{}_{m} \psi_{\alpha m'\ov{n}}+ \big(A_\mu^{\mr{S U}(N)^{(2)}}\big)^{\ov{n}'}{}_{\ov{n}}\psi_{\alpha m\ov{n}'} \nn
\hspace{-0.15in} &\quad +A_\mu^{\mr{U(1)}} \psi_{\alpha m\ov{n}}~, \label{eq:spinor-covder-psi}\\
\hspace{-0.15in} (D_{\CV\otimes \ov{\CV}})_\mu \wt{\chi}_{\dt\alpha}{}^{m\,\ov{n}}&=\partial_\mu \wt{\chi}_{\dt\alpha}{}^{m\,\ov{n}}-\frac{1}{2} \omega_\mu{}^{\ua\,\ub}(\wt{\sigma}_{\ua\, \ub})^{\dt\beta}{ }_{\dt\alpha} \wt{\chi}_{\dt\beta}{ }^{m\,\ov{n}}+ \big(A_\mu^{\mr{SU(N)}^{(1)}}\big)^{m}{}_{m'} \wt{\chi}_{\dt\alpha}{ }^{m'\,\ov{n}}+ \big(A_\mu^{\mr{S U}(N)^{(2)}}\big)^{\ov{n}}{}_{\ov{n}'}\wt{\chi}_{\dt\alpha}{ }^{m\,\ov{n}'} \nn
\hspace{-0.15in} &\quad+A_\mu^{\mr{U(1)}} \wt{\chi}_{\dt\alpha}{ }^{m\,\ov{n}}~, \label{eq:spinor-covder-chi-tilde}
\end{align}
Recall that for chiral spinors valued in a vector bundle $E \to \IX$ with  connection, the chiral Dirac operators are maps
\begin{align}\label{eq:DiracV}
\begin{split}
    \slashed{D}^{+}_{E} : \Gamma( S^{+} \otimes E) &\to \Gamma(S^{-} \otimes E) ~,\\
    \slashed{D}^{-}_{E} : \Gamma( S^{-} \otimes E) &\to \Gamma( S^{+} \otimes E) ~,
\end{split}
\end{align}
where $S^{\pm}$ are chiral spinor bundles on $\IX$.\footnote{\label{foot:adjoints}For a complex vector bundle $E$, the adjoint of $D^{+}_{E}$ is $(\slashed{D}^{+}_{E})^{\dagger} = \slashed{D}^{-}_{E}$.}
 The corresponding Dirac operator is
 %
 %
\begin{align}\label{eq:diracopmatrix}
    \slashed{D}_{E} &= \begin{pmatrix}
        0 & -\im\slashed{D}^{-}_{E} \\
        \im\slashed{D}^{+}_{E} & 0
    \end{pmatrix} ~,
\end{align}
in a basis determined by the chiral splitting. For us,
\begin{equation}
\slashed{D}^+_{\CV\otimes\ov{\CV}} := \wt{\sigma}^\mu \big(D_{\CV\otimes \ov{\CV}}\big)_\mu  ~,\quad\slashed{D}^-_{\CV\otimes \ov{\CV}}:= \sigma^\mu \big(D_{\CV\otimes \ov{\CV}}\big)_\mu~.    \label{eq:dslashed}
\end{equation}
The partition function of the theory is
\begin{align}
\wt{\mr{Z}}_T & =\int[\CD \wt{q} \CD q \CD \wt{\psi} \CD \psi \CD \wt{\chi} \CD \chi] e^{-\CS_{\mr{Hyper}}} \nonumber\\
& =  \mr{Det}\big(\im\slashed{D}_{\CV\otimes\ov{\CV}}^{+}\big) \mr{Det}\big(-\im\slashed{D}_{\CV\otimes\ov{\CV}}^{-}\big)\int[\CD \wt{q} \CD q ]e^{-S_{\mr{Hyper}}^{\mr{Bosonic}}} ~,
\label{eq:Zffsuntwisted-repeated}
\end{align}
where $S_{\mr{Hyper}}^{\mr{Bosonic}}$ is the bosonic part of the action \eqref{eq:hypaction}, and $\mr{Det}\big(\im\slashed{D}_{\CV\otimes\ov{\CV}}^{+}\big) \mr{Det}\big(-\im\slashed{D}_{\CV\otimes \ov{\CV}}^{-}\big)$ is a section of the determinant line bundle
\begin{equation}
\mr{DET}\big(\im\slashed{D}_{\CV\otimes\ov{\CV}}^{+}\big) \mr{DET}\big(-\im\slashed{D}_{\CV\otimes \ov{\CV}}^{-}\big)~. \label{eq:ut-det-prod-repeated}
\end{equation}

Recall that the determinant line bundle of a chiral Dirac operator coupled to a bundle $E$ with connection has fiber 
\be 
\left( \mr{\Lambda}^{\mr{max}} \Ker(\slashed{D}^\pm_{E})\right)^\vee  \otimes \mr{\Lambda}^{\mr{max}}\Cok(\slashed{D}^\pm_{E})\cong 
\left( \mr{\Lambda}^{\mr{max}} \Ker(\slashed{D}^\pm_{E})\right)^\vee  \otimes \mr{\Lambda}^{\mr{max}}\Ker(\slashed{D}^\pm_{E})^\dagger ~,
\ee
over the space of connections on $\wt{P}_T$.  
Note that $\mr{DET}\big(-\im\slashed{D}_{\CV\otimes \ov{\CV}}^{+}\big)$ is dual to $\mr{DET}\big(\im\slashed{D}_{\CV\otimes \ov{\CV}}^{-}\big)$, therefore \eqref{eq:ut-det-prod-repeated} is isomorphic to the trivial line bundle, and therefore, the partition function of a free hypermultiplet is just a $c$-number, as expected.

\subsection{Topological Twisting Of The $\mr{ffs}$ Theory}\label{app:top-twist-ffs-theory}

Upon twisting, $q_{A}$ becomes $M_{\dt{\alpha}}$ and $\wt{q}_{A}$ becomes $\wt{M}_{\dt{\alpha}}$. Note that $M_{\dt{\alpha}} \in \Gamma( S^{-} \otimes \CV \otimes \ov{\CV})$ and $\wt{M}_{\dt{\alpha}} \in \Gamma( S^{-} \otimes \ov{\CV}\otimes \CV)$.
After topological twisting, for $\CQ$-symmetry, and in particular, for $\CS_{\mr{Hyper}}$ to define a $\CQ$-exact energy-momentum tensor -- as we have already noted in section \ref{sec:backind} -- the fermions in the background vectormultiplet must be set to zero, and therefore, their $\CQ$-variations must be required to vanish for consistency. In particular, the vanishing of the $\CQ$-variation of $\chi_{\mu\nu}$ yields the condition
\begin{align}\label{eq:susy_back_constraint_twisted}
    D_{\mu\nu} &= F_{\mu\nu}^{+} ~, \text{ or equivalently } D_{\dt{\alpha}\dt{\beta}} = F_{\dt{\alpha}\dt{\beta}} ~,
\end{align}
where $F_{\mu\nu}$ is the composite curvature for the background $\mr{SU(N)} \times \mr{SU(N)} \times \mr{U(1)}$ connection (which, for purposes of writing the covariant derivatives above, was split into three connections for the three individual factors) and $D_{\mu\nu}$ is the self-dual 2-form auxiliary field of the background vectormultiplet. We also set 
\begin{align}\label{eq:Daux}
\mathscr{D} &= \frac{1}{6}\mathscr{R} ~,
\end{align}
to specialize to the twisted background.\footnote{See \cite{Karlhede:1988ax,Cushing:2023rha} for a derivation of this condition.}

To obtain the twisted action $S^{\mr{twisted}}_{\mr{hyp}}$ from the untwisted action \eqref{eq:hypaction} we use the constraints \eqref{eq:susy_back_constraint_twisted} and \eqref{eq:Daux} and the Lichnerowicz identity \cite{Lichnerowicz1963}\footnote{In standard mathematical literature, and indeed if the conventions/normalizations of \cite{Cushing:2023rha} were used, the Lichnerowicz identity would have the more familiar form 
\begin{align*} 
    \slashed{D}_{\CV\otimes\ov{\CV}}^{+}\slashed{D}_{\CV\otimes\ov{\CV}}^{-}M^{\dt{\alpha}} &= \n_{\mu}^{\mr{f}} \n_{\mr{f}}^{\mu} M^{\dt{\alpha}} - \frac{1}{4}\mathscr{R}M^{\dt{\alpha}} + \frac{1}{2}F^{\dt{\alpha}\dt{\beta}}M_{\dt{\beta}} ~.
\end{align*}
However, the conventions of this paper differ slightly from \cite{Cushing:2023rha}, leading to \eqref{eq:lichnerowicz}.}
\begin{align}\label{eq:lichnerowicz}
    \slashed{D}_{\CV\otimes\ov{\CV}}^{+}\slashed{D}_{\CV\otimes\ov{\CV}}^{-}M^{\dt{\alpha}} &= \n_{\mu}^{\mr{f}} \n_{\mr{f}}^{\mu} M^{\dt{\alpha}} - \frac{1}{16}\mathscr{R}M^{\dt{\alpha}} + F^{\dt{\alpha}\dt{\beta}}M_{\dt{\beta}} ~.
\end{align}
to combine terms in the twisted action containing $M_{\dt{\alpha}}$ and $\wt{M}_{\dt{\alpha}}$ as follows:
\begin{align}
     &\int_{\IX} \mr{vol}(g)\left(\n^{\mr{f}}_\mu \wt{M}_{\dt{\alpha}}\n_{\mr{f}}^\mu M^{\dt{\alpha}}-\frac{1}{16}\mathscr{R}\wt{M}_{\dt{\alpha}}M^{\dt{\alpha}} + \widetilde M_{\dt{\alpha}}F^{\dt{\alpha}\dt{\beta}}M_{\dt{\beta}}\right) \nonumber\\
    & \underset{\text{by parts}}{\overset{\text{integrate}}{=\joinrel=\joinrel=\joinrel=}} \int_{\IX} \mr{vol}(g)\left(\wt{M}_{\dt{\alpha}}\n^{\mr{f}}_\mu \n_{\mr{f}}^\mu M^{\dt{\alpha}}-\frac{1}{16}\mathscr{R}\wt{M}_{\dt{\alpha}}M^{\dt{\alpha}} + \widetilde M_{\dt{\alpha}}F^{\dt{\alpha}\dt{\beta}}M_{\dt{\beta}}\right) \nonumber\\
    &\underset{\text{\eqref{eq:lichnerowicz}}}{\overset{\text{Lichnerowicz}}{=\joinrel=\joinrel=\joinrel=\joinrel=}}\int_{\IX} \mr{vol}(g) \wt{M}\slashed{D}_{\CV\otimes\ov{\CV}}^{+}\slashed{D}_{\CV\otimes\ov{\CV}}^{-} M ~. \label{eq:identnew}
\end{align}

The path integral of the twisted theory is expressed as
\be\label{eq:partition_func_free_hyper} 
\mr{Z}_T^{\mr{twisted}} = \exp\left(-\frac{\mr{i}}{16\pi}\int_{\IX}\tau_{ab}~(F^a\wedge F^b)\right)\int [\CD\mr{HM}] e^{-S^{\mr{twisted}}_{\mr{hyp}}} ~,
\ee
where $S^{\mr{twisted}}_{\mr{hyp}}$ is the twisted version of the hypermulitplet part of the action of equation \eqref{eq:Stwisted-hyp}.
Thus the Gaussian path integral in \eqref{eq:partition_func_free_hyper} yields
\begin{equation}\label{eq:gaussian-twisted-repeated}
\begin{split}
    \int [\CD\mr{HM}] e^{-S^{\mr{twisted}}_{\mr{hyp}}}=\left(\mr{Det}\left(\slashed{D}_{\CV\otimes\ov{\CV}}^+\slashed{D}_{\CV\otimes\ov{\CV}}^-\right)\right)^{-1}\mr{Det}\big(\im\slashed{D}_{\CV\otimes\ov{\CV}}^{+}\big) \mr{Det}\big(-\im\slashed{D}_{\CV \otimes \ov{\CV}}^{-}\big) = 1 ~.
\end{split}    
\end{equation}
Because of $\CQ$-symmetry, the determinants cancel. The prefactor in \eqref{eq:partition_func_free_hyper} is a topological invariant, and in fact,
\be\label{eq:partition_func_free_hyper_twisted_final_repeated} 
\mr{Z}_T^{\mr{twisted}}  = \exp\left(-\frac{\mr{i}}{16\pi}\int_{\IX}\tau_{ab}~(F^a\wedge F^b)\right) ~.
\ee

\subsection{Central Action And Extension Of Domain }\label{app:central-action-and-extension}

 Since the two determinant line bundles in \eqref{eq:ut-det-prod-repeated} are dual, the action of the global symmetry associated to the center of $\widetilde{G}_T$ acts trivially on the product. Therefore we cannot use that expression to derive $C_T^{\mr{max}}$. However, a further extension of the domain  
 allows us to derive this group. 
Note that equation \eqref{eq:hypaction} just involves massless free bosons and fermions coupled to external fields. For trivial background fields, the flavor symmetry of the action is in fact larger than the product of flavor
symmetries associated to the separate punctures (as often happens in theories of class $\CS$). Let 
\be \label{eq:ffsflavorplus}
G_f:=  \mr{SU(N)} \times \mr{SU(N)} \times \mr{U(1)} ~.
\ee
There is (classically) a $G_f^+ \times G_f^-$ symmetry where $G_f^+$ acts trivially on $\wt{\chi}$, $\chi$ and acts on  $\wt{\psi}$, $\psi$ in the standard way. Similarly, $G_f^-$ acts trivially on 
$\wt{\psi}$, $\psi$. We introduce a background bundle with connection for $G_f^+ \times G_f^-$. The coupling to the bosonic fields breaks $\mr{SU(2)_R}$ symmetry by choosing a complex structure for the quaternionic target space:  We couple 
$G_f^+$ to the chiral multiplets $Q^{m\ov{n}}$ and $( \wt{Q}'^{m\ov{n}})^{\dagger}$, and $G_f^-$ to the anti-chiral multiplets $(\wt{Q}^{m \ov{n}})^{\dagger}$ and $\wt{Q}'^{ m \ov{n} }$.  
%
%
Then our full symmetry group is extended to: 
\be \label{eq:ffsflavorextended}
\wt{G}_T^{\mr{extended}} =  \mr{Spin(4)} \times \mr{SU(2)_R} \times G_f^+ \times G_f^- ~,
\ee
Now we can consider the chiral actions of subgroups of the center. Elements of $C_T^+$ 
are of the form 
\begin{equation}
(\zeta,1,\zeta,\omega^{k_1^+}\mathbf{1}_N,\omega^{k_2^+}\mathbf{1}_N,\xi^+,  1_{G_f^-} ) ~,   
\end{equation}
while elements of $C_T^-$ are of the form 
\begin{equation}
(1,\zeta,\zeta,1_{G_f^+}, \omega^{k_1^-}\mathbf{1}_N,\omega^{k_2^-}\mathbf{1}_N,\xi^- ) ~,   
\end{equation}
The action on the partition function is: 
\begin{equation}
\hspace{-0.1in}(\zeta,1,\zeta,\omega^{k_1^+}\mathbf{1}_N,\omega^{k_2^+}\mathbf{1}_N,\xi^+,  1_{G_f^-} )\cdot\wt{\mr{Z}}_T(\wt{P}_T, \wt{\n}_T)=(\zeta\omega^{k_1^+}\ov{\omega}^{k_2^+}\xi^+)^{\mr{Ind}(\slashed{D}_{\CV\otimes\ov{\CV}}^{+})}  \wt{\mr{Z}}_T(\wt{P}_T, \wt{\n}_T)~.    
\end{equation}
\begin{equation}
\hspace{-0.15in}(1,\zeta,\zeta,1_{G_f^+}, \omega^{k_1^-}\mathbf{1}_N,\omega^{k_2^-}\mathbf{1}_N,\xi^-,    ) \cdot\wt{\mr{Z}}_T(\wt{P}_T, \wt{\n}_T)= (\zeta\omega^{k_1^-}\ov{\omega}^{k_2^-}\xi^-)^{\mr{Ind}(\slashed{D}_{\CV\otimes\ov{\CV}}^{-})}\wt{\mr{Z}}_T(\wt{P}_T, \wt{\n}_T)~.    
\end{equation}
Demanding that these actions are trivial implies: 
\be \label{eq:xipm}
\xi^\pm = \zeta \omega^{-k_1^\pm + k_2^\pm} ~, 
\ee
and we thus produce invariance groups $C_T^{\pm, \mr{max}}$ isomorphic to $C_T^{\mr{max}}$  in  \eqref{eq:CTmax_freehyper}.  We can, of course, restrict to the 
vectorlike couplings used above. 
In general, we expect it is important to use the maximal symmetry group to find the most stringent conditions for defining $C_T^{\mr{max}}$.

\setcounter{table}{0}
\setcounter{figure}{0}
\renewcommand{\thetable}{G.\arabic{table}}
\renewcommand{\thefigure}{G.\arabic{figure}}
\section{Examples Of Cohomological Conditions For $A_{N-1}$ Class $\CS$ Theories With Full And Simple Punctures}\label{app:Class-S-Coho-Examples}

In this section, we will derive cohomological conditions for the existence of weakly coupled descriptions of $A_{N-1}$ class $\CS$ theories on general non-spin manifolds following the discussion in section \ref{sec:GenClassS-Twisting}. These cohomological conditions depend on the chosen pants decomposition for the weakly coupled description of the theory. For simplicity, we will restrict to the theories constructed from trinions with $\mr{ffs}$ and $\mr{fff}$ punctures and only full punctures being gauged.

Let us briefly recall the notation from section \ref{sec:GenClassS-Twisting}. Choose a pants decomposition of $C_{g,n}$ into a set $\CT$ of trinions. The external cutting curves are denoted by $c_i$ and internal cutting curves are denoted by $c_\alpha$. We will denote the external cutting curve bounding a full (resp. simple) puncture by $c_i^{\mr{f}}$ (resp. $c_i^{\mr{s}}$). 

For the cutting curves bounding a trinion $T$, we will write $c_{T,i}$, $c_{T,\alpha}$. The simply connected cover of the flavor group at an external cutting curve $c_i$ is denoted by $\wt{G}_{c_i}$ and the simply connected cover of the gauge group at an internal cutting curve $c_\alpha$ is denoted by $\mr{SU(N)}_{c_\alpha}$. Define the group 
\begin{equation}\label{eq:GS-An-1-class-S}
 G_{\mathsf{S}}=\bigg( \mr{Spin(4)} \times \mr{SU(2)_R} \times \prod_i \wt{G}_i  \times \prod_{\alpha} \mr{SU(N)}_\alpha  \bigg)\bigg/C''_{\CT}~,   
\end{equation}
where $C''_{\CT}$ is defined below \eqref{eq:Gphys-Class-S}. 
\par
The cohomological conditions for the theory are obtained from the existence of the principal $G_{\mr{S}}$ bundle $P_{\mr{S}}\to\IX$.
To determine these cohomological conditions, we will follow the same strategy as in section \ref{sec:cohcond}. Let $p_1,p_2,p_{3,c_i},p_{4,c_\alpha}$ be projection maps to various factors:
\begin{equation}\label{eq:p3-ci-and-p4-c-alpha}
    \begin{tikzcd}
	&& {G_{\mathsf{S}}} \\
	{\mathsf{Spin(4)}} & {\mathsf{SU(2)_R}} && {\widetilde{G}_{c_i}} & {\mathsf{SU(N)}_{c_\alpha}}
	\arrow["{p_1}"', from=1-3, to=2-1]
	\arrow["{p_2}", from=1-3, to=2-2]
	\arrow["{p_{3,c_i}}"', from=1-3, to=2-4]
	\arrow["{p_{4,c_\alpha}}", from=1-3, to=2-5]
\end{tikzcd}
\end{equation}
Let us also denote the transfer of structure groups of $P_{\mr{S}}$ along various projections by 
\begin{equation}\label{eq:bundles-for-An-1-class-S}
\begin{split}  
\mr{Fr}(\IX) & := (p_1)_*(P_{\mr{S}})  ~,\\ 
P^{\mr{R}} & := (p_2)_*(P_{\mr{S}}) ~,\\ 
P^{\mr{f},c_i} & := (p_{3,c_i})_*(P_{\mr{S}}) ~,\\
P^{\mr{gauge},c_\alpha} & := (p_{4,c_\alpha})_*(P_{\mr{S}}) ~.
\end{split} 
\end{equation}
Define a set of positive integers $n_{c_i^{\mr{s}}}$ by
\begin{equation}\label{eq:n-c-i-s}
p_{3,c_i^{\mr{s}}}(C''_{\CT})\cong\IZ/n_{c_i^{\mr{s}}}\IZ~.    
\end{equation}
Let us write 
\begin{equation}\label{eq:flavor_gauge_quotient_ZN}
    C_{c_i^{\mr{f}}}:= p_{3,c_i^{\mr{f}}}(C''_{\CT})\subseteq\IZ_N~,\quad  C_{c_\alpha}:=p_{4,c_\alpha}(C''_{\CT})\subseteq\IZ_N~.
\end{equation}
We can then introduce flat $C_{c_i^{\mr{f}}}$, $C_{c_\alpha}$-gerbes $b_{c_i^{\mr{f}}}$, $b_{c_\alpha}$ respectively. Their characteristic classes 
\begin{equation}\label{eq:mu-b-cif-and-cialpha}
\mu(b_{c_i^{\mr{f}}})\in H^2(\IX,C_{c_i^{\mr{f}}})~,\quad \mu(b_{c_\alpha})\in H^2(\IX,C_{c_\alpha})~,    
\end{equation}
can be identified with the obstruction to lifting the structure groups of $P^{\mr{f},c_i^{\mr{f}}}$ and $P^{\mr{gauge},c_\alpha}$ respectively from $\mr{SU(N)}/C^{\mr{flavor}}_{i,\mr{f}}$ and $\mr{SU(N)}/C^{\mr{grb}}_{\alpha}$, to $\mr{SU(N)}$. In particular, $\mu(b_{c_\alpha})$ can be identified with the 't Hooft flux of $\mr{SU(N)}_{c_\alpha}$ factor of the gauge group. Now choose transition functions 
\begin{equation}\label{eq:transition-func-for-GS-bundle}
 t_{pq}^{\mr{S}}:=\left[\left(\wt{t}^+_{pq},\wt{t}^-_{pq},\wt{t}^{\mr{R}}_{pq},\wt{t}^{\mr{f}}_{pq},\wt{t}^g_{pq}\right)\right]:U_{pq}\longrightarrow G_{\mr{S}}~,   
\end{equation}
for the $G_{\mr{S}}$-bundle $P_{\mr{S}}\to\IX$.
The cocycle on triple overlaps is given by
\begin{equation}
t_{pq}^{\mr{S}}t_{qr}^{\mr{S}}t_{rp}^{\mr{S}}=\left[\left(\zeta^+_{pqr},\zeta^-_{pqr},\zeta^{\mr{R}}_{pqr},\zeta^{\mr{f}}_{pqr},\zeta^g_{pqr}\right)\right]~.    
\end{equation}
Thus the cocycle condition is given by 
\begin{equation}
 \left(\zeta^+_{pqr},\zeta^-_{pqr},\zeta^{\mr{R}}_{pqr},\zeta^{\mr{f}}_{pqr},\zeta^g_{pqr}\right)\in C_{\CT}''~.   
\end{equation}
As before, the cocycles $\zeta^\pm_{pqr}$ and $\zeta^{\mr{R}}_{pqr}$ can be identified with the second Stiefel-Whitney class of $\IX$ and $P^{\mr{R}}$ respectively:
\begin{equation}
\begin{split}
    w_2(\IX)&= \left[\zeta_{pqr}^\pm\right] \in H^2\left(\IX, \mathbb{Z}_2\right) ~,\\ w_2\big(P^{\mr{R}}\big):&=\big[\zeta_{pqr}^{\mr{R}}\big] \in H^2\left(\IX, \mathbb{Z}_2\right) ~.
\end{split}
\end{equation}
The projections of the cocycles $\zeta^{\mr{f}}_{pqr}$ and $\zeta^g_{pqr}$ can be identified with representatives of the characteristic classes $\mu(b_{i,\mr{f}})$ and $\mu(b_{\alpha})$ by 
\begin{equation}
\begin{split}
\mu(b_{c_i^{\mr{f}}})&=\big[p_{3,c_i^{\mr{f}}}\left(\zeta_{pqr}^{\mr{f}}\right)\big]\in    H^2(\IX,C_{c_i^{\mr{f}}})~,
\\
\mu(b_{c_\alpha})&=\left[p_{4,c_\alpha}\left(\zeta_{pqr}^{g}\right)\right]\in H^2(\IX,C_{c_\alpha})~.
\end{split}
\end{equation}
Finally, let $c_1(c_i^{\mr{s}})\in H^2(\IX,\IZ)$ be the first Chern class 
of the $\mr{U(1)}$-bundle obtained by transfer of structure group of $P^{\mr{f},c_i^{\mr{s}}}$ along the isomorphism (see footnote \ref{foot:power_n_map}) 
\begin{equation}
\mr{U(1)}/\IZ_{n_{c_i^{\mr{s}}}}\stackrel{\cong}{\longrightarrow} \mr{U(1)}~.    
\end{equation}
The cocycle $p_{3,c_i^{\mr{s}}}\left(\zeta_{pqr}^{\mr{R}}\right)$ represents the $\bmod n_{i,\mr{s}}$ reduction of $c_1(c_i^{\mr{s}})$:
\begin{equation}
\left[p_{3,c_i^{\mr{s}}}\left(\zeta_{pqr}^{\mr{R}}\right)\right]= r_{n_{c_i^{\mr{s}}}}(c_1(c_i^{\mr{s}}))~.   
\end{equation}
We are now equipped to write the cohomological conditions. Due to the definition of $C_{\CT}''$, we must have 
\begin{equation}
    \zeta^-_{pqr}=\zeta^+_{pqr}=\zeta^{\mr{R}}_{pqr}~,
\end{equation}
which implies 
\begin{equation}
    w_2(\IX)=w_2(P^{\mr{R}})~.
\end{equation}
Next, if two trinions $T,T'\in\CT$ are bounded by a common internal cutting curve $c_{T,\alpha}=c_{T',\alpha}$, then we must have 
\begin{equation}
 p_{4,c_{T,\alpha}}\left(\zeta_{pqr}^{g}\right)=p_{4,c_{T',\alpha}}\left(\zeta_{pqr}^{g}\right)~, 
\end{equation}
which implies the cohomological condition
\begin{equation}
    \mu(b_{c_{T,\alpha}})=\mu(b_{c_{T',\alpha}})~.
\end{equation}
Next, we derive cohomological conditions for the two types of trinions.
\paragraph{$\mr{fff}$ Trinion.}
For every trinion $T\in\CT$ of type $\mr{fff}$ with two external cutting curves $c^{\mr{f}}_{T,i}$, $c^{\mr{f}}_{T,j}$ and one internal cutting curve $c^{\mr{f}}_{T,\alpha}$, equations \eqref{eq:ClassS-condition1},\eqref{eq:CT'_general} and the explicit form \eqref{eq:CTmax_TN_theory} implies that 
\begin{align}
\left(\zeta^\pm_{pqr}\right)p_{3,c_{T,i}^{\mr{f}}}\left(\zeta_{pqr}^{\mr{f}}\right) p_{3,c_{T,j}^{\mr{f}}}\left(\zeta_{pqr}^{\mr{f}}\right) p_{4,c_{T,\alpha}^{g}}\left(\zeta_{pqr}^{\mr{f}}\right) &=1~,\quad N~~\text{even}~,
\\
p_{3,c_{T,i}^{\mr{f}}}\left(\zeta_{pqr}^{\mr{f}}\right) p_{3,c_{T,j}^{\mr{f}}}\left(\zeta_{pqr}^{\mr{f}}\right) p_{4,c_{T,\alpha}^{g}}\left(\zeta_{pqr}^{\mr{f}}\right) &=1~,\quad N~~\text{odd}~. 
\end{align}
This gives the cohomological condition
\begin{align}
\hspace{-0.1in}\rho_s(w_2(\IX))+\iota(\mu(b_{c_{T,i}^{\mr{f}}}))+\iota(\mu(b_{c_{T,j}^{\mr{f}}}))+\iota(\mu(b_{c^{\mr{f}}_{T,\alpha}}))&=0\in H^2(\IX,\IZ_{[2,N]})~,\quad N~~\text{even}~,
\\ 
\hspace{-0.1in}\iota(\mu(b_{c_{T,i}^{\mr{f}}}))+\iota(\mu(b_{c_{T,j}^{\mr{f}}}))+\iota(\mu(b_{c^{\mr{f}}_{T,\alpha}}))&=0\in H^2(\IX,\IZ_N)~,\quad N~~\text{odd}~,
\end{align}
where $[2,N]=\mr{lcm}(2,N)$, $\rho_s$ is the map in cohomology induced from the inclusion map $\rho_s:\IZ_2\to\IZ_{[2,N]}$, and $\iota$ is the map in cohomology obtained from inclusions in \eqref{eq:flavor_gauge_quotient_ZN}. If the trinion is bounded by two internal cutting curves $c_{T,\alpha}^{\mr{f}}, c_{T,\beta}^{\mr{f}}$ and one external cutting curve $c_{T,i}^{\mr{f}}$, the cohomological condition we get is 
\begin{align}
 \hspace{-0.1in}   \rho_s(w_2(\IX))+\iota(\mu(b_{c_{T,i}^{\mr{f}}}))+\iota(\mu(b_{c_{T,\alpha}^{\mr{f}}}))+\iota(\mu(b_{c^{\mr{f}}_{T,\beta}}))&=0\in H^2(\IX,\IZ_{[2,N]})~,\quad N~~\text{even}~,
\\ 
\hspace{-0.1in} \iota(\mu(b_{c_{T,i}^{\mr{f}}}))+\iota(\mu(b_{c_{T,\alpha}^{\mr{f}}}))+\iota(\mu(b_{c^{\mr{f}}_{T,\beta}}))&=0\in H^2(\IX,\IZ_N)~,\quad N~~\text{odd}~.
\end{align}
If the trinion is bounded by three internal cutting curves $c_{T,\alpha}^{\mr{f}}, c_{T,\beta}^{\mr{f}}$ and $c_{T,\gamma}^{\mr{f}}$, the cohomological condition we get is 
\begin{align}
\hspace{-0.1in}    \rho_s(w_2(\IX))+\iota(\mu(b_{c_{T,\alpha}^{\mr{f}}}))+\iota(\mu(b_{c^{\mr{f}}_{T,\beta}}))+\iota(\mu(b_{c_{T,\gamma}^{\mr{f}}}))&=0\in H^2(\IX,\IZ_{[2,N]})~,\quad N~~\text{even}~,
\\ 
\hspace{-0.1in} \iota(\mu(b_{c_{T,\alpha}^{\mr{f}}}))+\iota(\mu(b_{c^{\mr{f}}_{T,\beta}}))+\iota(\mu(b_{c_{T,\gamma}^{\mr{f}}}))&=0\in H^2(\IX,\IZ_N)~,\quad N~~\text{odd}~.
\end{align}
\paragraph{$\mr{ffs}$ Trinion.}
For every trinion $T\in\CT$ of type $\mr{ffs}$ with two external cutting curves $c^{\mr{f}}_{T,i},c^{\mr{s}}_{T,j}$ and one internal cutting curve $c^{\mr{f}}_{T,\alpha}$, equations \eqref{eq:ClassS-condition1}, \eqref{eq:CT'_general} and the explicit form \eqref{eq:CTmax_freehyper} implies that 
\begin{align}
\left(\zeta^\pm_{pqr}\right)\left(p_{3,c_{T,i}^{\mr{f}}}\left(\zeta_{pqr}^{\mr{f}}\right)\right)^{-1}  p_{4,c_{T,\alpha}^{\mr{f}}}\left(\zeta_{pqr}^{g}\right)p_{3,c_{T,j}^{\mr{s}}}\left(\zeta_{pqr}^{\mr{f}}\right) =1~.
\end{align}
This gives the cohomological condition
\begin{align}
    \rho_s(w_2(\IX))-\iota(\mu(b_{c_{T,i}^{\mr{f}}}))+\iota(\mu(b_{c^{\mr{f}}_{T,\alpha}}))+\iota\circ r_{n_{c_{T,j}^{\mr{s}}}}(c_1(c_j^{\mr{s}}))&=0\in H^2(\IX,\IZ_{[2,N]})~,
\end{align}
where $r_{n}$ denotes reduction $\bmod~n$. If the trinion is bounded by two internal cutting curves $c_{T,\alpha}^{\mr{f}}, c_{T,\beta}^{\mr{f}}$ and one external cutting curve $c_{T,i}^{\mr{s}}$, the cohomological condition we get is 
\begin{align}
    \rho_s(w_2(\IX))+\iota(\mu(b_{c_{T,\alpha}^{\mr{f}}}))-\iota(\mu(b_{c^{\mr{f}}_{T,\beta}}))+\iota\circ r_{n_{c_{T,i}^{\mr{s}}}}(c_1(c_{T,i}^{\mr{s}}))&=0\in H^2(\IX,\IZ_{[2,N]})~.
\end{align}
Note that in this case, we cannot have all three internal cutting curves.

\cleardoublepage
\phantomsection
\addcontentsline{toc}{section}{References}
\bibliographystyle{ytamsalpha} 
\bibliography{toptwist}

\end{document}